\documentclass{scrartcl}

\usepackage[utf8]{inputenc}
\usepackage[T1]{fontenc}
\usepackage{authblk}

\usepackage{graphicx}
\usepackage{amsmath,amssymb,amsthm}
\usepackage{esint} % double contour integral symbols and all that
\usepackage{enumitem,color}
\usepackage{comment}
\usepackage{hyperref}
\usepackage{makecell}
\numberwithin{equation}{section}

\usepackage{xstring}

\newtheorem{thm}{Theorem}

\theoremstyle{definition}

\theoremstyle{remark}
\newtheorem{rem}[thm]{Remark}

\newcommand{\fig}[2]{\includegraphics[width=#1\textwidth]{#2}}
\renewcommand{\O}{\mathcal{O}}

\newcommand{\N}{\mathbb{N}}

\newcommand{\crit}{\hbox{\tiny cr}}
\newcommand{\gs}{\gamma_{\hbox{\tiny{S}}}}
\newcommand{\gsp}{\gamma'_{\hbox{\tiny{S}}}}
\newcommand{\gst}{\gamma_{\hbox{\tiny{str}}}}
\newcommand{\cat}{{\hbox{Cat}}}
\newcommand{\SM}{\mathcal{S}}
\newcommand{\sm}{\varsigma}
\newcommand{\CM}{\mathcal{C}}

\title{Liouville Quantum Duality \\ and Random Planar Maps}
\author[$$]{Bertrand Duplantier\thanks{bertrand.duplantier@ipht.fr}}
\author[$$]{Emmanuel Guitter\thanks{emmanuel.guitter@ipht.fr}}
\affil[$$]{{\normalsize Université Paris-Saclay, CEA, CNRS, Institut de Physique
    Théorique, \par 91191 Gif-sur-Yvette,  \textsc{France} }}

\date{\today}

\begin{document}
\maketitle 
\begin{abstract} We consider models of block-weighted random planar maps in which possibly decorated
maps are decomposed canonically into blocks, each block receiving the weight $u$. These maps present a transition 
at some critical value $u=u_{\crit}$ above which the maps degenerate into Brownian trees.
We show that the enumerative properties and critical exponents of the maps at $u=u_{\crit}$ and those for 
$u<u_{\crit}$ are connected by duality relations which are precisely those expected in the context of the Liouville quantum gravity 
description of random surfaces. We illustrate this result by various instances of block-weighted maps:
random planar quadrangulations decomposed into simple blocks, Hamiltonian cycles on cubic or bicubic planar maps  
decomposed into irreducible blocks, and meandric systems.
  \end{abstract}

\section{Introduction}
\label{sec:intro}
In a recent paper \cite{FS24}, W. Fleurat and Z. Salvy exhibited an interesting phase transition for random planar maps
counted with a weight $u$ per 2-connected block, or, equivalently,  for random planar quadrangulations with a weight $u$ per simple block. These two problems are instances of the general class of \emph{block-weighted random maps} in which planar random maps, possibly decorated by some underlying statistical system, are canonically decomposed into simpler, more regular, elementary blocks, attached to each other by pinch points or small bottleneck so as to form a tree-like structure.
\begin{figure}[h!]
  \centering
  \fig{.8}{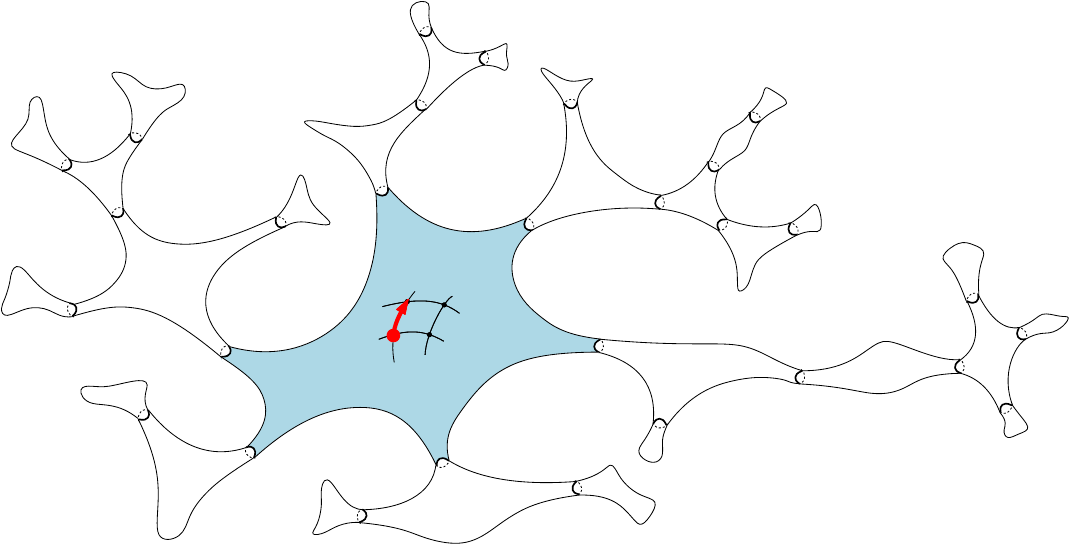}
   \caption{A schematic picture of a map decomposed into blocks separated by small bottlenecks.
   The map is made of $30$ blocks connected to each other into a tree-like structure. If the map is rooted (with a marked edge, here in red), this singularizes a root block, here in light blue.}
  \label{fig:schemaquad}
\end{figure}
Figure~\ref{fig:schemaquad} presents a schematic picture of a such a decomposition of a map. Various \emph{block decomposition} schemes were considered \cite{ZS23,ZSPhD,AFZ24}, including decomposition for maps equipped with a spanning tree and it was found
that their statistics, as obtained by assigning a weight $u$ per block, all follow the same common pattern. Consider maps of large size (or mass) $n$. The system presents a \emph{subcritical regime} at small block weight $u<u_{\crit}$ where the maps are formed of a single macroscopic block, which encloses a finite portion $\O(n)$ of the total mass, to which smaller blocks are attached, forming a tree-like structure made of small outgrowths. The size of the latter increases
with increasing $u$ to the expense of the macroscopic block size, until one reaches the \emph{critical regime} at $u=u_{\crit}$, at which the largest block has a size $\O(n^{\alpha})$ for some critical exponent $\alpha<1$. 
For $u>u_{\crit}$, one enters the \emph{supercritical regime} for which the largest block is of size $\O(\log(n))$.

\medskip
As well explained in \cite{ZSPhD}, the transition from subcritical to critical to supercritical is characterized by a number of other changes in the statistics of the random planar maps, for instance in the asymptotic law for their enumeration at large size, or in their metric properties: the typical distances are of order $n^\nu$ with different exponents $\nu=1/4$, $1/3$ and $1/2$
for the successive regimes in the case of quadrangulations \cite{FS24}. While maps in the supercritical regime mainly share the 
characteristics of a \emph{Continuum Random Tree} (CRT), a.k.a.\ Brownian tree, their statistics is much more interesting in the subcritical and critical regimes, characterized by a number of non-trivial critical exponents, depending on the possible decoration of the maps. 

\medskip
The purpose of this paper is (i) to prove that the subcritical and critical regimes in block-weighted planar maps
are related by a number of universal relations between their respective critical exponents and (ii) 
to show that these relations are exactly those predicted by the so-called \emph{duality}
in Liouville quantum gravity (LQG). 

\medskip
The study of certain natural probability measures on the space of
two-dimensional Riemannian manifolds (and singular limits of these
manifolds) is often called ``two-dimensional quantum gravity.''
These models have been very thoroughly studied in the physics
literature, starting in the mid-eighties with the modeling of discrete random surfaces
 \cite{1978CMaPh..59...35B,1986PhLB..174...87B,1985PhLB..159..303D}, followed by that of geometrical statistical models on discrete random surfaces (see \cite{1986PhLA..119..140K,
1988PhRvL..61.1433D,1989MPLA....4..217K,Daul-1995,MR1666816} for early references, and for reviews, see \cite{Ginsparg-Moore, MR1320471}). These studies were motivated in part by deep connections to  random matrix theory, as well as to string theory and
conformal field theory (CFT). Polyakov \cite{MR623209} suggested in 1981 that the summation
over random two-dimensional Riemannian metrics 
 could be represented canonically by the now celebrated \textit{Liouville theory of quantum gravity}. It was followed in 1988 by the Knizhnik-Polyakov-Zamolodchikov  (KPZ) breakthrough, relating critical exponents in the plane and on a two-dimensional random surface \cite{MR947880,MR981529,MR1005268}.
 
 \medskip
The Liouville quantum gravity random measure \cite{2008arXiv0808.1560D} comes equipped  with a parameter $\gamma$, which describes the strength of the underlying Gaussian free field (GFF). All the so-called minimal models solved in the matrix approach  and studied via LQG had a Liouville parameter $\gamma \leq 2$. This  corresponds to a central charge $c\leq 1$ of the CFT  borne by the random surface, and to a so-called \textit{string susceptibility exponent} $\gs \leq 0$, that characterizes the scaling behavior of the partition function of the random surface.
 
\medskip
Perhaps too little attention was then paid to the observations made  in Refs.  \cite{1990MPLA....5.1041D,1992PhLB..286..239J,1992PhLB..296..323K,1992MPLA....7.3081K,1993NuPhB.394..383A,1994NuPhB.426..203D,1994MPLA....9.1221A}, which studied modifications of  discrete matrix minimal models, resulting in a non-standard positive string susceptibility exponent $\gsp>0$; this happened  
at special values of coupling parameters for which the critical occurrence of bottlenecks lead to a proliferation of ``bubbles'', often called ``baby universes''', and very similar to the ``blocks'' mentioned above. A general  relation  between string susceptibility exponents,  $\gsp=1-1/(1-\gs)$  was then conjectured.

\medskip
In the sequel, the continuum Liouville CFT  was tentatively extended to values $\gamma'>2$ of the Liouville parameter in a series of works by Klebanov and coworkers   \cite{1995PhRvD..51.1836K,1995NuPhB.434..264K,1995NuPhB.440..189B,1996NuPhS..45..135K}, which defined the so-called 
``other gravitational dressing'' of the Liouville potential,  
 or ``dual branch of gravity''. In this setting, the associated $\gamma'$-LQG critical exponents were related to the Euclidean ones by a non-standard KPZ formula. However, as we shall see below,  this approach, though certainly of heuristic and predictive value, was purely \textit{formal}, because the Liouville measure used there was identically vanishing, once  properly regularized.

\medskip
A way to mathematically construct  the singular LQG measures for $\gamma'>2$, has been proposed in Refs. \cite{2009arXiv0901.0277D,2008ExactMethodsBD,pre06228485,MR4504639}. Standard $\gamma$-LQG measures 
with the \textit{dual} value of the LQG parameter, $\gamma:=4/\gamma'<2$,  are modified 
by the additional randomness of sets of point masses where  \textit{finite} amounts of  quantum area
 are \textit{localized}. The latter are ment to  result from the Riemann conformal mapping of the critical outgrowths of the random surface onto a principal component. In this setting, the string susceptibility exponent $\gsp=\gs(\gamma')$ and the quantum scaling exponents respectively obey the duality relation to $\gs=\gs(\gamma)$ and the non-standard KPZ relation seen above. 
 
  \medskip
A first study of the geometry of outgrowths (``baby universes'') in random quadrangulations was performed in \cite{Minbus}, limited however to a situation where these outgrowths remain small. (SLE) duality in Boltzmann 
maps has been studied in \cite{MR4245624}.
To date, it does not seem that the relevance of LQG duality has been thoroughly explored in the planar map setting, which is the main aim of the present  work.
More precisely, we will use the \emph{analytic combinatorics} approach, as described by Flajolet and Sedgewick in \cite{FSbook}, to relate the properties of block-weighted maps at $u=u_{\crit}$ to those for $u<u_{\crit}$.
These relations are then identified as the duality relations obtained in the LQG description
of random surfaces. In particular, block-weighted maps at $u=u_{\crit}$ provide a combinatorial 
realization of $\gamma'$-LQG in the regime $\gamma'>2$. 

\medskip
The paper is organized as follows: Section~\ref{sec:LQG} presents a rapid survey of the Liouville theory of quantum 
gravity. Section~\ref{sec:LQGhistory} gives a brief history of $\gamma$-LQG with arbitrary real parameter $\gamma$ and introduces the KPZ relations. Section~\ref{LQGGFF} provides the definition of the Liouville quantum measure in terms of the Gaussian free field.
We then study in Section~\ref{QBS} the properties of so-called quantum balls and show how to use them to measure dimensions of fractals in terms of the KPZ relations.

Section~\ref{duality} is devoted to duality in LQG. We first show in Section~\ref{gg4} how $\gamma'$-LQG for $\gamma'>2$ can be constructed from $\gamma$-LQG with $\gamma<2$ thanks to invariant properties of the measure under $\gamma \to 4/\gamma=\gamma'$, leading in particular to a ``dual version'' of the KPZ relations. Section~\ref{sec:dualmeasure} gives in particular duality relations for quantum dimensions 
or for the string suceptibility exponent. Section~\ref{sec:dualqmeasure} eventually presents a construction of the quantum measure with $\gamma'>2$ from the dual measure with parameter $4/\gamma'=\gamma$  and random sets of point masses where finite amounts of the quantum area are localized. 

We then turn in Section~\ref{sec:substitution} to general properties obtained from analytic combinatorics in a problem 
whose generating function is defined via a certain \emph{substitution scheme}.
We show in particular in Section~\ref{sec:pfasymp} how three possible regimes may emerge, judiciously named as \emph{subcritical}, \emph{critical} and \emph{supercritical}\footnote{Note that this nomenclature must be distinguished from the identical one used in \cite{MR4103975,MR4634685}, where the Liouville parameter $\gamma$ can take complex values.}, according to the dominant singularity under substitution. 
These regimes are characterized by different, but related, string susceptibility exponents for the asymptotics of the partition function 
of the underlying enumeration problem. The corresponding asymptotics for $2$-point correlators is studied in Section~\ref{sec:2point}.

Section~\ref{sec:blockweighted} aims at using the previous results in the contexts of map enumeration problems.
We show in Section~\eqref{sec:gensetting} that the above-studied substitution scheme applies precisely to 
block-weighted planar maps for which the thee subcritical, critical and supercritical regimes are reached by 
letting the weight $u$ per block lie below, at, or above some critical value $u_{\crit}$. We then deduce in
Section~\ref{sec:duality} a number of duality relations connecting the critical exponents in the
subcritical and critical regimes, which we identify as duality relations of LQG.
 
Sections~\ref{sec:quads}, \ref{sec:Hamcubic} and \ref{sec:meandres} present various examples of block-weighted
decorated maps. Sections~\ref{sec:quads} deals with the original problem in \cite{FS24} of quadrangulations with a weight $u$ per simple block. We recover in Section~\ref{sec:quadpf} the results of \cite{FS24} which we analyze in light of our general duality predictions. Section~\ref{sec:pure2p} displays our results for a particular case of $2$-point correlator, namely
maps with two marked edges. We also show in Section~\ref{sec:numcheck} how all the asymptotic 
results may be recovered numerically by a precise analysis of \emph{exact enumerations} for maps of finite size.  
We finally discuss in Section~\ref{sec:profile} the universal \emph{distance profile} between two random points \emph{in the same block}
at $u=u_{\crit}$. We show that the distance in a block scales as $n^{1/6}$ for maps of total size $n$.

Section~\ref{sec:Hamcubic} deals with cubic maps equipped with a Hamiltonian cycle and decomposed
into \emph{irreducible blocks}, a problem in bijection with that of tree-rooted maps studied in \cite{AFZ24}. 
We recover in Section~\ref{sec:irredblocks} the expected asymptotics for their partition function.
We then study two instances of $2$-point correlators: maps with two marked edges in Section~\ref{sec:HP2points}
and \emph{open} Hamiltonian paths in Section~\ref{sec:open}, for which we again test the duality relations.
We finally consider in Section~\ref{sec:bicubic} block-weighted \emph{bicubic} maps equipped with a Hamiltonian cycle,
a problem known to be in a different universality class as for cubic maps \cite{GKN99}.

We end our study by considering meandric systems in Section~\ref{sec:meandres}. We analyze in 
Section~\ref{sec:irredmeandres} meandric sytems with a weight $u$ per irreducible components and in 
Section~\ref{sec:weightq} with an additional weight $q$ per connected component. 

All the results of Sections~\ref{sec:Hamcubic} and \ref{sec:meandres} are tested against exact numerical data.
We gather in Section~\ref{sec:conclu} a few remarks on duality and Hausdorff dimensions.

\section{Liouville quantum gravity (LQG)}
\label{sec:LQG}
\subsection{A brief history of LQG}
\label{sec:LQGhistory}
Consider a  planar domain $D\subset \mathbb C$ as the parameter domain of the random
 surface, and  $h$  an instance of the (zero boundary for now) massless \textit{Gaussian free field} (GFF), a random distribution 
on $D$, associated with the  Dirichlet energy 
$(h,h)_{\nabla}:=\frac{1}{2\pi}
\int_D [\nabla h(z)]^2 dz,$ 
and whose two point correlations are given by Green's function on $D$.  
(Critical) Liouville quantum gravity consists of changing the Lebesgue area measure $dz$ on $D$ to the \textit{quantum area measure}, formally defined as 
 $\mu_\gamma(dz):=e^{\gamma h(z)} dz$, where $\gamma$ is a real parameter. The GFF $h$ is a random distribution, not a function, but the measure $\mu_\gamma$
can be constructed, for $\gamma \in [0,2)$,  
as the limit of regularized quantities.  
 
 \medskip
In Liouville conformal field theory, the complete  action  is 
$S(h)=\frac{1}{2}(h,h)_{\nabla}+\lambda\, \mathcal \mu_\gamma(D),$
where the so-called ``cosmological constant'', $\lambda \geq 0$, weights the partition function according to the area of the random manifold. The corresponding (Boltz\-mann-Gibbs) statistical weight, $\exp[-S(h)]$,  is to be integrated over with a ``flat'' uniform functional measure $\mathcal D h$ on $h$  (which makes sense \emph{a priori} for finite-dimensional approximations to $h$).  The existence of the Liouville CFT was recently rigorously established \cite{MR3465434}.

 \medskip
Liouville quantum gravity is expected to describe random surfaces meant to be the scaling limit of
random simply connected maps. For instance, the pullback, via the Riemann conformal map to the plane, of the random area measure of the infinite triangulation 
 (rooted at the origin, and  viewed as a Riemann surface) was precisely conjectured in  \cite{2008arXiv0808.1560D} to take in the scaling limit the Liouville form, $\exp[{\gamma(h(z)-\gamma \log|z|)}] dz$, where $h$ is an instance of the whole plane GFF, and 
 $\gamma^2=8/3$.  

\medskip
{A purely  combinatorial
and probabilistic (and mathematically rigorous) approach to random planar maps as random metric spaces, starting with the Cori-Vauquelin-Schaeffer bijection \cite{CoriVauquelin,schaeffer1998},  witnessed  tremendous developments. A milestone was the result by Le Gall and Miermont  that many random planar map models,  like random triangulations or quadrangulations, when equipped with the graph distance, have for scaling limit a universal random metric space, the \textit{Brownian map or sphere}  \cite{pre06199563,pre06200784}.

\medskip
In parallel, the mathematical theory of LQG, resting on the so-called \textit{mating of trees} approach \cite{DMS14} (see also \cite{MR4619311}), allowed Miller and Sheffield,  building on \cite{zbMATH06669112}, to equip the $\sqrt{8/3}$--LQG sphere with a metric space structure  \cite{MR4050102,MR4348679}, as well as the Brownian map of Refs. \cite{pre06199563,pre06200784} with a  conformal structure  \cite{MR4242633}, to finally show that these two objects have the same law. The embedding 
into the plane defined in the latter work for the Brownian map was proved to exist only abstractly, and the natural question whether it can be described in an explicit way was solved in  \cite{MR4072229,MR4448682}, whereas another approach to the conformal embedding of a random planar map was developed in \cite{MR4567714}  based on \cite{MR4390048}.

\medskip
Various values of $\gamma \in (0,2]$ are expected 
when weighting the random map by the 
partition function of a (critical) statistical physical model defined on that
map (e.g., Ising model, $O(n)$ or Potts model). Building on \cite{MR3572324}, the mating of trees framework allowed one to  establish such a convergence to LQG in the so-called  peanosphere topology \cite{DMS14}, with random paths becoming \textit{Schramm-Loewner evolution} (SLE$_\kappa$) curves of parameter $\kappa \in [0,\infty)$, with a canonical LQG--SLE correspondance  \cite{MR3551203,2010arXiv1012.4800D,DMS14}
\begin{equation}\label{LQGSLE}
\gamma^2=\kappa \wedge 16/\kappa \in (0,4].
\end{equation}

By the usual  conformal invariance \emph{Ansatz} in physics, it is natural to expect
that if one conditions on the infinite map, and then
samples the loops or clusters in these critical models (as mapped into the
plane, say), their law, in the scaling limit, will be {\em independent}
of the random measure. This independence in turn leads  to the famous \emph{KPZ formula}. 

\medskip
One of the most influential papers in this field is indeed a 1988 one by
Knizhnik, Polyakov, and Zamolodchikov \cite{MR947880} (see also \cite{MR981529,MR1005268}).  They derive a
relationship between scaling dimensions (\emph{i.e.}, conformal weights $x$) of
scaling fields defined using Euclidean geometry and analogous dimensions ($\Delta_\gamma$)
 hereafter defined via the Liouville quantum gravity measure $\mu_\gamma$ with $\gamma < 2$,  
\begin{equation}
x=\frac{\gamma^2}{4}\Delta_\gamma^2+\left(1-\frac{\gamma^2}{4}\right)\Delta_\gamma,
\label{KPZrelation}
\end{equation}
where $\Delta_\gamma$ is the positive inverse to relation (\ref{KPZrelation}) 
\begin{equation}
\Delta_{\gamma}=\Delta_{\gamma}(x):=\frac{1}{\gamma}\left(\sqrt{4x+a^2_\gamma}-a_\gamma\right),\,\,\,a_\gamma:={2}/{\gamma}-{\gamma}/{2}.
\label{InvKPZ}
\end{equation}
In the continuum limit, the statistical system is
described at criticality by a \emph{conformal field theory} (CFT), whose universality class is characterized by the so-called \emph{central charge} $c$ ($c\leq 1$).  KPZ determine $\gamma$ as \cite{MR947880,MR981529,MR1005268}
\begin{equation}
\gamma=\frac{1}{\sqrt{6} }\left(\sqrt{25-c}-\sqrt{1-c}\right) \leq 2, \,\,\, c\leq 1.
\label{gamma-c}
\end{equation}
With this parametrization, one recovers the usual form of the KPZ relation
\begin{equation}
\Delta_\gamma(x)=\frac{\sqrt{{24}{}x+1-c}-\sqrt{1-c}}{\sqrt{25-c}-\sqrt{1-c}}.
\label{InvKPZ-c}
\end{equation}

The heuristic value of this formalism was checked against manifold instances of exactly solved lattice models, 
  using the random matrix theory
 approach (e.g., in  \cite{1986PhLA..119..140K,1988PhRvL..61.1433D,1989MPLA....4..217K,
Daul-1995}); some direct comparison  to
  correlation functions calculated in Liouville field theory 
\cite{1994NuPhB.429..375D,1995PhLB..363...65T,zamolodchikov-1996-477,2001JHEP...11..044H,MR1877816} was also possible \cite{MR2057108}. The KPZ relation was further used to predict Brownian intersection exponents \cite{PhysRevLett.61.2514,MR1666816} (rigorously proved using the Schramm--Loewner Evolution (SLE) \cite{OS} 
 in \cite{MR2002m:60159a,MR2002m:60159b}), or multifractal properties of SLE \cite{2000PhRvL..84.1363D,MR2112128,2006math.ph...8053D}, rigorously established later on \cite{BS,MR3638311,gwynne2018}. A  rigorous description of the KPZ formalism for multiple SLE paths in terms of the  welding of wedges in LQG is provided in \cite[Appendix B]{DMS14}.

\medskip
A mathematical proof of the KPZ relation, based on the stochastic properties of the GFF, first appeared in  \cite{2008arXiv0808.1560D} (see also \cite{2009arXiv0901.0277D,zbMATH06797761}); it  was then also proved for multiplicative cascades \cite{2009CMaPh.tmp...46B} in the framework of  Gaussian multiplicative chaos  \cite{PSS:8474530}, as well as in other probabilistic settings \cite{BNGRV,MR4089487}.

One  important consequence is that KPZ appears to hold in a broader context than the original CFT realm (which relates
$\gamma$ to $c$), \emph{i.e.}, for any
fractal structure sampled \emph{independently} of the GFF,  and measured with the quantum random measure $\mu_\gamma$, and  for any $0\leq
\gamma <2$. 

Note that Eq. \eqref{gamma-c} only yields values of  $\gamma$ in the range $\gamma \leq 2$.  A different probabilistic approach is needed to construct measures for a Liouville parameter $\gamma'>2$ \cite{2009arXiv0901.0277D,2008ExactMethodsBD,pre06228485}; it is based on 
the \textit{duality} property of Liouville quantum gravity:  for $\gamma' > 2$, a singular (purely atomic) quantum
measure can be properly defined in terms of the $\gamma$-quantum measure, for the dual value
$\gamma=4/\gamma' <2$.  
This establishes the existence of the 
``dual version'' of the KPZ relation, as 
argued long ago in Ref. \cite{1995PhRvD..51.1836K}. 

 For the critical value $\gamma=2$, the construction of the corresponding Liouville measure $\mu_{\gamma=2}$ requires special care: it involves a so-called derivative martingale \cite{10.1214/13-AOP890}, or, equivalently, a logarithmic renormalization \cite{DRSV12}. The resulting measure is atom-free, and a KPZ relation can be proved for it.  Similar results hold for critical Mandelbrot cascades  \cite{pre06249794}.
 
 \subsection{Liouville quantum measure}\label{quantum}
\label{LQGGFF}
Let us now relate LQG and GFF. Let $h$ be an instance of a centered GFF on a
bounded simply connected domain $D$ with Dirichlet boundary conditions.
This means that $h = \sum_n \alpha_n f_n$, where the $\alpha_n$ are
i.i.d.\ zero mean unit variance normal random variables and the
$f_n$ are an orthonormal basis with respect to the Dirichlet inner product,
$
(f_1, f_2)_\nabla := \frac{1}{2\pi} \int_D \nabla f_1(z) \cdot \nabla f_2(z) dz,
$
of the Hilbert space closure $H(D)$ of the space $H_s(D)$ of
$C^\infty$ real-valued functions compactly supported on $D$.
This sum diverges pointwise a.s., but it does converge
a.s. in the space of distributions on $D$, and one can also
make sense of the mean value of $h$ on various sets \cite{MR2322706}. 
Special care is then required
to make sense of the quantum gravity measure, because the GFF is a distribution
and not a function (it typically oscillates between $\pm \infty$).

Given an instance $h$ of the Gaussian free field on $D$, let 
$h_\varepsilon(z)$ denote the mean value of $h$ on the circle of
radius $\varepsilon$ centered at $z$ (where $h(z)$ is defined to be
zero for $z \in \mathbb C \setminus D$).   The regularized Liouville quantum measure is defined as \cite{2008arXiv0808.1560D}

\begin{equation}
\label{liouvillemeasure}
\mu_{\gamma,\varepsilon}(dz):=\varepsilon^{\gamma^2/2} e^{\gamma h_\varepsilon(z)}dz,
\end{equation}
in a way reminiscent of so-called Wick normal ordering
 (see also Ref. \cite{MR0292433} 
 for earlier work on the so-called H{\o}egh-Krohn model). 
 One can show that for $\gamma \in [0,2)$
it is
(almost surely) the case that as $\varepsilon \to 0$, the measures \eqref{liouvillemeasure}
converge weakly to a limiting 
non-degenerate measure,  denoted by $\mu_{\gamma} (dz):= e^{\gamma h(z)}dz$.

For each $z \in D$, denote now by $C(z;D)$ the \emph{conformal radius} of $D$
viewed from $z$.  That is, $C(z;D) = |\phi'(z)|^{-1}$ where $\phi:D
\to \mathbb D$ is the conformal map to the unit disc with $\phi(z) = 0$.
A  geometrical analysis of GFF properties \cite{2008arXiv0808.1560D} yields   
 the variance 
 $
 \hbox{Var} \, h_\varepsilon (z)=\log [C(z; D)/\varepsilon].
$
By standard expectation of the exponential of a (centered) Gaussian variable, we then have 
$\mathbb E\, e^{\gamma h_\varepsilon(z)}=e^{ \gamma^2{\hbox{\footnotesize Var}}[
h_\varepsilon(z)]/2}
= \left[ {C(z;D)}/{\varepsilon} \right]^{\gamma^2/2}.$
The expectation of the quantum measure \eqref{liouvillemeasure} 
is therefore $\mathbb E\, \mu_{\gamma,\varepsilon}(A) = \int_A C(z;D)^{\frac{\gamma^2}{2}}dz=\mathbb E\, \mu_{\gamma}(A)$,
independently of $\varepsilon$, and for each Borel subset $A \subset D$.

In the $\gamma=2$ case, it has been shown that the properly regularized measure is
\begin{equation}
\label{liouvillemeasure2}
\mu_{\gamma=2,\varepsilon}(dz):=\sqrt{\log \left({1}/{\varepsilon}\right)}\varepsilon^{2} e^{\gamma h_\varepsilon(z)}dz,
\end{equation}
which weakly converges as $\varepsilon \to 0$ to an atom-free non-vanishing mesure  \cite{DRSV12}.

\subsection{Quantum balls, scaling exponents and KPZ}
\label{QBS}
For the quantum area measure $\mu_\gamma$ on $D$, we let $B^\delta(z)$ be the Euclidean ball $B_\varepsilon(z)$ centered at $z$
whose radius $\varepsilon$ is chosen so that $\mu_\gamma(B_\varepsilon(z)) = \delta$.  (If
 not  unique,  take the
radius to be $\sup \{ \varepsilon: \mu_\gamma(B_\varepsilon(z)) \leq \delta
\}$.)
We refer to $B^\delta(z)$ as the \emph{quantum ball}
of area $\delta$ centered at $z$. In particular, if $\gamma = 0$
then $\mu_0$ is Lebesgue measure and $B^\delta(z)$ is
$B_{\varepsilon}(z)$ where $\delta = \pi \varepsilon^2$.
 
When $\varepsilon$ is small,
the limiting quantum measure, $\mu_{\gamma}(B_{\varepsilon}(z))$, of the Euclidean ball $B_{\varepsilon}(z)$, is very well
approximated by the simple form
\begin{equation}
\label{mu}
\mu_{\gamma\odot}(B_{\varepsilon}(z)):=\varepsilon^{2+\gamma^2/2} e^{\gamma h_\varepsilon(z)},
\end{equation}
so that, in expectation, $\mathbb E[\mu_{\gamma}(B_{\varepsilon}(z))|h_\varepsilon(z)]\sim \pi \mu_{\gamma\odot}(B_{\varepsilon}(z))$ for $\varepsilon \to 0$. In this simplified perspective,
one views $\mu_{\gamma\odot}$ as a function on balls, defined by
(\ref{mu}), rather than the fully defined measure on $D$.
It will be proved to be convenient to rewrite it as
\begin{eqnarray}
\label{Q}
\mu_{\gamma\odot}(B_{\varepsilon}(z))=\varepsilon^{\gamma Q_\gamma} e^{\gamma h_\varepsilon(z)},\quad Q_\gamma= 2/\gamma+\gamma/2.
\end{eqnarray}
By similarity to the definition of quantum balls $B^\delta$, one  defines the  quantum ball $\tilde B^\delta(z)$
 centered at $z$  as the (largest) Euclidean ball $B_{\varepsilon}(z)$
such that, 
\begin{equation}
\label{delta}
\mu_{\gamma\odot} (B_\varepsilon(z)) = \delta. 
\end{equation}
Pairs $(z,h)$ are then sampled under the law
\begin{eqnarray}
e^{\gamma h_\varepsilon(z)} dh dz=\exp{\left[{-\frac{1}{2}(h,h)_\nabla +\gamma h_\varepsilon(z)}\right]} \mathcal D h dz,
\label{randommetricbis}
\end{eqnarray}
where
$dh$ represents the whole GFF measure and $\mathcal D h$ the flat functional measure. 
\begin{figure}[tb]
\begin{center}
\includegraphics[angle=0,width=.573290\linewidth]{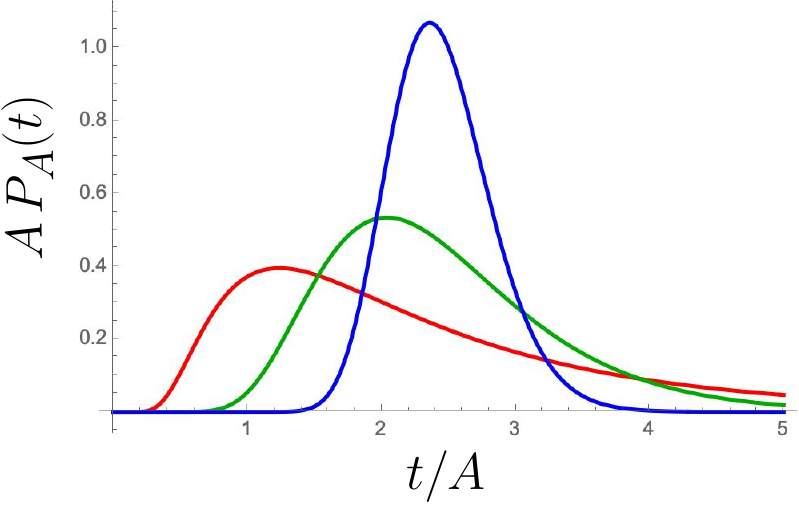}
\caption{Distribution \eqref{PA} of $t=-\log \varepsilon_A$ getting sharper  for increasing values of $A=-(\log \delta)/\gamma$, $A=5$ (red), $20$ (green) and $100$ (blue).}
\label{f.tphi3ter}
\end{center}
\end{figure}

For a given $\delta <1$, let us define in logarithmic coordinates $A:=-(\log \delta)/\gamma$, and $t:=-\log \varepsilon_A$, where $\varepsilon_A$ is the \emph{random} Euclidean radius of the quantum ball $\tilde B^\delta(z)=B_{\varepsilon_A}(z)$. 
  The probability density  $P_A(t)$ 
yields access to geometrical properties of the Liouville quantum measure.  
It has the explicit form  \cite{2009arXiv0901.0277D,2008arXiv0808.1560D}
\begin{equation} \label{PA} P_A(t)=\frac{A}{\sqrt{2\pi t^3}} \exp\left[-\frac{1}{2t}\left({A}
-a_\gamma {t}\right)^2\right],\quad a_\gamma=\frac{2}{\gamma}-\frac{\gamma}{2}. 
\end{equation}
For  $A\to +\infty$, \emph{i.e.},  $\delta \to 0$,
the distribution \eqref{PA}  gets localized,  as illustrated in Fig. \ref{f.tphi3ter},  and yields the typical scaling of quantum balls in $\gamma$-Liouville quantum gravity,  
$\ \delta \asymp \varepsilon^{\gamma a_\gamma}=\varepsilon^{2-\gamma^2/2}$ (see also Ref. \cite{2009arXiv0902.3842H}).
  Its integral equals
% \begin{eqnarray} \nonumber \label{intPA} \int_0^\infty dt P_A(t)&=&\exp\left[-{A}(|a_\gamma|-a_\gamma)
%\right]\\ 
%&=&\left\{
%\begin{matrix}
%1,& a_\gamma \geq 0\\
%&&\\
%\exp\left(2{A}\,a_\gamma\right)= \delta^{1-4/\gamma^2} < 1,& a_\gamma < 0. \label{intPA<1}\\
%\end{matrix}
%\right.
%\end{eqnarray}
\begin{eqnarray} \nonumber \label{intPA} \int_0^\infty dt P_A(t)&=&\exp\left[-{A}(|a_\gamma|-a_\gamma)
\right]\\ \nonumber \label{intPA1}
&=&1,\,\,\,\,\,\,\,\,\,\,\,\,\,\,\,\,\,\,\,\,\,\,\,\,\,\,\,\,\,\,\,\,\,\,\,\,\,\,\,\quad\quad\quad\quad\quad a_\gamma \geq 0\\
\label{intPA<1}
&=&\exp\left(2{A}\,a_\gamma\right)= \delta^{1-4/\gamma^2} < 1,\,\,\,\, a_\gamma < 0.
\end{eqnarray}
The first case, $a_\gamma \geq 0$, corresponds to $\gamma \in[0,2]$, so that $P_A$ is a normalized probability density, as expected. 
The interesting non-normalizable case  $a_\gamma < 0$ \eqref{intPA<1} for $\gamma >2$ will be reinterpreted in Section \ref{duality}. (See Fig. \ref{f.tphi4quater}.) 
\begin{figure}[tb]
\begin{center}
\includegraphics[angle=0,width=.573290\linewidth]{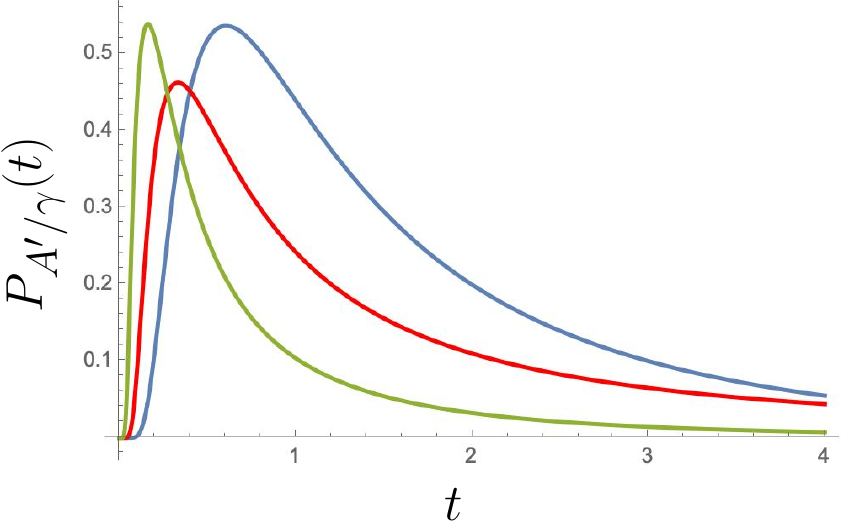}
\caption{
Distributions \eqref{PA} for  dual values $\gamma=\sqrt{2}$ (blue) and $\gamma=\sqrt{8}$ (green), and for self-dual  $\gamma=2$ (red),  with a common parameter $A'=-\log \delta=2$. The $\gamma=\sqrt{8}$ case does not integrate to a unit probability.}
\label{f.tphi4quater}
\end{center}
\end{figure}

\medskip
Recall that a (deterministic or random) fractal subset $X$ of
$D$ has Euclidean scaling exponent $x$ (and {Euclidean dimension} $2-2x$) if,
for $z$ chosen uniformly in $D$ and independently
of $X$, the probability
${\mathbb P}\{B_\varepsilon(z) \cap X \not = \emptyset \} \asymp \varepsilon^{2x},$
in the sense
 that
 $\lim_{\varepsilon \to 0} {\log {\mathbb P}
}/{\log \varepsilon} =
2x.$

The same fractal $X$ has \textit{quantum scaling exponent $\Delta$} if when
$X$ and $(z,h)$, sampled with weight (\ref{randommetricbis}), are chosen independently, we have
\begin{equation}
\label{quantumP}
{\mathbb P}\{\tilde B^\delta(z) \cap X \not = \emptyset \}\asymp \delta^{\Delta},\quad \delta \to 0.
\end{equation}
Let $N_\gamma(\delta, X)$ be the
number of $(\mu_{\gamma\odot}, \delta)$ balls intersected by a fractal $X$  
and
$N_0(\varepsilon^2, X)$ the number of balls intersecting $X$
that have radius $\varepsilon$, \emph{i.e.}, Euclidean area $\pi \varepsilon^2$.  
Then  we have the equivalent scaling dimension definitions, 
$$\lim_{\varepsilon \to 0} { \log \mathbb E [N_0(\varepsilon^2, X)] }/{\log \varepsilon^2} = x-1,$$
for the Euclidean exponent $x$, and 
\begin{equation}\label{Ndelta}
\lim_{\delta \to 0} { \log \mathbb E[N_\gamma(\delta, X)] }/{\log \delta} = \Delta-1,
\end{equation}
for the quantum scaling exponent $\Delta$.

\medskip
We then have the following KPZ result \cite{2008arXiv0808.1560D}: 
Fix $\gamma \in [0,2)$ and a compact subset $\tilde D$ of $D$ to  avoid boundary effects.  If
$\tilde X=X \cap \tilde D$ has Euclidean scaling exponent $x \geq 0$ then it
has quantum scaling exponent $\Delta=\Delta_\gamma(x)$, where $\Delta_\gamma$ is the
non-negative solution \eqref{InvKPZ} to \eqref{KPZrelation}. In a nutshell, the reason is that under the law (\ref{randommetricbis}) one has the chain of identities \cite{2008arXiv0808.1560D} 
\begin{equation}
\label{quantumP'}
{\mathbb P}\{\tilde B^\delta(z) \cap \tilde X \not = \emptyset \}={\mathbb P}\{B_{\varepsilon_A}(z) \cap \tilde X \not = \emptyset \}={\mathbb E}[\varepsilon_A^{2x}]\asymp \delta^{\Delta_\gamma},\quad \delta \to 0.
\end{equation}
This relation still holds in the $\gamma=2$ case for the measure \eqref{liouvillemeasure2} \cite{DRSV12}. 

\section{Liouville quantum duality}
\label{duality}
\subsection{$\boldsymbol{\gamma\gamma'=4}$  duality}
\label{gg4}
The preceding section described $\gamma$-LQG for $\gamma\leq 2$. Liouville quantum gravity ${\gamma'}-$LQG is expected to describe for $\gamma'>2$ random surfaces meant to be the scaling limit of
random simply connected maps with large amounts of area cut off
by small bottlenecks. As first observed in Refs.  \cite{1990MPLA....5.1041D,1992PhLB..286..239J,1992PhLB..296..323K,1992MPLA....7.3081K,1993NuPhB.394..383A,1994NuPhB.426..203D,1994MPLA....9.1221A}, there indeed exist discrete matrix models where  a non-standard string susceptibility exponent $\gsp>0$ appears; this happens precisely 
at special values of the coupling parameters for which the critical occurrence of bottlenecks leads to a proliferation of ``bubbles'', also called ``baby universes''. 
   The corresponding continuous surface is a tree-like foam of
Liouville quantum bubbles of \textit{dual} parameter
\begin{equation}
\label{gamma'}
\gamma :=4/\gamma', \,\,\,\gamma <2< \gamma', 
\end{equation}
 connected to each other at ``pinch
points'' and rooted at a ``principal bubble'' confor\-mally  parameterized by a domain $D$. 

\medskip
The Liouville measure  was first formally considered for $\gamma'>2$ in Refs.  \cite{1995PhRvD..51.1836K,1995NuPhB.434..264K,1995NuPhB.440..189B,1996NuPhS..45..135K}, in the so-called 
``other gravitational dressing'' of the Liouville potential,  
 or ``dual branch of gravity'', leading in particular to a \emph{dual} version of the KPZ relation.
 However, the limit of the regularized measure \eqref{liouvillemeasure} actually \emph{vanishes} for $\gamma'\geq 2$:  $\lim_{\varepsilon \to 0} \mu_{\gamma' \geq 2,\varepsilon}(z)=0$. This is a quite general phenomenon, first observed by Kahane  in the eighties \cite{MR829798} in the case of the so-called \emph{Gaussian multiplicative chaos} (GMC), inspired by  Mandelbrot's cascade model.  A precise mathematical description of Liouville duality therefore requires additional probabilistic machinery \cite{2009arXiv0901.0277D,2008ExactMethodsBD,pre06228485}, to be described below.  
 We first relate $\gamma'$ to $\gamma$ \emph{heuristically}, following Ref. \cite{2009arXiv0901.0277D}, by using the regularized measure \eqref{liouvillemeasure}. 

\medskip
The definitions of the (pseudo)  measure \eqref{mu} on balls and of quantum balls \eqref{delta} still make 
sense for $\gamma'>2$.  Observe that 
$Q_\gamma$ in \eqref{Q} is \emph{invariant} under duality, so that one has  the remarkable duality relation  \begin{equation}
 \label{mugammagamma'}
\mu_{\gamma\odot }^{1/\gamma}=\mu_{\gamma'\odot }^{1/\gamma'},\,\,\,\gamma\gamma'=4, 
\end{equation} 
 \emph{i.e.}, a
$\gamma'$-quantum ball $\tilde B^{\delta'}(z)$ \eqref{delta} of size $\delta'$ has $\gamma$-quantum size
$\delta={\delta^{\,\prime}}^{\, 4/\gamma'^2}.$ 
  (Intuitively, the ball
contains about a $\delta$ fraction of the total $\gamma$-quantum
area, but only a $\delta' < \delta$ fraction of the $\gamma'$-quantum
area because the latter also includes
points on
non-principal bubbles.) The expected number of $\gamma$-quantum size-$\delta$ balls (\emph{i.e.}, $\gamma'$-quantum size-$\delta'$ balls)
 needed to cover the principal bubble $D$ thus scales as
$N_{\gamma}(\delta,D)\asymp \delta^{\,-1} = {\delta^{\,\prime}}^{-4/\gamma'^2}.$ 

\medskip
If a fractal random subset $X\subset D$ has Euclidean scaling
exponent $x$, it has by KPZ  a quantum scaling exponent $\Delta_{\gamma}$  for the \emph{standard}  $\gamma<2$ Liouville measure. This essentially says  
by \eqref{Ndelta} that the expected number
$N_{\gamma}(\delta, X)$ of $\gamma$-quantum size-$\delta$ balls (\emph{i.e.}, number
$N_{\gamma'}(\delta', X)$ of $\gamma'$-quantum size-$\delta'$ balls)
required to cover $X$, scales as 
\begin{eqnarray} \label{quantumP4}
 {N_{\gamma}(\delta, X)} \asymp \delta^{\Delta_{\gamma}-1}= {\delta^{\,\prime}}^{\, \Delta_{\gamma'}-1} \asymp  N_{\gamma'}(\delta',X) ,
\end{eqnarray}
where we define for $\gamma' >2$ the \emph{dual} quantum exponent $\Delta_{\gamma'}$  by 
 \begin{eqnarray}
 \label{dualdim}\gamma'(\Delta_{\gamma'} - 1) :=
\gamma(\Delta_{\gamma}-1), \,\,\, \gamma\gamma'=4.
\end{eqnarray}
 By hypothesis, the triple $(\gamma,x,\Delta_{\gamma})$ satisfies  the KPZ relation (\ref{KPZrelation}),  valid for $\gamma<2$, and it is easy to see that 
 the \textit{dual} triple $(\gamma',x,\Delta_{\gamma'})$ does also for $\gamma'>2$, with the further duality property  $x=\Delta_{\gamma}\Delta_{\gamma'}$ \cite{MR2112128}.  One can also check that the dual quantum exponent $\Delta_{\gamma'}$ coincides with the continuation of  \eqref{InvKPZ}  
to the range $\gamma' >2$. Notice however, that for $\gamma'> 2$ the coefficient  $a_{\gamma'}$ in \eqref{InvKPZ}   
is \emph{negative}. As a consequence, the dual versions of the conformal field theoretic KPZ expressions \eqref{gamma-c} and \eqref{InvKPZ-c} in terms of the central charge $c$ are \begin{align}\label{dualKPZc}
\gamma'&=\left(\sqrt{25-c}+\sqrt{1-c}\right)/\sqrt{6}\,\, \geq 2, \,\,\, c\leq 1,\\ \nonumber
\Delta_{\gamma'}&=\frac{\sqrt{{24}{}x+1-c}+\sqrt{1-c}}{\sqrt{25-c}+\sqrt{1-c}}.
\end{align}
These dual quantum exponents appear in Refs. \cite{1995PhRvD..51.1836K,1995NuPhB.434..264K}.
 They were independently introduced in the natural context of \emph{boundary} exponents associated with the
\textit{dense phase} of the so-called $O(n)$ model, \emph{i.e.}, with the chordal Schramm-Loewner Evolution $\textrm{SLE}_{\kappa}$ for $\kappa > 4$ \cite{MR2112128}.  In the dense critical phase, the \emph{dual} boundary quantum exponents obey \textit{additivity} laws for mutually-avoiding paths, whereas  in the dilute phase, the \emph{standard} quantum boundary exponents are those that are additive \cite{MR2112128}. 

Observe finally that the well-known SLE \emph{duality}, $\kappa\kappa'=16$ \cite{2000PhRvL..84.1363D,MR2112128,MR2439609,dub_dual}, which describes the outer boundaries of non-simple SLE$_{\kappa'\geq 4}$ curves as simple SLE$_{\kappa\leq 4}$ ones, is in direct relation to LQG duality. Indeed, by the canonical LQG--SLE correspondence \eqref{LQGSLE}, the Liouville $\gamma\gamma'=4$ duality yields the dual relations,
\begin{equation} 
\gamma^2=\kappa\wedge 16/\kappa,\quad \gamma'^{\,2}=\kappa\vee 16/\kappa. 
\end{equation}}

\subsection{Dual measure}
\label{sec:dualmeasure}
If we choose the pair
$(z,h)$ from the weighted measure $\mu_{\gamma',\varepsilon}(dz)dh$ as in
(\ref{randommetricbis}), we can repeat for $\gamma'>2$ the analysis of Section \ref{QBS} 
 of $\tilde B^{\delta'}(z)$ balls \eqref{delta}; however one finds that for $a_{\gamma'} <0$,
 $\varepsilon_A=0$ a.s. for large $A=-(\log \delta')/\gamma'$ \cite{MR2112128}. The weighted measure is thus \emph{
singular}, \emph{i.e.}, there is a quantum area of at least $\delta'$
\textit{localized} at point $z$ for small enough $\delta'$.

We may define a
$\delta'$\textit{-regularized}
 measure 
 by restricting to the event $\varepsilon_A >0$, which means that
  there is \textit{strictly less than  $\delta'$ {localized} at  $z$}. With this conditioning, the 
 result  for the intersection probability \eqref{quantumP'} still holds for $\gamma' >2$, 
\begin{equation}\label{fullexpmartingalekpz'}
{\mathbb E}[\varepsilon_A^
{2x} 1_{\varepsilon_A>0}]= {\delta^{\,\prime}}^{\Delta_{\gamma'}},
\end{equation}
with the \emph{dual} exponent $\Delta_{\gamma'}$,  continuation of \eqref{InvKPZ} to $\gamma'>2$. This gives in particular for $x=0$,
\begin{equation}\label{Delta0}
\Delta_{\gamma'}(x=0)
=1-4/\gamma'^2,
\end{equation} 
which yields for $\gamma' >2$ the
probability for a given point $z$ to be $\delta$-regular, 
${\mathbb E}[1_{\varepsilon_A>0}]={\delta^{\,\prime}}^{\,1-4/\gamma'^2}=
\delta'/\delta,$ 
recovering \eqref{intPA<1} and 
 vanishing when $\delta'\to 0$. 
The ratio of  the expectation (\ref{fullexpmartingalekpz'}) to  
this probability then gives the $\delta'$-regular conditional   
 version of  probability  \eqref{quantumP}.  
  It indeed scales as 
${\delta^{\,\prime}}^{\, \Delta_{\gamma'}}\times({\delta}/{\delta'})=
\delta^{\Delta_{\gamma}},$ \emph{i.e.}, 
   with now the \textit{standard} quantum exponent $\Delta_{\gamma}$.
   
 \paragraph{Duality and string susceptibility exponents}
 In  Liouville field theory, the asymptotic behavior of the partition function of a random surface of fixed area $\mathcal A$ is given by $\mathcal Z \propto e^{\lambda \mathcal A} {\mathcal A}^{\gs-3}$, 
with the so-called cosmological constant $\lambda$ and string susceptibility exponent $\gs$, given for $\gamma \leq 2$ by 
   \cite{Ginsparg-Moore,MR1320471}
\begin{eqnarray}
\label{string} \gs=2-2Q_\gamma/\gamma=1-4/\gamma^2.
\end{eqnarray}
 The latter thus obeys the duality relation  \cite{1995PhRvD..51.1836K,MR2112128}
 \begin{eqnarray}
\label{dualstring}  (1-\gs)(1-\gsp)=1,
\end{eqnarray}
to the dual suceptibility exponent,
 \begin{equation}\label{dualgammastring}
 \gsp=1-4/{\gamma'}^{\, 2},
 \end{equation}
with
\begin{equation}\nonumber
\gs<  0<\gsp,\,\,   \gamma <2<\gamma'=4/\gamma. 
\end{equation}
In terms of the central charge $c\leq 1$, the string susceptibility exponent  and its dual take the form
\begin{align}\nonumber
\gs&=\frac{1}{12}\left(c-1-\sqrt{(1-c)(25-c)}\right)\,\, \leq 0,\\  \label{dualgammaS}
\gsp&=\frac{1}{12}\left(c-1+\sqrt{(1-c)(25-c)}\right)\,\, \geq 0\,.
\end{align}
 Note finally that the duality equation \eqref{dualdim} for quantum scaling exponents can simply be recast as \cite{MR2112128} 
\begin{equation} \label{dualdimbis}
\Delta_{\gamma'}=\frac{\Delta_{\gamma}-\gs}{1-\gs},
\end{equation} 
or, by involution, as
\begin{equation} \label{dualdimter}
\Delta_\gamma=\frac{\Delta_{\gamma'}-\gsp}{1-\gsp}.
\end{equation} 
Notice that for $\gamma<2$, one has for a vanishing Euclidean scaling exponent $x=0$, $\Delta_{\gamma}(x=0)=0$, hence from \eqref{dualdimter} $\Delta_{\gamma'}(x=0)=\gsp$, in agreement with \eqref{Delta0} and \eqref{dualgammastring}.

\subsection{Construction of the dual quantum measure}
\label{sec:dualqmeasure}
A way to rigorously construct  the singular quantum measures $\mu_{\gamma'>2}$, which would reproduce all the scaling properties seen above,  has been presented in Refs. \cite{2009arXiv0901.0277D,2008ExactMethodsBD,pre06228485}. In the first approach,
one uses the dual measure $\mu_{\gamma<2}$
and the additional randomness of sets of point masses where  \textit{finite} amounts of  quantum area
 are \textit{localized}.  
More precisely, conditionally on the  quantum measure $\mu_{\gamma}$, with $\gamma=4/\gamma' <2$, one considers a Poisson random measure $\mathcal N_{\gamma'}(dz,d\eta)$ distributed on $D\times (0,\infty)$,  of intensity 
$\mu_{\gamma}(dz) \times \Lambda^{\alpha'}(d\eta)$, where 
$\Lambda^{\alpha'}(d \eta) := d\eta/\eta^{1+\alpha'}, \,\,\,\alpha':=4/\gamma'^2 =\gamma^2/4 \in (0,1),$ letting each point $(z,\eta)$ represent an atom of size $\eta$ located at $z$. 
The dual measure for $\gamma'>2$ is then \emph{purely atomic},
\begin{equation}\label{dualmeasure}
\mu_{\gamma'}(dz):=\int_0^{\infty}\eta\, \mathcal N_{\gamma'} (dz,d\eta),\,\, \gamma' >2.
\end{equation}
From this follows, for  any Borelian  $A\subset D$, the characteristic Laplace transform, 
\begin{equation}\label{Laplace}\mathbb E\exp[-\lambda\,\mu_{\gamma'}(A)]=\mathbb E\exp\left[\Gamma(-\alpha')\lambda^{\alpha'}\mu_{\gamma}(A)\right],
\end{equation}
 a L\'evy-Khintchine formula valid for all $\lambda \in \mathbb R_+$, where $\Gamma(-\alpha')=-\Gamma(1-\alpha')/\alpha' <0$ is the usual Euler $\Gamma$-function. 
  From \eqref{Laplace} follows the relation between  moments  of  dual measures 
 $$
 \mathbb E\left[\big(\mu_{\gamma'}(A)\big)^p\right]=\frac{\Gamma(1-p/\alpha')[-\Gamma(-\alpha')]^{p/{\alpha'}}}{\Gamma(1-p)}\mathbb E\left[\big(\mu_{\gamma}(A)\big)^{p/{\alpha'}}\right],\,\,\,p<\alpha=4/{\gamma'}^2,
 $$
showing
 the characteristic  scaling of $\mu_{\gamma'}$
 as $\mu_{\gamma}^{1/\alpha'}$, in agreement
with  (\ref{mugammagamma'}). 
In Ref.\cite{pre06228485}, a slightly different (and more general) construction was proposed, which the authors called ``atomic Gaussian multiplicative chaos''. The random measure there constructed  
and the dual measure $\mu_{\gamma>2}$ \eqref{dualmeasure} then obey the same law \eqref{Laplace}  in terms of $\mu_{\gamma'}$,  thus coincide, even though the two constructions a priori differed. A form of the dual KPZ relation \eqref{KPZrelation} for $\gamma >2$ (in the Hausdorff dimension sense)  is then proved in Ref. \cite{pre06228485} for the atomic GMC and for dual exponents defined exactly as in Eq. \eqref{dualdim}. 

\section{Substitution and asymptotics} 
\label{sec:substitution}
Duality is intimately linked to the \emph{branching structure} of maps which allows us
to preserve the connected nature of a map while cutting off large amounts of area enclosed in outgrowths
(the, possibly large, baby universes). At the combinatorial level,
this branching structure can be made fully explicit by considering a \emph{block decomposition of the maps}.
Indeed, many families of, possibly decorated, planar maps have a canonical decomposition into \emph{elementary blocks}, which are more regular (\emph{e.g.} $2$-connected, simple, irreducible, ...) 
decorated submaps attached to each other by small bottlenecks or attaching points so as to form a tree-like pattern; see Figure~\ref{fig:schemaquad} for an illustration. 
The statistics of maps therefore results from the superposition of two critical phenomena, one describing the elementary blocks, with its own statistics, and the other dealing with the tree-like arrangement of these blocks. 

More precisely, the elementary blocks form a subfamily of possibly decorated maps, characterized by their generating function
\begin{equation}
\phi(\tau)=\sum_{i\geq 0} \phi_i \tau^i\ ,
\label{eq:phioftau}
\end{equation}
where $\tau$ is a fugacity per unit block size. Here, the quantity $\phi_i$ is the number of elementary block
configurations of size $i$. In particular, the large $i$ asymptotics of $\phi_i$ is encoded in the \emph{singularity} of 
$\phi(\tau)$ as $\tau$ approached some critical value $\tau_\phi$. 

As for the tree describing the arrangement of blocks, it is characterized by its generating function 
\begin{equation}
y(z)=\sum_{n\geq 1} y_n z^n\ ,
\label{eq:yofz}
\end{equation}
where $z$ is now a fugacity per block, \emph{i.e.}, $y_n$ is the number of tree configurations with $n$ attached blocks (including ``blocks of size $i=0$'' corresponding to potential attaching points where no block of positive size $i>0$ is attached). 
If the average number of outgrowths attached to a single block exceeds a certain value, the tree becomes critical, \emph{i.e.}, its size ($=$ number of blocks) becomes arbitrarily large. This critical behavior is now described by the singularity of $y(z)$ when $z$ approaches some critical value $z_c$. 
If the number of possible attaching points on a block is equal or proportional to its size, the generating functions $y(z)$ and $\phi(\tau)$ are linked by a simple \emph{substitution relation} (see below). At the critical point $\tau=\tau_\phi$, the tree may then be below, at or above criticality. 

While Liouville quantum gravity with $\gamma<2$ describes the case \emph{below criticality}, with small outgrowths attached to a block of macroscopic size, ${\gamma'}-$LQG with $\gamma'>2$ describes instead the case \emph{at criticality}, where the size of the largest block is only a vanishing fraction of the total size of the map. 
The duality relations \eqref{dualstring}, \eqref{dualdimbis} and \eqref{dualdimter}, that we saw in Section \ref{duality}, connect these two situations and can be recovered combinatorially, 
as we shall now see by analyzing the link between the singular behaviors of $\phi$ and $y$ resulting from the substitution relation.

\subsection{Generating function singularities and partition function asymptotics}
\label{sec:pfasymp}
Following \cite{FSbook}, we consider a combinatorial problem, with ordinary generating function
$y(z)$ as in \eqref{eq:yofz} above, which can be related to some other combinatorial problem, with generating fonction
$\phi(\tau)$ as in \eqref{eq:phioftau}, via a simple \emph{substitution} procedure. By this, we mean that we may write
\begin{equation}
y(z)=z\, \phi(y(z))\ .
\label{eq:yphi}
\end{equation}
By a simple expansion in powers of $z$, we see that $y(z)$ enumerates planted plane trees with a weight $z$ per
vertex of arbitrary degree and a weight $\phi_i$ per vertex with \emph{descending degree} (number of children) $i$. 
The vertices here are supposed to describe the blocks in the block decomposition of a map 
(including possible blocks of size $i=0$), where a block of size $i\geq 0$ gives rise to $i$ attached descending 
blocks in the tree structure formed by the blocks.  
As discussed in  \cite{FSbook}, we may now look for a \emph{critical point} of the tree generating function $y(z)$, \emph{i.e.}, a value $z= z_c$ for which the mapping $z\mapsto y$ is not locally invertible , namely when
\begin{equation*}
1= z_c\, \phi'(y(z_c))\ .
\end{equation*}
Setting $y(z_c)=y_c$, a critical point $(z_c,y_c)$  is thus characterized by the two identities
\begin{align} 
\label{eq:ycrit}
& \phi(y_c)=y_c\, \phi'(y_c)\ , \\
\label{eq:ycrit2}
& z_c=\frac{y_c}{\phi(y_c)}\ .
\end{align}
This in general selects at most one single \emph{combinatorial} critical point $(z_c,y_c)$ which is such that the mapping
$z\mapsto y$  is invertible (and strictly increasing) for $z\in [0,z_c)$. 

\medskip
In this paper, we furthermore assume that 
\begin{equation*}
\phi_i\propto \tau_\phi^{-i}/i^{1+\alpha}
\end{equation*}
at large $i$, with
\begin{equation}
1<\alpha<2\ ,
\label{eq:alpharange}
\end{equation}
so that the mapping $\tau\mapsto \phi$ is itself singular at $\tau=\tau_\phi$, with the following expansion for $\tau\to \tau_\phi^{-}$:
\begin{equation}
\phi(\tau)=\phi(\tau_\phi)-(\tau_\phi-\tau)\phi'(\tau_\phi)+K(\tau_\phi-\tau)^\alpha+ o((\tau_\phi-\tau)^\alpha)\ .
\label{eq:phiexp}
\end{equation}
We may now encounter three situations.

\paragraph{Supercritical case.} Equation \eqref{eq:ycrit} has a combinatorial solution
with $y_c<\tau_\phi$. In this case, letting $z$ increase from $0$, we reach the singularity at $y=y_c$ before 
encountering the singularity of $\phi$. Since $\phi$ is regular at $y_c$, we may write
\begin{equation*}
z_c-z=\frac{y_c}{\phi(y_c)}- \frac{y}{\phi(y)}=: H(y) 
\end{equation*}
with
\begin{equation*}
H(y_c)=0\ , \quad H'(y_c)=-\frac{1}{\phi(y_c)}+y_c \frac{\phi'(y_c)}{\phi(y_c)^2}=0
\end{equation*}
so that 
\begin{equation*}
z_c-z=\frac{1}{2} H''(y_c)(y_c-y)^2 +o((y_c-y)^2)\ ,
\end{equation*}
namely
\begin{equation*}
y=y_c-\left(\frac{2}{H''(y_c)}(z_c-z)\right)^{1/2}+o((z_c-z)^{1/2}).
\end{equation*}
We therefore deduce the large $n$ behavior
\begin{align}
&y_n\propto \frac{z_c^{-n}}{n^{3/2}}&\hbox{(supercritical)\ .}
\label{eq:supcr1}
\end{align}
\paragraph{Critical case.}  Equation \eqref{eq:ycrit} has a combinatorial solution with a confluence of singularities,
with $y_c=\tau_\phi$.  In this case, we may write the expansion \eqref{eq:phiexp} for $\tau=y$ as
\begin{equation*}
\begin{split}
\phi(y)&=\phi(y_c)-(y_c-y)\phi'(y_c)+K(y_c-y)^\alpha+ o((y_c-y)^\alpha)\\
&= \frac{y}{z_c}+K(y_c-y)^\alpha+ o((y_c-y)^\alpha)\\
\end{split}
\end{equation*}
where we used the identity \eqref{eq:ycrit} to remove the constant term. 
We therefore have
\begin{equation*}
\begin{split}
z_c-z&=z_c- \frac{y}{\phi(y)}=z_c-\frac{y_c-(y_c-y)}{\frac{1}{z_c}\left[y_c-(y_c-y)+z_cK(y_c-y)^\alpha+ o((y_c-y)^\alpha)\right]}\\
&=\frac{z_c^2}{y_c}K(y_c-y)^\alpha+ o((y_c-y)^\alpha)\ ,\\
\end{split}
\end{equation*}
which yields
\begin{equation*}
y=y_c-\left(\frac{y_c}{z_c^2K}(z_c-z)\right)^{1/\alpha}+o\left((z_c-z)^{1/\alpha}\right)\ .
\end{equation*}
We now obtain the large $n$ asymptotic behavior
\begin{align}
&y_n\propto \frac{z_c^{-n}}{n^{1+1/\alpha}}&\hbox{(critical)\ .}
\label{eq:crit1}
\end{align}
\paragraph{Subcritical case.}  Equation \eqref{eq:ycrit} has no solution  
with $y_c\leq \tau_\phi$. In this case, letting $z$ increase from $0$, we reach the singularity of $\phi$ at the value $z=z_\phi=\tau_\phi/\phi(\tau_\phi)$ for which $y(z_\phi)=\tau_\phi$. We may now use \eqref{eq:phiexp} for $\tau=y$ to write
\begin{equation*}
\begin{split}
z_\phi-z&=z_\phi- \frac{y}{\phi(y)}=z_\phi-\frac{\tau_\phi-(\tau_\phi-y)}{\phi(\tau_\phi)-(\tau_\phi-y)\phi'(\tau_\phi)+K(\tau_\phi-y)^\alpha+ o((\tau_\phi-y)^\alpha)}\\
&= 
z_\phi-\frac{\tau_\phi}{\phi(\tau_\phi)}\left(1-(\tau_\phi-y)\left(\frac{1}{\tau_\phi}-\frac{\phi'(\tau_\phi)}{\phi(\tau_\phi)}
\right)-\frac{K}{\phi(\tau_\phi)}(\tau_\phi-y)^\alpha)+o((\tau_\phi-y)^\alpha))\right)
\\
&=K_\phi(\tau_\phi-y)+
\frac{K\tau_\phi}{\phi(\tau_\phi)^2}(\tau_\phi-y)^\alpha+o((\tau_\phi-y)^\alpha)\\
& \hspace{5.cm} \hbox{with\ \ }
K_\phi=\frac{1}{\phi(\tau_\phi)}\left(1-\tau_\phi\frac{\phi'(\tau_\phi)}{\phi(\tau_\phi)}\right)\ .\\
\end{split}
\end{equation*}
This leads to
\begin{equation*}
y=\tau_\phi-\frac{1}{K_\phi}(z_\phi-z)+\frac{K\tau_\phi}{\phi(\tau_\phi)^2}\frac{1}{K_\phi^{1+\alpha}}
(z_\phi-z)^\alpha+o((z_\phi-z)^\alpha)
\end{equation*}
so that we now get the large $n$ behavior
\begin{align}
&y_n\propto \frac{z_\phi^{-n}}{n^{1+\alpha}}&\hbox{(subcritical)\ .}
\label{eq:subcr1}
\end{align}
\begin{rem}
\label{rem:remark1}
In the following, we will also consider the case $\alpha=2$ with a logarithmic singularity, namely with
\begin{equation*}
\phi(\tau)=\phi(\tau_\phi)-(\tau_\phi-\tau)\phi'(\tau_\phi)-K(\tau_\phi-\tau)^2\log(\tau_\phi-\tau)+ 
o\left((\tau_\phi-\tau)^2\log(\tau_\phi-\tau)\right)\ .
\end{equation*}
It is easily seen that nothing changes for the supercritical  case while the subcritical case now
yields
\begin{equation*}
y=\tau_\phi-\frac{1}{K_\phi}(z_\phi-z)-\frac{K\tau_\phi}{\phi(\tau_\phi)^2}\frac{1}{K_\phi^{3}}
(z_\phi-z)^2\log(z_\phi-z)+o\left((z_\phi-z)^2\log(z_\phi-z)\right)
\end{equation*}
so that 
\begin{align*}
&y_n\propto \frac{z_\phi^{-n}}{n^{3}}&\hbox{(subcritical)\ .}
\end{align*}
The critical case now leads to
\begin{equation*}
y=y_c-\left(-\frac{2y_c}{z_c^2K}\frac{z_c-z}{\log(z_c-z)}\right)^{1/2}+o\left(\left(-\frac{z_c-z}{\log(z_c-z)}\right)^{1/2}\right)
\end{equation*}
so that 
\begin{align*}
&y_n\propto \frac{z_c^{-n}}{n^{3/2}(\log n)^{1/2}}&\hbox{(critical)\ .}
\end{align*}
 \end{rem}
 
 \medskip
 The following table summarizes the various asymptotic behaviors for $y_n$ that we have obtained:
 \begin{center}
\begin{tabular}{||c | c c c||} 
 \hline
 & subcritical & critical & supercritical \\ [0.5ex] 
 \hline\hline
$1< \alpha<2$ & $\displaystyle{\frac{z_\phi^{-n}}{n^{1+\alpha}}}$ &$ \displaystyle{\frac{z_c^{-n}}{n^{1+1/\alpha}}}$ & $\Bigg.\displaystyle{\frac{z_c^{-n}}{n^{3/2}}}$ \\ 
 \hline
$\alpha=2 \ (\times \log)$ & $\displaystyle{\frac{z_\phi^{-n}}{n^{3}}}$ &$\displaystyle{\frac{z_c^{-n}}{n^{3/2}(\log n)^{1/2}}}$ & $\Bigg.\displaystyle{\frac{z_c^{-n}}{n^{3/2}}}$ \\
\hline
\end{tabular}
\end{center}

\subsection{Asymptotics for 2-point correlators}
\label{sec:2point}
Let us now consider some other observable enumerated by $w_n$ whose associated generating function
$w(z)=\sum_{n\geq 0}w_n z^n$ is linked to $y(z)$ via the relation 
\begin{equation}
w(z)=\psi(y(z))
\label{eq:psisubst}
\end{equation}
with $\psi(\tau)=\sum_{i\geq 0}\psi_i\tau^i$ such that $\psi_i\propto \tau_\phi^{-i}/i^{1+\beta}$ at large $i$, 
where $0<\beta< 1$.
Otherwise stated, $\psi$ is singular \emph{at the same value $\tau=\tau_\phi$} as $\phi$, now with expansion
\begin{equation*}
\psi(\tau)=\psi(\tau_\phi)-K_\psi(\tau_\phi-\tau)^\beta+ o((\tau_\phi-\tau)^\beta)\ .
\end{equation*}
Using the results of previous section, we immediately deduce the following large $n$ asymptotics for $w_n$ in the three
possible situations.
\paragraph{Supercritical case.}  The quantity $w(z)$ is singular at $z=z_c$ (as given by Equations \eqref{eq:ycrit} and \eqref{eq:ycrit2}) with 
\begin{equation*}
w(z)=w(z_c)-\left(\frac{2}{H''(y_c)}(z_c-z)\right)^{1/2}\psi'(y_c)+o((z_c-z)^{1/2}).
\end{equation*}
In particular, we deduce the large $n$ behavior
\begin{align}
&w_n\propto \frac{z_c^{-n}}{n^{3/2}}&\hbox{(supercritical)\ .}
\label{eq:SCasym}
\end{align}

\paragraph{Critical case.} The quantity $w(z)$ is singular at $z=z_c$ where $y(z_c)=y_c=\tau_\phi$, with 
\begin{equation*}
w(z)=w(z_c)-K_\psi\left(\frac{y_c}{z_c^2K}(z_c-z)\right)^{\beta/\alpha}+o\left((z_c-z)^{\beta/\alpha}\right)\ .
\end{equation*}
We now obtain the large $n$ asymptotic behavior
\begin{align}
&w_n\propto \frac{z_c^{-n}}{n^{1+\beta/\alpha}}&\hbox{(critical)\ .}
\label{eq:crit2}
\end{align}

\paragraph{Subcritical case.} 
The quantity $w(z)$ is now singular at $z=z_\phi$ where $y(z_\phi)=\tau_\phi$, with
 \begin{equation*}
w(z)=w(z_\phi)-K_\psi \left(\frac{1}{K_\phi}(z_\phi-z)\right)^\beta+o((z_\phi-z)^\beta)
\end{equation*}
so that
\begin{align}
&w_n\propto \frac{z_\phi^{-n}}{n^{1+\beta}}&\hbox{(subcritical)\ .}
\label{eq:sCasym}
\end{align}

\begin{rem}
\label{rem:remark2}
In the case $\alpha=2$ with a logarithmic singularity, the supercritical \eqref{eq:SCasym} and subcritical \eqref{eq:sCasym} asymptotics 
above for $w_n$  remain valid, while, in the critical case, we now have
\begin{align}
&w_n\propto \frac{z_c^{-n}}{n^{1+\beta/2}(\log n)^{\beta/2}} &\hbox{(critical)\ .}
\label{eq:critlog}
\end{align}
If, moreover, $\beta=1$ with a logarithmic singularity, \emph{i.e.},
\begin{equation}
\psi(\tau)=\psi(\tau_\phi)+K_\psi(\tau_\phi-\tau)\log(\tau_\phi-\tau)+ 
o\left((\tau_\phi-\tau)\log(\tau_\phi-\tau)\right)\ ,
\label{eq:betaonelog}
\end{equation}
the supercritical asymptotics \eqref{eq:SCasym} above for $w_n$ remains valid, while the 
subcritical one becomes simply
\begin{align*}
&w_n\propto \frac{z_\phi^{-n}}{n^{2}}  &\hbox{(subcritical)}\ .
\end{align*}
The critical case requires more attention: we know have 
\begin{equation*}
w(z)-w(z_c)\sim K_\psi\!\left(\!-\frac{2y_c}{z_c^2K}\frac{z_c-z}{\log(z_c-z)}\right)^{1/2}\log\left(-\frac{2y_c}{z_c^2K}\frac{z_c-z}{\log(z_c-z)}\right)^{1/2}\ ,
\end{equation*}
so that 
\begin{equation*}
w(z)=w(z_c)+ \frac{K_\psi}{2}\!\left(\!-\frac{2y_c}{z_c^2K}\right)^{1/2}\!\!(z_c-z)^{1/2}\left(\log(z_c-z)\right)^{1/2}+o\left((z_c-z)^{1/2}
\log(z_c-z)^{1/2}\right).
\end{equation*}
We then deduce that
\begin{align}
\label{eq:critlog2}&w_n\propto \frac{z_c^{-n} (\log n)^{1/2} }{n^{3/2}}  &\hbox{(critical)}\ .
\end{align}
Note that \eqref{eq:critlog2} is not the $\beta\to 1$ limit of \eqref{eq:critlog}.
\end{rem}

\medskip
The following table provides a summary of the various asymptotic behaviors for $w_n$:
\begin{center}
\begin{tabular}{||c | c c c||} 
 \hline
 & subcritical & critical & supercritical \\ [0.5ex] 
 \hline\hline
\makecell{$1< \alpha<2$\\$0< \beta<1$} &$\displaystyle{\frac{z_\phi^{-n}}{n^{1+\beta}}}$ &  $\displaystyle{\frac{z_c^{-n}}{n^{1+\beta/\alpha}}}$ &$\Bigg.\displaystyle{\frac{z_c^{-n}}{n^{3/2}}}$ \\
 \hline
\makecell{$\alpha=2\ (\times \log)$\\$0< \beta<1$ }& $\displaystyle{\frac{z_\phi^{-n}}{n^{1+\beta}}}$ & $\displaystyle{\frac{z_c^{-n}}{n^{1+\beta/2}(\log n)^{\beta/2}} }$ & $\Bigg.\displaystyle{\frac{z_c^{-n}}{n^{3/2}}}$ \\
 \hline
\makecell{$\alpha=2\ (\times \log)$\\$\beta=1\ (\times \log)$}  & $\displaystyle{\frac{z_\phi^{-n}}{n^{2}}}$ & $\displaystyle{\frac{z_c^{-n} (\log n)^{1/2} }{n^{3/2}}}$ & $\Bigg.\displaystyle{\frac{z_c^{-n}}{n^{3/2}}}$ \\ [1ex] 
 \hline
\end{tabular}
\end{center}

\section{Substitution and duality for random maps} 
\label{sec:blockweighted}
\subsection{General setting}
\label{sec:gensetting}
\subsubsection{Generating functions}
\begin{figure}
  \centering
  \fig{.8}{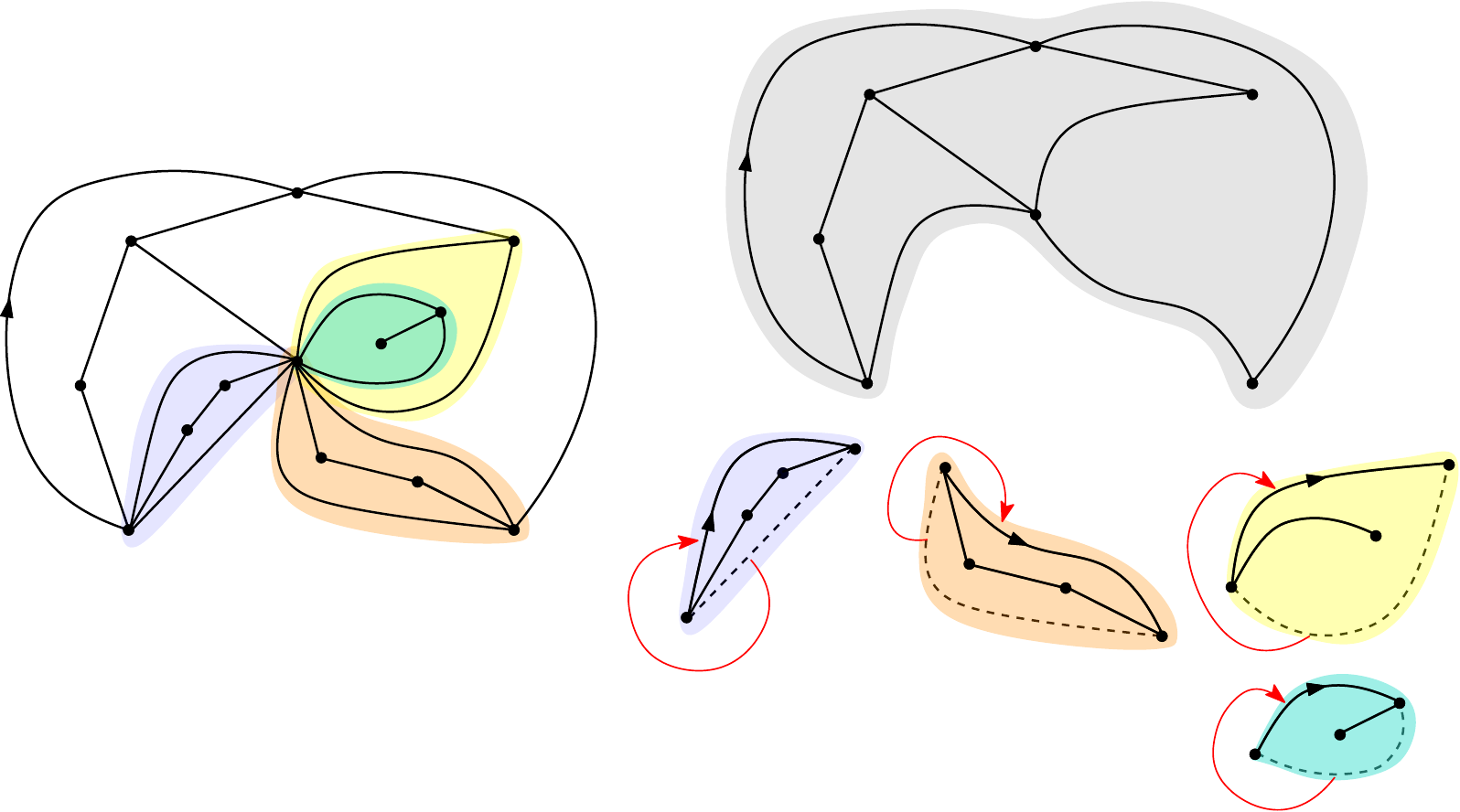}
   \caption{Decomposition of a rooted planar quadrangulation (left, including an infinite external face of degree $4$) into $5$ blocks which are rooted planar \emph{simple} (\emph{i.e.}, without multiple edges) quadrangulations. The blocks are obtained by cutting out the maximal cycles of length $2$ 
   and transforming the extracted rooted maps with a boundary of length $2$ into rooted quadrangulations
   by identifying their two sides (red arrows). Proceeding recursively decomposes the map into blocks which are themselves rooted planar simple quadrangulations.}
  \label{fig:blockquad}
\end{figure}
As we will see in the next sections, all the cases of block-weighted decorated maps that we will study in this article correspond to a substitution of the form
\begin{equation}
M_u(g)=1+u\left[B\left(g\, M_u(g)^2\right)-1\right]
\label{eq:mapsubst}
\end{equation}
with $B(t)=1+\sum_{j\geq 1}b_j t^j$, where $b_j$ enumerates rooted decorated maps of size $j$ \emph{formed of a single
block}. Expanding $M_u(g)$ as $M_u(g)=\sum_{n\geq 0}m_n^{(u)} g^n$, the quantity $m_n^{(u)}$ then enumerates
rooted decorated maps of size $n$, now formed of possibly several blocks, \emph{with a weight $u$ per block},
with a conventional initial term $m_0^{(u)}=1$. The reader is asked to accept at this stage the precise form of
the relation \eqref{eq:mapsubst} or to refer to \cite{FS24,ZSPhD} for explicit examples\footnote{Note that, in these references, a more general setting where the argument of $B$ is $g\, M_u(g)^d$ is considered. Here we will restrict to
the case $d=2$ but our results are easily extended to other integer values of $d$.}.  We will study in detail in the next 
sections several instances of block-weighted decorated maps for which $B$ can be made explicit. Figure~\ref{fig:blockquad} gives an illustration of a block decomposition in the particular case of a quadrangulation.

To make the connection with the analysis of Section \ref{sec:substitution}, we note that we may rewrite \eqref{eq:mapsubst} as our generic substitution equation \eqref{eq:yphi} upon setting
\begin{equation}
\begin{split}
& z=\sqrt{g}\ , \quad y(z)=z M_u(z^2)\ , \\
& \phi(\tau)=1+u\left[B(\tau^2)-1\right]=1+u\sum_{j\geq 1} b_j \tau^{2j}\ .\\
\end{split}
\label{eq:substcor}
\end{equation}
The tree described by the generating function $y(z)$ is nothing but the so-called \emph{block tree} introduced
in \cite{FS24,ZSPhD}.
Equations \eqref{eq:ycrit} and \eqref{eq:ycrit2} for the tree critical point translate, in the variables $g$ and $t:=\tau^2=g\, M_u(g)^2$, into
\begin{align}
\label{eq:Mcrit} 
& 1+u(B(t_c)-1)=2u\, t_c B'(t_c)\quad \hbox{with\ \ }t_c:=g_cM_u(g_c)^2\ ,\\
&g_c=\frac{t_c}{\left(1+u(B(t_c)-1)\right)^2}\ . \nonumber
\end{align} 
Let us write Equation \eqref{eq:Mcrit} equivalently as
\begin{equation}
u=\frac{1}{1-B(t_c)+2 t_c\, B'(t_c)}\ .
\label{eq:equ}
\end{equation}
It is easily seen by a Taylor expansion of $1/u$ that, for positive $b_j$'s, the quantity $u$, as defined by \eqref{eq:equ}, is a decreasing function of the non-negative variable $t_c$ (starting at $u=+\infty$ for $t_c=0$), or equivalently that $t_c$ is a decreasing function of $u$ (reaching $0$ when $u\to+\infty$). 

Assume now that $B$ has a singularity at $t_{\crit}$ of the form
\begin{equation}
B(t)=B(t_{\crit})-(t_{\crit}-t)B'(t_{\crit})+K_B(t_{\crit}-t)^\alpha+ o((t_{\crit}-t)^\alpha)
\label{eq:Bsing}
\end{equation}
with $1<\alpha<2$, so that $\phi$ has a singularity of the form \eqref{eq:phiexp} for $\tau_\phi=\sqrt{t_{\crit}}$.
Then we deduce that \eqref{eq:equ} has a solution $t_c\leq t_{\crit}$ for
$u\geq u_{\crit}$  with
\begin{equation}
u_{\crit}=\frac{1}{1-B(t_{\crit})+2 t_{\crit}\, B'(t_{\crit})}\ .
\label{eq:equcrit}
\end{equation}
More precisely, we are in the critical (resp. subcritical and supercritical regime) when $u=u_{\crit}$ (resp. $u<u_{\crit}$ and $u>u_{\crit}$).
In particular, we deduce from \eqref{eq:subcr1}, \eqref{eq:crit1} and \eqref{eq:supcr1} the large $n$ asymptotics
\begin{align}
\label{eq:subcr3}&m_n^{(u)}\propto \frac{g_{\crit}(u)^{-n}}{n^{1+\alpha}}&\hbox{for\ \ } u<u_{\crit}\ ,\\
\label{eq:crit3}&m_n^{(u)}\propto \frac{g_c(u)^{-n}}{n^{1+1/\alpha}} &\hbox{for\ \ } u=u_{\crit}\ ,\\
\label{eq:supcr3}&m_n^{(u)}\propto \frac{g_c(u)^{-n}}{n^{3/2}}&\hbox{for\ \ } u>u_{\crit}\ ,
\end{align}
where $g_c(u)$ is given by $g_c M_u(g_c)^2=t_c$ with $t_c=t_c(u)$ obtained by inverting \eqref{eq:equ} , while
$g_{\crit}(u)$ is given by $g_{\crit} M_u(g_{\crit})^2=t_{\crit}$, \emph{i.e.},
\begin{equation}
g_{\crit}(u)=\frac{t_{\crit}}{(1+u(B(t_{\crit})-1))^2}\ .
\label{eq:guB}
\end{equation}
Note in particular that, even though $t_{\crit}$ is independent of $u$,  both $g_c$ and $g_{\crit}$
depend on $u$ and that $g_c(u)=g_{\crit}(u)$ when $u=u_{\crit}$.

\bigskip
\subsubsection{Two-point correlators} 
Let us now discuss 2-point correlators for block-weighted maps. Consider the generating function $S_u(g)=\sum_{n\geq 0}s_n^{(u)} g^n$ where $s_n^{(u)}$ enumerates rooted decorated maps of size $n$, with a weight $u$ per block, and \emph{with an additional marking in the same block as the root}. We will discuss later in more details via concrete examples what we mean by marking and how to extend the notion of blocks in the presence of such a marking. 
As it turns out, $S_u(g)$  generally satisfies a substitution relation of the form
\begin{equation}
S_u(g)=M_u(g)^a\, C\left(g\, M_u(g)^2\right)
\label{eq:substSC}
\end{equation}
with $a$ an integer (typically $a=0, 1$ or $2$ in the examples discussed below) with $M_u$ defined
via \eqref{eq:mapsubst}. Note that our asymptotic results below do not depend on the particular value of $a$. Writing $C(t)=\sum_{j\geq 0}c_j t^j$, the coefficient $c_j$ (which does not depend on $u$) now enumerates rooted decorated maps of size $j$ formed of a single block, and with an additional marking.
The above substitution may be rewritten in the canonical form \eqref{eq:psisubst} upon setting
$z$ and $y(z)$ as in \eqref{eq:substcor} and
\begin{equation*}
w(z)=z^a\, S_u(z^2)\ , \quad \psi(\tau)=\tau^a\, C(\tau^2)\ .
\end{equation*}
Assume now that $C$ has a singularity at $t_{\crit}$ of the form
\begin{equation}
C(t)=C(t_{\crit})-K_C(t_{\crit}-t)^\beta+ o((t_{\crit}-t)^\beta), 
\label{eq:Csing}
\end{equation}
with $0<\beta<1$. We now deduce from \eqref{eq:sCasym}, \eqref{eq:crit2} and \eqref{eq:SCasym} in Section \ref{sec:2point} the large $n$ asymptotics
\begin{align}
&s_n^{(u)}\propto \frac{g_{\crit}(u)^{-n}}{n^{1+\beta}}&\hbox{for\ \ } u<u_{\crit}\ ,\nonumber \\
&\label{eq:ucritcor}
s_n^{(u)}\propto \frac{g_c(u)^{-n}}{n^{1+\beta/\alpha}} &\hbox{for\ \ } u=u_{\crit}\ ,\\
&s_n^{(u)}\propto \frac{g_c(u)^{-n}}{n^{3/2}}&\hbox{for\ \ } u>u_{\crit}\ .\nonumber
\end{align}

\subsection{Duality: comparing $\boldsymbol{u=1}$ and $\boldsymbol{u=u_{\crit}}$}
\label{sec:duality}
\subsubsection{Generating functions}
\label{sec:gfduality}
Let us now assume that we know (or at least have some knowledge of) 
the combinatorics of the problem at $u=1$, corresponding to the statistics of decorated maps where we do not pay attention to their block structure. The corresponding generating function 
$M_1(g)$ is related to $B$ via \eqref{eq:mapsubst} with $u=1$, namely
\begin{equation}
M_1(g)=B\left(g\, M_1(g)^2\right)\ .
\label{eq:M1}
\end{equation}
In this paper, we will always consider the case where
\begin{equation}
m_n^{(1)}\propto \frac{(g_1)^{-n}}{n^{2-\gs}}
\label{eq:gsdef}
\end{equation}
for some $g_1$ and with a string susceptibility exponent $\gs$ satisfying 
$2<2-\gs<3$. In particular, the problem at $u=1$ can be neither in the supercritical regime \eqref{eq:supcr3},
nor in the critical regime \eqref{eq:crit3} since $1+1/\alpha<2$, 
which means that $u_{\crit}>1$ with the asymptotics of $m_n^{(1)}$ being of the form \eqref{eq:subcr3}. As a consequence, we deduce that the singularity of $B$ and that of $M_1$ are characterized by
\emph{the same exponent}, namely that \eqref{eq:Bsing} holds with the exponent
\begin{equation*}
\alpha=1-\gs
\end{equation*}
and, moreover, that $g_1=g_{\crit}(u=1)$ is such that $g_1 M_1(g_1)^2=t_{\crit}$. From \eqref{eq:M1}, we deduce that $M_1(g_1)=B(t_{\crit})$
while, upon differentiating \eqref{eq:M1} with respect to $g$, then setting $g=g_1$, we obtain that
\begin{equation*}
B(t_{\crit})=M_1(g_1)\ , \qquad B'(t_{\crit})=\frac{M_1'(g_1)}{M_1(g_1)^2+2 g_1 M_1(g_1)M_1'(g_1)}\ .
\end{equation*}
Altogether, we deduce from \eqref{eq:equcrit} that the value of $u_{\crit}$ is fixed by
\begin{equation}
u_{\crit}=\frac{M_1(g_1)+2 g_1M_1'(g_1)}{M_1(g_1)(1-M_1(g_1))+2 g_1M_1'(g_1)}
\label{eq:ucritval}
\end{equation} 
and may thus be estimated from the knowledge of the combinatorics at $u=1$ only. 
At $u=u_{\crit}$, we have 
\begin{equation}
m_n^{(u_{\crit})}\propto \frac{g_c(u_{\crit})^{-n}}{n^{2-\gsp}}
\label{eq:mnucritasymp}
\end{equation}
with, from \eqref{eq:guB},
\begin{equation}
 g_c(u_{\crit})=g_1\left(1+\frac{M_1(g_1)(1-M_1(g_1))}{2 g_1 M_1'(g_1)}\right)^2
 \label{eq:gucr1}
\end{equation}
and
\begin{equation}
1-\gsp= \frac{1}{\alpha}\quad\hbox{\emph{i.e.},\ \ }(1-\gs)(1-\gsp)=1.
\label{eq:dualgamma}
\end{equation}
This later identity is precisely the expected \emph{duality relation} \eqref{dualstring} for the string susceptibility 
exponent. We recover it here combinatorially directly from our singularity analysis.

\medskip
\begin{rem}
\label{rem:remark3}
The case $2-\gs=3$ in \eqref{eq:gsdef} corresponds to a singularity of $M_1(g)$ proportional 
to $(g_1-g)^2\log(g_1-g)$. From Remark~\ref{rem:remark1}, the function $B(t)$ has a similar singularity
proportional to $(t_{\crit}-t)^2\log(t_{\crit}-t)$ and \eqref{eq:mnucritasymp} is replaced by
\begin{equation*}
m_n^{(u_{\crit})}\propto \frac{g_c(u_{\crit})^{-n}}{n^{2-\gsp}(\log n)^{1/2}}\quad \hbox{with}\quad \gsp=\frac{1}{2}\ .
\end{equation*}
Again we have $(1-\gs)(1-\gsp)=1$.
\end{rem}

\bigskip
\subsubsection{Two-point correlators} 
\label{sec:2pointduality}
\begin{figure}
  \centering
  \fig{.8}{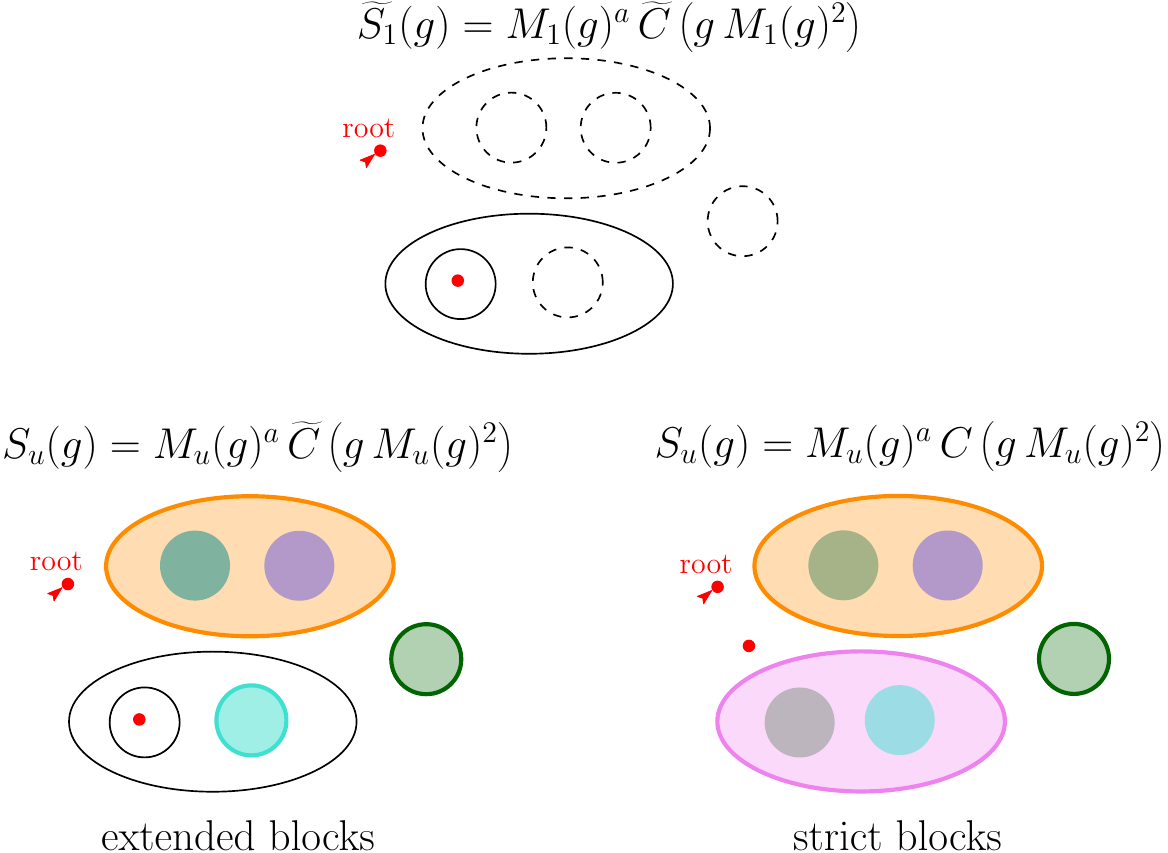}
   \caption{Schematic picture of the block structure of a map with two markings: we indicated the root (red dot with an arrow), the additional marking (isolated red dot) and the boundaries of the blocks. Note that any two blocks are either disjoint or included one into the other. Top: the generating function $\widetilde{S}_1$ corresponds to configurations for which the additional marking is anywhere so that the related function $\widetilde{C}$ corresponds to configurations where only blocks \emph{not containing the additional marking} are shrunk (blocks with a dashed boundary). In particular, a sequence of nested blocks containing the additional marking may remain in $\widetilde{C}$ (blocks with a solid boundary). Bottom left: if we define $S_u$ by substitution upon using the same function $\widetilde{C}$, then only those blocks which do not contain the additional marking 
receive the weight $u$ (colored blocks); this is the extended block prescription. Bottom right: if instead we define $S_u$ using the generating function
 $C$ corresponding to configuration with a single block and two markings, the substitution imposes that, in $S_u$, the additional marking \emph{belongs to the same block as the root}; this is the strict block prescription. In that case, all blocks
(colored blocks) but the root block receive the weight $u$.}
  \label{fig:extended}
\end{figure}

If we now consider 2-point correlators, we usually have access to the quantity 
$\widetilde{S}_1(g)=\sum_{n\geq 0}{\widetilde{s}^{(1)}}_n g^n$ where $\widetilde{s}^{(1)}_n$ enumerates rooted decorated maps of size $n$ with an additional marking \emph{anywhere}, ignoring blocks. We then \emph{define} the quantity $\widetilde{C}(t)$ by 
\begin{equation}
\widetilde{S}_1(g)=M_1(g)^a\,\widetilde{C}\left(g\, M_1(g)^2\right)
\label{eq:barS1C}
\end{equation}
for some appropriate $a$ depending on the problem at hand (our asymptotic results below do not depend on $a$). We may then consider two prescriptions, illustrated in Figure~\ref{fig:extended}. 

\paragraph{\emph{Extended block prescription.}} In this prescription, we set 
\begin{equation*}
C(t)=\widetilde{C}(t)
\end{equation*}
in the relation \eqref{eq:substSC}, which now defines $S_u(g)$ for arbitrary $u$. In particular, we identify $S_1(g)$ with 
$\widetilde{S}_1(g)$. In that case, the quantity $S_u(g)$ enumerates configurations with a weight $u$ per block as wanted, but with an \emph{extended notion of blocks} which is modified by the additional marking.
More precisely, the substitution \eqref{eq:barS1C} does not allow to remove blocks that would contain the additional marking so that configurations enumerated $\widetilde{C}(t)$ may contain sequences of nested blocks if the deepest block 
in the nested sequence contains the marking. Using then \eqref{eq:substSC} for arbitrary $u$ with 
$C=\widetilde{C}$, such a nested sequence containing the marking is thus considered as a single block in $S_u(g)$ and the weight $u$ is assigned only to those blocks not in this sequence (\emph{i.e.}, those blocks which can be removed without affecting the marking).

\medskip
\paragraph{\emph{Strict block prescription.}} In this prescription, we instead insist on keeping the same notion of blocks, irrespectively of the marking. In that case, we impose that the (marked) configurations enumerated by $C(t)$ have a single block 
(\emph{i.e.}, do not contain sequences of nested blocks as in the previous prescription). With this new prescription, 
the quantity $S_u(g)$ given by \eqref{eq:substSC} enumerates configurations with an arbitrary number of blocks,
each weighted by $u$, and with an additional marking \emph{in the same block as the root} (\emph{i.e.}, going from the root to the marking does not require entering nested blocks). In particular, the quantity  $S_1(g)$
is no longer equal to $\widetilde{S}_1(g)$, as $C(t)$ is not equal to $\widetilde{C}(t)$. Still it
is expected that $\widetilde{C}(t)$ has the same singular behavior as $C(t)$, given by \eqref{eq:Csing},   
namely that
\begin{equation}
\widetilde{C}(t)=\widetilde{C}(t_{\crit})-K_{\widetilde{C}}(t_{\crit}-t)^\beta+ o((t_{\crit}-t)^\beta). 
\label{eq:Cbarsing}
\end{equation}
In the following, we will consider in detail a simple 2-point correlator for which we can show that
\begin{equation}
C(t)=1-\frac{1}{\widetilde{C}(t)}
\end{equation}
so that \eqref{eq:Cbarsing} is a direct consequence of \eqref{eq:Csing}.

\bigskip
Let us now assume that 
\begin{equation*}
\widetilde{s}_n^{(1)}\propto \frac{(g_1)^{-n}}{n^{1+2\Delta-\gs}}
\end{equation*} 
with $0<2\Delta-\gs<1$. Then, from its connection with $\widetilde{S}_1(g)$ via \eqref{eq:barS1C}
the generating function $\widetilde{C}$ has a singularity of the form \eqref{eq:Cbarsing} with
\begin{equation}
\beta=2\Delta-\gs
\label{eq:Deltaval}
\end{equation}
so that, in both the extended and strict block prescriptions, the singularity of $C$ is of the form \eqref{eq:Csing}
with the same value of $\beta$.
For $u=u_{\crit}$, we obtain from \eqref{eq:ucritcor} that
\begin{equation*}
s_n^{(u_{\crit})}\propto \frac{g_c(u_{\crit})^{-n}}{n^{1+2\Delta'-\gsp}} 
\end{equation*} 
with
\begin{equation}
2\Delta'-\gsp=\frac{\beta}{\alpha}=\frac{2\Delta-\gs}{1-\gs}
\label{eq:deltadeltaprime0}
\end{equation}
\emph{i.e.},
\begin{equation*} 
\Delta'=\frac{\Delta-\gs}{1-\gs}.
\end{equation*}
Again we obtain via analytic combinatorics the duality relation \eqref{dualdimbis} expected from Liouville quantum duality (with the identification $\Delta=\Delta_\gamma$
and $\Delta'=\Delta_{\gamma'}$). Note that both the extended block prescription (with a weight $u$ assigned only to those block not containing the additional marking) and the strict block prescription (with the constraint that the additional marking is in the same block as the root) are, at $u=u_{\crit}$, realizations of the dual $\gamma'$-LQG. 

\section{Block-weighted quadrangulations}
\label{sec:quads}
In the coming three sections, we will describe in details three families of maps, with possible decorations, for which there are
natural notions of blocks: planar quadrangulations, cubic or bicubic planar maps decorated by Hamiltonian paths
and meandric systems.

\subsection{Rooted planar quadragulations and simple blocks}
\label{sec:quadpf}
Our first family of maps is that considered in \cite{FS24,ZSPhD}, namely rooted planar quadrangulations with a weight 
$g$ per face. As it is well known, their generating function is given by
\begin{equation}
M_1(g)=\sum_{n\geq 0} \underbrace{2\frac{3^n}{n+2}\cat(n)}_{\displaystyle{m_n^{(1)}}} g^n= \frac{18 g-1+(1-12g)^{3/2}}{54 g^2}
\label{eq:exprM1}
\end{equation}
with
\begin{equation}
\cat(n)=\frac{1}{n+1}{2n\choose n}
\label{eq:catalan}
\end{equation}
the $n$-th Catalan number, with large $n$ asymptotics
\begin{equation}
\cat(n)\sim \frac{4^n}{\sqrt{\pi}\, n^{3/2}}\ .
\label{eq:asympcat}
\end{equation}
We therefore have the asymptotics
\begin{equation*}
m_n^{(1)}\propto \frac{(g_1)^{-n}}{n^{2-\gs}}
\quad \hbox{with}\quad g_1=\frac{1}{12}\ \hbox{and}\quad \gs=-\frac{1}{2}\ .
\end{equation*}

The generating function $M_1(g)$ is related via \eqref{eq:M1} to the generating function $B(t)$ of rooted planar \emph{simple quadrangulations}, \emph{i.e.}, quadrangulations without multiple edges and with a weight $t$ per face, namely
\cite{Minbus,FS24,ZSPhD}
\begin{equation}
B(t)=1+\sum_{j\geq 1} \underbrace{2\frac{(3j-3)!}{(2j-1)!j!}}_{\displaystyle{b_j}} t^j= 1+
\frac{9 t-2 \sqrt{3t} \sin \left(\frac{1}{3} \arcsin\left(\frac{3 \sqrt{3t}}{2}\right)\right)}{1+2 \cos
   \left(\frac{2}{3} \arcsin \left(\frac{3 \sqrt{3t}}{2}\right)\right)}\ .
   \label{eq:Bpure}
\end{equation}
More precisely, as explained in \cite{Minbus}, we may transform a general rooted planar quadrangulations (enumerated by $M_1$) into a simple one (enumerated by $B$) by cutting it along its maximal (for the inclusion) cycles of length two (the so-called minimal necks in \cite{Minbus}) and sewing these cycles into edges. Upon sewing its boundary of length two into a root edge, each removed part is itself a rooted planar quadrangulations enumerated by $M_1$. Such a removed part can be attached to each edge of the simple quadrangulation enumerated by $B$ and since a planar quadrangulation has twice as many edges as faces, this immediately leads to the substitution relation \eqref{eq:M1},
see Figure~\ref{fig:blockquad}. 

From the above explicit formula for $b_j$, we immediately read its large $j$ asymptotics, $b_j\propto t_{\crit}^{-j}/j^{1+\alpha}$, with $t_{\crit}=4/27=g_1 M_1(g_1)^2$ and 
$\alpha=3/2=1-\gs$ as expected. The value of $\alpha$ can also been obtained via the expansion 
\eqref{eq:Bsing} of the right hand side of \eqref{eq:Bpure} with $B(t_{\crit})=4/3$ and $B'(t_{\crit})=K_B=3$.

Consider now the the quantity $M_u(g)$ given by the substitution \eqref{eq:mapsubst}. As explained in \cite{FS24,ZSPhD}, its coefficient $m_n^{(u)}$ enumerates rooted planar quadrangulations with $n$ faces, with a weight $u$ per block in a decomposition of the quadrangulations into \emph{simple blocks}. More precisely, as we just saw, a general rooted quadrangulation can be 
transformed by a cutting/sewing procedure into a simple one plus a number of attached parts which may themselves be viewed as rooted quadrangulations. Repeating the cutting/sewing operation recursively within the removed parts  results into a canonical decomposition of initial quadrangulation into simple blocks; see Figure~\ref{fig:blockquad} for an illustration. Each such block receives the weight $u$ in $M_u$.

From the explicit expression \eqref{eq:exprM1}, we have $M_1(g_1)=4/3$ and $M_1'(g_1)=16$, and it is found via \eqref{eq:ucritval} that
\begin{equation*}
u_{\crit}=\frac{9}{5}
\end{equation*}
while, from \eqref{eq:mnucritasymp}, \eqref{eq:gucr1} and \eqref{eq:dualgamma}, we get
\begin{equation}
m_n^{(u_{\crit})}\propto \frac{g_c(u_{\crit})^{-n}}{n^{2-\gsp}}\quad \hbox{with}\quad g_c(u_{\crit})
=\frac{5^2}{2^4 3^3}=\frac{25}{432}\quad \hbox{and}\quad \gsp=1-\frac{1}{\alpha}=\frac{1}{3}\ ,
\label{eq:53} 
\end{equation}
as found in \cite{FS24,ZSPhD,BONZOM2015161}. The values of $\gs$ and $\gsp$ that we have found here correspond precisely to
the expected values \eqref{string} and \eqref{dualgammastring} for $\gamma$-LQG and ${\gamma'}-$LQG with $\gamma=\sqrt{8/3}=4/\gamma'$, as expected
for a model in the universality class of pure quantum gravity, characterized by a central charge $c=0$.  
We also verify, here in the particular case of planar quadrangulations,
the validity of the Liouville quantum duality relation $(1-\gs)(1-\gsp)=1$.

\subsection{Rooted quadrangulations with a second marked edge}
\label{sec:pure2p}
\begin{figure}[h!]
  \centering
  \fig{.6}{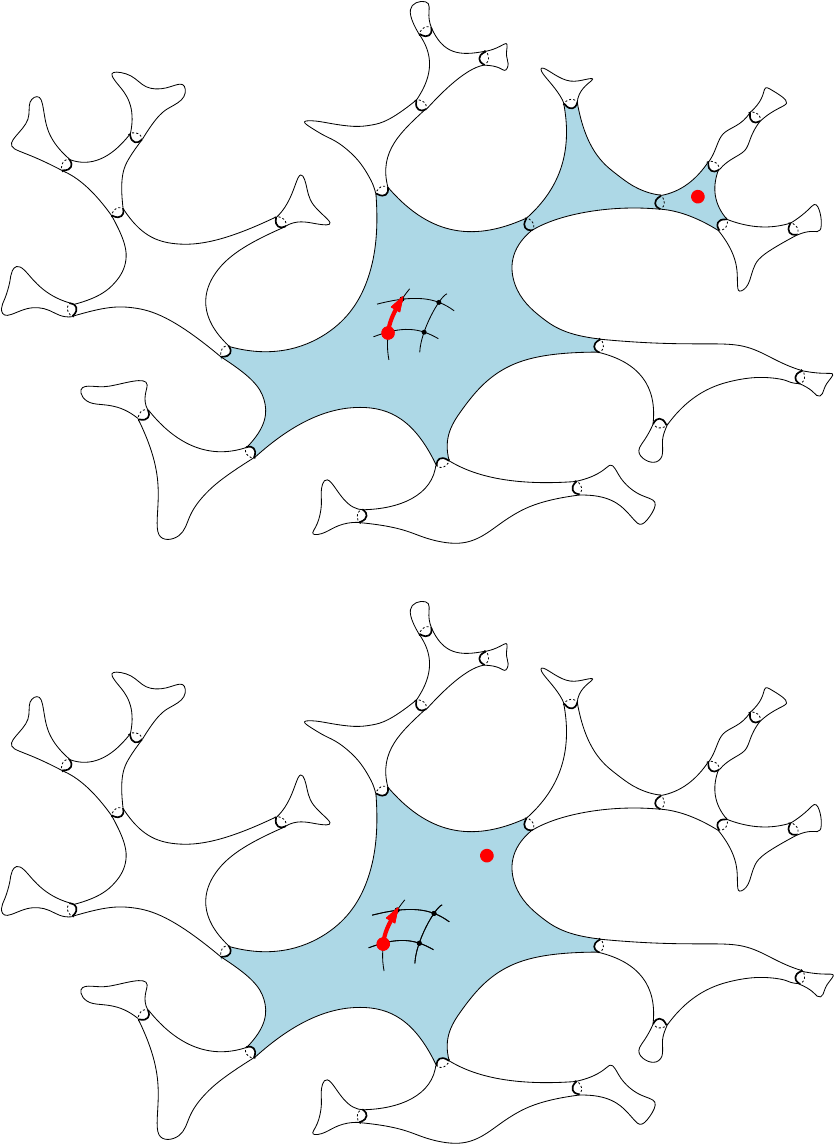}
   \caption{A schematic picture of a quadrangulation enumerated by $S_u(g)$. Top: the
   extended block prescription $S_u(g)=M_u(g)^2\,\widetilde{C}\left(g\, M_u(g)^2\right)$ corresponds to an additional
   marked edge (isolated red dot) anywhere in the map but the sequence of blocks in blue, separating it from the root edge (red arrow), is considered as a single extended block. Bottom: the strict block prescription $S_u(g)= C\left(g\, M_u(g)^2\right)$ corresponds to an additional
   marked edge (isolated red dot) in the same (blue) block as the root edge. In both cases, a weight $u$ is attached only to the uncolored blocks.  At $u=u_{\crit}$, both prescriptions lead to an asymptotic behavior \eqref{eq:43} for 
   $s^{(u_{\crit})}_n=[g^n]S_{u_{\crit}}(g)$, characterized by the same exponent 
 $\Delta'=1/3$.}
  \label{fig:twopoint}
\end{figure}
As in \cite{Minbus}, let us now consider planar rooted quadrangulations with \emph{an additional marked edge}
and call $\widetilde{S}_1(g)$ the associated generating function with a weight $g$ per face. Note that
the additional marked edge may be identical to the root edge itself but, in that case, we decide for
convenience to enumerate the associated configuration twice\footnote{This ``doubling'' convention can be
thought of as a doubling of the root edge, with an extra copy on the left or on the right.}. 
With this convention, it is easily seen (see \cite{Minbus} for details\footnote{The correspondence with \cite{Minbus}
is made by identifying the variable $z$ in this reference with our variable $t$, the function $R_1(g)$ 
with our function $M_1(g)$ and the quantity $\hat{h}(1,z)$ with $\widetilde{C}(t)$.}) that 
\begin{equation}
\widetilde{S}_1(g)=M_1(g)^2\,\widetilde{C}\left(g\, M_1(g)^2\right)\quad \hbox{with \ }
\widetilde{C}(t)=\left(\frac{2 \sin \left(\frac{1}{3} \arcsin\left(\frac{3 \sqrt{3t}}{2}\right)\right)}{\sqrt{3t}}\right)^2\ .
\label{eq:barCpure}
\end{equation}
A simple way to recover this relation is by noting that a quadrangulation on the Riemann sphere with $n$ faces has $2n$ edges, so that we may write
\begin{equation}
\widetilde{S}_1(g)=\left(2g\frac{\partial\, }{\partial g}+1\right)M_1(g)\ , 
\label{eq:S1deriv}
\end{equation}
where the additional term $1$ in the operator comes from the fact that the root edge is counted twice if identical to the additional marked edge.
Applied formally to the identity \eqref{eq:M1}, $M_1(g)=B(g\, M_1(g)^2)$, Eq.~\eqref{eq:S1deriv} allows us, after
some computations, to recover \eqref{eq:barCpure} with the identification
\begin{equation}
\widetilde{C}(t)=\frac{1}{B(t)-2t\, B'(t)}\ .
\label{eq:CbarB}
\end{equation}
As discussed in \cite{Minbus}, the quantity $\widetilde{C}(t)$ enumerates rooted quadrangulations with an additional marked edge
different from the root edge, with a weight $t$ per face and without multiple edges \emph{except for
possible double edges separating the root edge from the additional marked edge} (note that $\widetilde{C}(t)$ also contains a conventional constant term $\widetilde{C}(0)=1$.
This term is responsible for the factor $2$ when the additional marked edge in $\widetilde{S}_1(g)$ is the root edge itself).

The double edges in the configurations enumerated by $\widetilde{C}(t)$ necessarily form a nested sequence separating successive simple blocks, with the additional marked edge lying inside the deepest block. As explained in Section~\ref{sec:2pointduality}, choosing $C(t)=\widetilde{C}(t)$ to define $S_u(g)$ corresponds to what we called the extended block prescription, see Figure~\ref{fig:twopoint}-top.
Here we prefer to adopt the strict block prescription by taking for $C(t)$ the generating function for rooted quadrangulations with an additional marked edge different from the root edge, with a weight $t$ per face and \emph{without multiple edges} (with now a conventional constant term $C(0)=0$). From the nested structure of the double edges in $\widetilde{C}(t)$, it can be shown (see \cite[Eq.~(2.35)]{Minbus}) that the generating function 
$\widetilde{C}(t)$ is related to $C(t)$ via\footnote{The correspondence with \cite{Minbus}
is now via the identification of the quantity $\hat{g}(1,z)$ with $C(t)$.}
\begin{equation}
\widetilde{C}(t)=1+ C(t)\, \widetilde{C}(t)\quad \hbox{\emph{i.e.}, } C(t)=1-\frac{1}{\widetilde{C}(t)}\ .
\label{eq:CbarC}
\end{equation}
Using \eqref{eq:CbarB}, we get the identification 
\begin{equation}
C(t)=1-B(t)+2t\, B'(t)\ .
\label{eq:CB}
\end{equation}
With this definition of $C(t)$, the quantity $S_u(g)$ defined via
\begin{equation*}
S_u(g)=C\left(g\, M_u(g)^2\right)\ ,
\end{equation*} 
namely \eqref{eq:substSC} with $a=0$, enumerates
quadrangulations with an additional marked edge, with a weight $g$ per face and $u$ per simple block,
and \emph{such that the additional marked edge is in the same block as the root edge}, see Figure~\ref{fig:twopoint}-bottom. Note that, in a rooted planar quadrangulation, any edge can be canonically oriented from its closest extremity
to the origin of the root edge to its other extremity (the two extremities cannot be at the same distance since the map is bipartite). If an edge separates two blocks, we take the convention that it belongs to
the block on its right. Demanding that the root edge and the additional marked edge are in the same block means
therefore that one can connect the right side of both edges while staying in the same block. The value $a=0$ (instead of $a=2$ for $\widetilde{S}_1$) in \eqref{eq:substSC} reflects the fact that, when going from a configuration enumerated by $C$ to one enumerated by $S_u$, we cannot graft blocks immediately to the right of the root edge, nor of the additional marked edge.  

\medskip
Now it follows from \eqref{eq:CbarB} and \eqref{eq:CB} that $C(t)$ and $\widetilde{C}(t)$ have the same singular behavior
at $t=t_{\crit}=4/27$, which is that of $B'(t)$, characterized by the exponent $\beta=\alpha-1=1/2$ which, using
\eqref{eq:Deltaval}, amounts to $2\Delta-\gs=1/2$, \emph{i.e.} $\Delta=0$ since $\gs=-1/2$. 
To summarize, the marking of an additional edge corresponds to a 2-point correlator with quantum scaling exponent
$\Delta=0$, as expected. 

\medskip
At the phase transition point $u=u_{\crit}$, we then obtain from \eqref{eq:deltadeltaprime0} that
\begin{equation}
s_n^{(u_{\crit})}\propto \frac{g_c(u_{\crit})^{-n}}{n^{1+2\Delta'-\gsp}} \quad\hbox{with
\ } 
2\Delta'-\gsp=-\frac{\gs}{1-\gs}=\frac{1}{3}\ ,
\label{eq:43} 
\end{equation}
corresponding to 
\begin{equation*}
\Delta'=\gsp=\frac{1}{3}\ .
\end{equation*}

The obtained exponent may be understood by the following heuristic argument: the quantity 
$s_n^{(u_{\crit})}$ corresponds to the marking of an additional edge (among $2n$) in the same block
as the root edge. We may therefore estimate it asymptotically as $s_n^{(u_{\crit})}\sim 2n\times
m_n^{(u_{\crit})}\times p_n$ where $p_n$ is the probability that the additional marked edge is in the
same block as the root edge. Now it is known \cite{FS24} (see also \cite[Appendix A with $\lambda=2/3$]{BFSS})
that, at $u=u_{\crit}$, the largest blocks have size 
of order $n^{2/3}$. Then $p_n$ is dominated by configurations where both the root edge and the additional 
marked edge are in the same large block, \emph{i.e.}, $p_n\propto (n^{2/3}/2n)^2\propto n^{-2/3}$. We eventually get 
\begin{equation*}
s_n^{(u_{\crit})}\propto n\times \frac{g_c(u_{\crit})^{-n}}{n^{5/3}}\times n^{-2/3}=\frac{g_c(u_{\crit})^{-n}}{n^{4/3}}\ .
\end{equation*}
\medskip\null
\begin{rem}
\label{rem:remark4}
The results \eqref{eq:53} and \eqref{eq:43} can be obtained alternatively by an expansion of 
$M_{u_{\crit}}(g)$ for $g$ around $g_c(u_{\crit})=25/432$. It is easily shown (see for instance \cite{FS24}) that 
$B(t)$ is algebraic, solution of the equation
\begin{equation*}
B(t)^3-B(t)^2-18t\, B(t)+27 t^2+16 t=0 \ .
\end{equation*}
We then deduce from \eqref{eq:mapsubst} that $M_u(g)$ is also algebraic, solution of
\begin{equation}
\begin{split}
27u^3 g^2\, M_u(g)^4+(1-18 u^2 g)M_u(g)^3&-(3-2u+2u^3g-18 u^2g)M_u(g)^2\\
&+(3-4u+u^2)M_u(g)-(1-u)^2=0\ .\\
\label{eq:MuEq}
\end{split}
\end{equation}
Setting $u=u_{\crit}=9/5$ and 
\begin{equation*}
g=\frac{25}{432}(1-\epsilon^3)\ ,
\end{equation*}
we deduce from \eqref{eq:MuEq} that
\begin{equation}
M_{u_{\crit}}(g)=\frac{8}{5}\left(1-\left(\frac{27}{32}\right)^{1/3}\epsilon^2 +\O(\epsilon^3)\right)
\label{eq:Mucritexp}
\end{equation}
while 
\begin{equation}
S_{u_{\crit}}(g)= C(g\, M_{u_{\crit}}(g)^2)=\frac{5}{9}\left(1-\left(\frac{256}{125}\right)^{1/3}\epsilon+\O(\epsilon^2)\right)\ .
\label{eq:Sucrexp}
\end{equation}
Since $\epsilon=(g_c(u_{\crit})-g)^{1/3}$, the generating function $M_{u_{\crit}}(g)$ (resp. the 2-point generating function $S_{u_{\crit}}(g))$ has therefore a leading singularity
of the form $(g_c(u_{\crit})-g)^{2/3}$ (resp. $(g_c(u_{\crit})-g)^{1/3}$) responsible for the
exponent $2-\gsp=5/3$ in \eqref{eq:53} (resp. $1+2\Delta'-\gsp=4/3$ in \eqref{eq:43}).
\end{rem}

\subsection{Numerical check} 
\label{sec:numcheck}
\begin{figure}
  \centering
  \fig{.6}{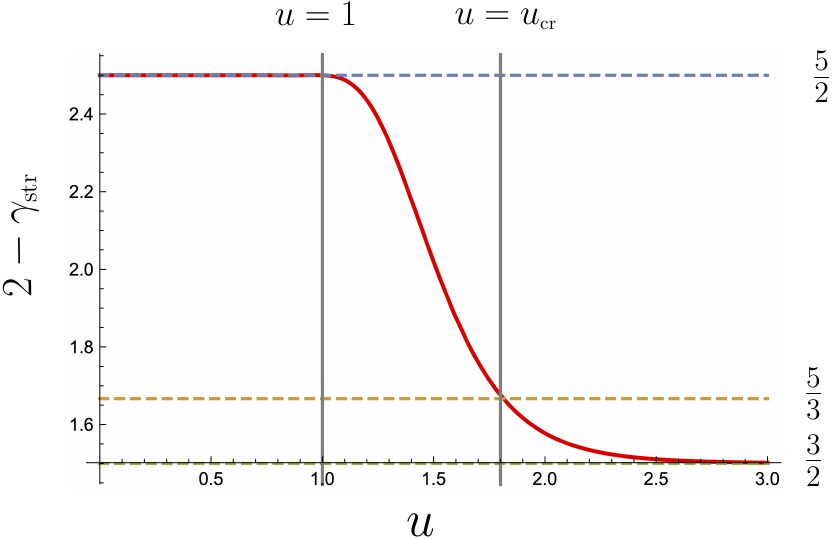}
   \caption{The $(50,5)$-estimate of the exponent $2-\gst$ for the asymptotics \eqref{eq:genasymp} of $m_n^{(u)}$ as a function of $u$ (see Appendix~\ref{app:numerics} for a precise definition). We indicated by vertical lines the positions $u=1$ and $u=u_{\crit}=9/5$ and by horizontal dashed lines the
   expected values $5/2$ ($u<u_{\crit}$), $5/3$ ($u=u_{\crit}$) and $3/2$ ($u>u_{\crit}$).}
  \label{fig:quadgamma}
\end{figure}
It is interesting to have a numerical check of our results. The goal here is not to corroborate our exact results
by numerics but rather to see how these results could have been reached numerically from
the data of, say $m_n^{(1)}$ for  the first values $n=1,2\ldots, N$. Indeed, we will consider below other
decorated map problems for which we do not know the generating function $M_1(g)$ exactly,
but still know the values of $m_n^{(1)}$ for $n$ up to $N$ (with $N$ typically of order $20$ to $35$).
From these values, we can extract, by expanding the substitution equation \eqref{eq:M1} up to order
$N$ in $g$, the value of $b_j$ for $j=1,2,\ldots, N$. Then, from the expansion of the substitution equation
\eqref{eq:mapsubst} up to order $N$ in $g$, we obtain the values $m_n^{(u)}$ for $n$ up to $N$.
For instance, in the case of rooted quadrangulations, we get
\begin{equation*}
\begin{split}
&m_0^{(u)}= 1,\  m_1^{(u)}=2 u,\ m_2^{(u)}=u (1 + 8 u), \
m_3^{(u)}= 2 u (1 + 6 u + 20 u^2),\\
&m_4^{(u)}= 2 u (3 + 18 u + 56 u^2 + 112 u^3),\ 
m_5^{(u)}=2 u (11 + 70 u + 225 u^2 + 480 u^3 + 672 u^4), \\
&m_6^{(u)}= u (91 + 624 u + 2134 u^2 + 4840 u^3 + 7920 u^4 + 8448 u^5), \\
&m_7^{(u)}= 2 u (204 + 1512 u + 5551 u^2 + 13468 u^3 + 24024 u^4 + 
    32032 u^5 + 27456 u^6), \\
&m_8^{(u)}= 2 u (969 + 7752 u + 30600 u^2 + 79590 u^3 + 152880 u^4 + 
    227136 u^5 + 256256 u^6\\ & \hfill + 183040 u^7), \\
&m_9^{(u)}=2 u (4807 + 41382 u + 175389 u^2 + 488784 u^3 + 1006740 u^4 + 
    1622208 u^5 \\ & \hfill + 2079168 u^6 + 2036736 u^7 + 1244672 u^8), \\
&m_{10}^{(u)}=u (49335 + 455400 u + 2067010 u^2 + 6160560 u^3 + 
    13566969 u^4 + 23473056 u^5 \\ & \hfill+ 32868480 u^6 + 37209600 u^7 + 
    32248320 u^8 + 17199104 u^9)\\
    &\vdots
\end{split}
\end{equation*}
We can summarize our results for the large $n$ asymptotics of $m^{(u)}_n$ as
\begin{equation}
 m^{(u)}_n\propto \frac{g_*^{-n}}{n^{2-\gst}}
 \label{eq:genasymp}
 \end{equation}
with $g_*$ equal to $g_{\crit}(u)$ for $u< u_{\crit}$ and to $g_c(u)$ for $u\ge u_{\crit}$,
and with $\gst=\gs=-1/2$ for $u< u_{\crit}$, $\gst=\gsp=1/3$ for $u= u_{\crit}$ and
$\gst=1/2$ for $u> u_{\crit}$. As explained in Appendix~\ref{app:numerics}, we may extract from the 
first values of $m_n^{(u)}$ a numerical estimate of $2-\gst$.
Figure~\ref{fig:quadgamma} displays the $(50,5)$-estimate of $2-\gst$, obtained from the $5$-th iteration
of some convergence algorithm applied on some appropriate sequence built from $m_n^{(u)}$ for $n$ up to
$50$ (see Appendix~\ref{app:numerics} for details). We note that this estimate reproduces perfectly the
expected value $5/2$ for low $u$, up to a value of $u$ slightly above $1$ and then decreases significantly
to reach the expected value $3/2$ for large $u$. The finite size crossover between these two values is  
numerically not sharp but, remarkably, the estimated value found for $u=u_{\crit}=9/5$ is fully consistent
with the expected value $5/3$.

\begin{figure}
  \centering
  \fig{.6}{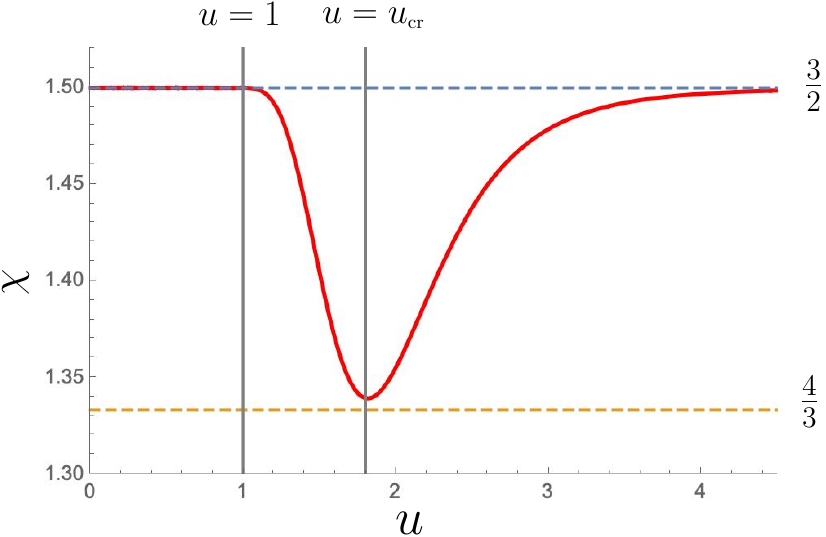}
   \caption{The $(35,6)$-estimate of the exponent $\chi$ for the asymptotics \eqref{eq:chiasymp} of $s_n^{(u)}$ as a function of $u$. We indicated by vertical lines the positions $u=1$ and $u=u_{\crit}=9/5$ and by horizontal dashed lines the
   expected values $3/2$ ($u<u_{\crit}$ and $u>u_{\crit}$) and $4/3$ ($u=u_{\crit}$).}
  \label{fig:quadDelta}
\end{figure}
We may repeat our analysis to check numerically the asymptotics of $s_n^{(u)}$, namely
\begin{equation}
s_n^{(u)}\propto \frac{g_*^{-n}}{n^{\chi}} 
\label{eq:chiasymp}
\end{equation} 
with $\chi=1+2\Delta-\gs=3/2$ for $u< u_{\crit}$, $\chi=1+2\Delta'-\gsp=4/3$ for $u=u_{\crit}$
and $\chi=3/2$ again for $u> u_{\crit}$. Figure~\ref{fig:quadDelta} displays the $(35,6)$-estimate of $\chi$: 
this estimate confirms perfectly the expected value $3/2$ for low $u$ (including the value $u=1$) and
large $u$ and decreases in the vicinity of $u=u_{\crit}$ towards a value fully consistent 
with the expected value $4/3$ at $u=u_{\crit}$.

\subsection{Distance profile in a simple block at $\boldsymbol{u=u_{\crit}}$}
\label{sec:profile}
\begin{figure}
  \centering
  \fig{.7}{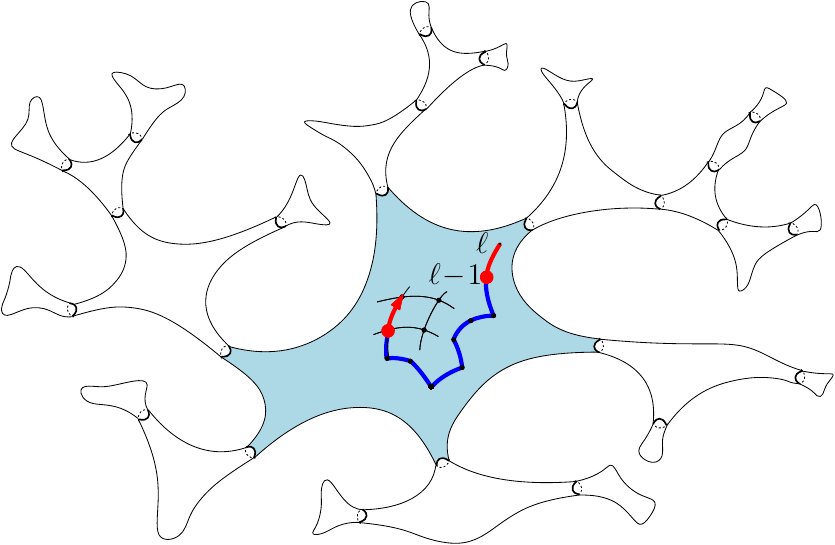}
   \caption{A schematic picture of a quadrangulation enumerated by $s^{(u_{\crit})}_{\ell;n}$.  The additional marked edge (non-arrowed red) lies in the same block as the root edge (arrowed red), and its extremities are at distance
   $\ell-1$ and $\ell$ from the origin of the root edge. The blue path represents a geodesic path between its extremities.}
  \label{fig:geodesic}
\end{figure}

In this section, we consider again rooted quadrangulations with $n$ faces, with a weight $u_{\crit}=9/5$ per simple block, and with an additional marked edge \emph{in the same block as the root edge}. Recall that the corresponding
generating function is $s^{(u_{\crit})}_n$, with the asymptotics \eqref{eq:43}.
We now wish to control the distance on the quadrangulation between the additional marked edge and the root edge.
More precisely, we introduce the generating function $s^{(u_{\crit})}_{\ell;n}$ for the configurations counted by 
$s^{(u_{\crit})}_n$ where \emph{the additional marked edge is at distance $\ell$ from the root edge}. By this, we mean that the graph distances\footnote{Recall that the graph distance between two vertices is the length ($=$ number of edges) of a shortest path linking these vertices.}
from the origin vertex of root edge to the two extremities of the additional marked edge are $\ell-1$ and $\ell$ (these distances must differ by $1$ since a planar quadrangulation is bipartite);
see Figure~\ref{fig:geodesic} for an illustration. 
We have of course the sum rule
\begin{equation*}
s^{(u_{\crit})}_n=\sum_{\ell\geq 1}s^{(u_{\crit})}_{\ell;n}\ .
\end{equation*}

At large $n$, a non-trivial value of $s^{(u_{\crit})}_{\ell;n}$ is obtained in the \emph{scaling regime} where we set
\begin{equation}
\ell=n^{\nu}\, r \quad \hbox{with} \quad \nu=\frac{1}{6} \ ,
\label{eq:scalingell}
\end{equation}
keeping $r$ finite. This particular scaling can be understood heuristically as follows: as we already mentioned, at $u=u_{\crit}$, the largest blocks in quadrangulations with $n$ faces have a number of faces of order $N\propto n^{2/3}$. Even though the number of large blocks is small compared to that of small blocks of finite size, the requirement that the root edge and the additional marked edge are in the same block imposes at large $n$ that these two edges lie in the same large block. Indeed the two edges are in a given large block with a probability proportional to $(n^{2/3}/2n)^2\propto n^{-2/3}$ while they are in the same arbitrary small block with   
 a probability proportional to $n\times (1/2n)^2\propto n^{-1}$ (the number of small blocks is proportional to $n$), hence this latter situation is asymptotically negligible. Now, in a simple quadrangulation with $N$ faces, the typical graph distance is of order $N^{1/4}$ (see for instance \cite{Minbus}), so that the typical distance in configurations enumerated by $s^{(u_{\crit})}_{n}$ is of order $(n^{2/3})^{1/4}=n^{1/6}$. Note that the scaling \eqref{eq:scalingell} concerns the distance of an additional marked edge in the same block as the root edge. If we had allowed the marked edge to be anywhere
 in the quadrangulation, the relevant scaling at $u=u_{\crit}$ would be $\ell\propto n^{1/3}$ instead \cite{FS24}.

\bigskip
In order to have a quantitative description of the distance statistics, let us introduce the continuous \emph{distance profile} in a simple block at $u=u_{\crit}$, defined as
\begin{equation*}
\rho(r)=\lim_{n\to \infty} n^{1/6} \frac{s^{(u_{\crit})}_{\lfloor n^{1/6}r\rfloor;n}}{s^{(u_{\crit})}_n}\ . 
\end{equation*}
The distance profile $\rho(r)$ is such that $\rho(r)dr$ measures, among the edges in the same block as the root edge, 
the proportion of those which lie at rescaled distance between $r$ and $r+dr$ from this root edge (\emph{i.e.}, with $\ell$ between $\lfloor n^{1/6}r\rfloor$ and $\lfloor n^{1/6}(r+dr)\rfloor$).

We can use the results of \cite{Minbus} for the statistics of distances in large simple quadrangulations to obtain
an explicit expression for $\rho(r)$. A rapid derivation of this  expression is given in Appendix~\ref{app:profile}, anticipating results from \cite{BGM25}.
\begin{figure}
  \centering
  \fig{.6}{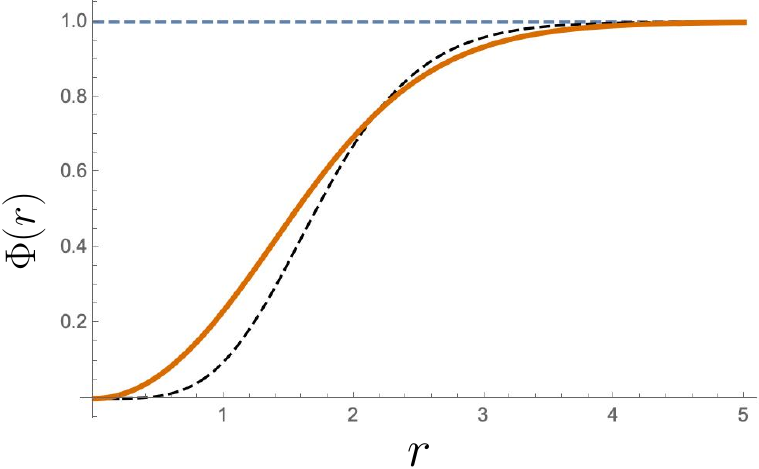}
   \caption{Plot of the cumulative distance profile $\Phi(r)$ given by \eqref{eq:secphiexp} (thick line). 
   We also indicated (dashed black line) for comparison the cumulative distance profile of ordinary
   quadrangulations, as found in \cite{BDG03}.}
  \label{fig:intprofildualIPE}
\end{figure}
We find
\begin{equation}
\rho(r)=\frac{d\, }{dr}\Phi(r)\ ,
\label{eq:rhophi}
\end{equation}
where the \emph{cumulative distance profile $\Phi(r)$} is given by
\newcommand{\co}{{\mathrm{c}}}
\newcommand{\coh}{{\mathrm{ch}}}
\newcommand{\si}{{\mathrm{s}}}
\newcommand{\sih}{{\mathrm{sh}}}
\begin{equation}
\begin{split}
\Phi(r) &=\frac{3}{\Gamma\left(\frac{4}{3}\right)}\int_0^\infty e^{-\mu^3} d\mu\, \mu^3 
\left\{1-6\frac{1-\co(\mu r^2)\, \coh(\mu r^2)+\frac{1}{\sqrt{3}}\si(\mu r^2)\, \sih(\mu r^2)}{\left(\co(\mu r^2)-\coh(\mu r^2)\right)^2}
\right\}
\\
& \hbox{with}\quad \co(x)=\cos\left(\frac{\sqrt{3x}}{2^{2/3}}\right) \ , \quad \coh(x)=\cosh\left(\frac{3\sqrt{x}}{2^{2/3}}\right) \ ,\\
& \phantom{with} \quad \si(x)=\sin\left(\frac{\sqrt{3x}}{2^{2/3}}\right) \ , \quad  \sih(x)=\sinh\left(\frac{3\sqrt{x}}{2^{2/3}}\right) \ .\\
\end{split}
\label{eq:secphiexp}
\end{equation}
A plot of the $\Phi(r)$ is presented in Figure~\ref{fig:intprofildualIPE}.
At small $r$, we have in particular 
\begin{equation}
\Phi(r)\underset{r\to 0}{\sim}
\frac{3 \Gamma \left(\frac{5}{3}\right)}{5\times 2^{4/3}
   \Gamma \left(\frac{4}{3}\right)}\, r^2\ .
\label{eq:phirsmall}
\end{equation}
\begin{figure}
  \centering
  \fig{.6}{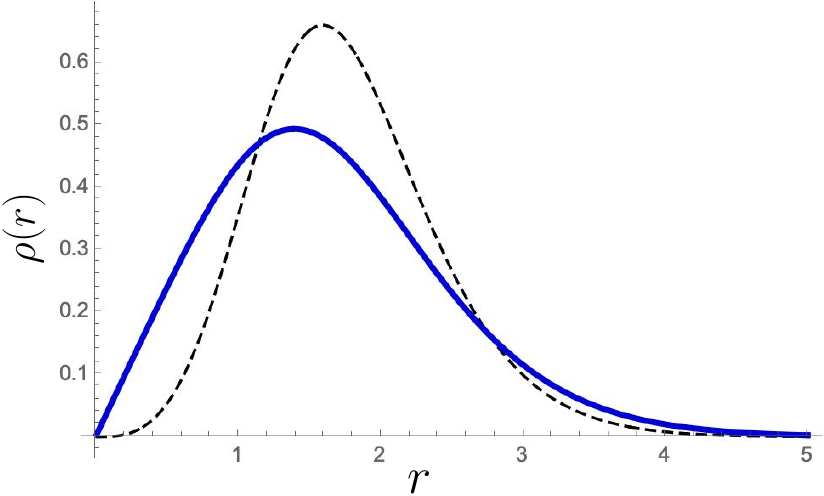}
   \caption{Plot of the distance profile $\rho(r)$ given by \eqref{eq:secrhoexp} (thick line). We also indicated (dashed line) for comparison the distance profile of ordinary
   quadrangulations; see \cite{BDG03}.}
  \label{fig:profildualIPE}
\end{figure}
Note that this behavior differs from that of ordinary quadrangulations for which $\Phi(r)\propto r^4$ (in the relevant
scaling regime $\ell=r n^{1/4}$) \cite{BDG03}. Again, a heuristic explanation for the behavior \eqref{eq:phirsmall} can be given as follows:
fix a distance $\ell$ of order $n^{1/6}$, the ball of radius $\ell$ centered at the root edge in the quadrangulation 
contains a number of edges of order $\ell^4\propto n^{2/3}$. These edges may be in the same block
as the root edge or not. At $u=u_{\crit}$, the statistic is governed by configuration where the root edge 
is in a large block of size of order $n^{2/3}$ and the probability that an edge is in the same large block as the root edge scales as $(n^{2/3}/2n)\propto n^{-1/3}$.
We thus have a total number of $n^{2/3}\times n^{-1/3}=n^{1/3} \propto \ell^2$ of edges in the same block as the root edge and at distance at most $\ell$.  
   
\bigskip   
As for $\rho(r)$, we obtain from \eqref{eq:rhophi} the integral expression
\begin{equation}
\rho(r) =\frac{3^2 2^{7/3}}{\Gamma\left(\frac{4}{3}\right)}\int_0^\infty e^{-\mu^3} d\mu\, \mu^{7/2} \sih(\mu r^2)
\frac{\co(\mu r^2)\left(\co(\mu r^2)+ \coh(\mu r^2)\right)-2}{\left(\co(\mu r^2)-\coh(\mu r^2)\right)^3}
\ ,
\label{eq:secrhoexp}
\end{equation}
so that
\begin{equation*}
\rho(r)\underset{r\to 0}{\sim}
\frac{3 \Gamma \left(\frac{5}{3}\right)}{5\times 2^{1/3}
   \Gamma \left(\frac{4}{3}\right)}\, r\ .
\end{equation*}
At large $r$, a saddle point estimate of the integral in \eqref{eq:secrhoexp} shows that
\begin{equation*}
\log(\rho(r))\propto - k\, r^\delta \quad\hbox{with}\quad \delta=\frac{6}{5}\ ,
\end{equation*}
for some positive constant $k$, in agreement with Fisher's law \cite{Fisher66} which states that $\delta=1/(1-\nu)$. 
A plot of the distance profile $\rho(r)$ is presented in Figure~\ref{fig:profildualIPE}.

\section{Hamiltonian cycles on cubic or bicubic planar maps} 
\label{sec:Hamcubic}
\subsection{Irreducible blocks for cubic maps}
\label{sec:irredblocks}
Consider now the problem of cubic planar maps equipped with a Hamiltonian cycle \cite{EKG98,DDGG23,DGG23}. 
By cubic planar map, we mean a map on the Riemann sphere 
whose all vertices have degree $3$. We denote by $2n$ the (necessarily even) number of vertices of the map, so that its number of edges is $3n$ and its number of faces $n+2$. A Hamiltonian cycle is a closed path along the edges of the map which visits each vertex of the map exactly once. We will assume that this cycle is moreover rooted, \emph{i.e.}, one of the edges it passes through is marked and oriented. Note that the cycle has length (number of visited edges) equal to $2n$ and it 
leaves exactly $n$ edges unvisited. We denote by $m_n^{(1)}$ the number of cubic planar maps with $2n$ vertices equipped with a rooted Hamiltonian cycle.
\begin{figure}
  \centering
  \fig{.8}{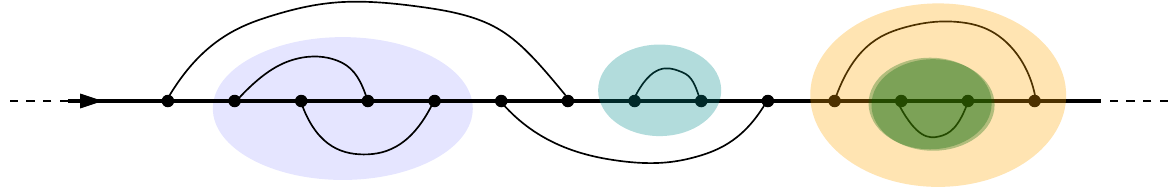}
   \caption{Representation of a (rooted) Hamiltonian cycle on a cubic planar map as a system of non-crossing arches 
   linking vertices along a straight segment. We indicated by colors its block decomposition into $5$ irreducible arch configurations (including the uncolored one containing the root). This configuration receives a weight $u^5 g^7$ in $M_u(g)$.}
  \label{fig:archescubic}
\end{figure}

By cutting the root edge in the middle and stretching the path into a straight segment with the (cut) root edge oriented from left to right, a cubic planar map with a rooted Hamiltonian cycle can be bijectively represented as a configuration of $n$ non crossing arches (corresponding to the unvisited edges of the cubic map) on both sides of a straight segment connecting by pairs $2n$ points along the segment; see Figure~\ref{fig:archescubic} for an illustration. The quantity $m_n^{(1)}$ is therefore the number of such arch configurations with $2n$ points, hence $n$ arches, so that we 
have 
\begin{equation*}
m_n^{(1)}=\sum_{k=0}^n {2n \choose 2k} \cat(k)\cat(n-k)=\cat(n)\cat(n+1)
\end{equation*}
with $\cat(n)$ as in \eqref{eq:catalan}.
We deduce in particular, at large $n$,
\begin{equation*}
m_n^{(1)}\propto \frac{(g_1)^{-n}}{n^{2-\gs}}\quad \hbox{with}\quad g_1=\frac{1}{16}\quad \hbox{and} \quad
 \gs=-1\ ,
\end{equation*}
which corresponds precisely to the case discussed in Remark~\ref{rem:remark3}.

The associated generating function (with a weight $g$ per arch)
\begin{equation*}
M_1(g)=\sum_{n\geq 0}m_n^{(1)}g^n=\frac{1-\, _2F_1\left(-\frac{1}{2},\frac{1}{2};2;16 g\right)}{2 g}
\end{equation*}
is not algebraic and has a singularity at $g=g_1=1/16$ of the form
\begin{equation*}
M_1(g)=M_1(g_1)-(g_1-g)M_1'(g_1)-\frac{512}{\pi}(g_1-g)^2\log(g_1-g)+ 
o\left((g_1-g)^2\log(g_1-g)\right)
\end{equation*}
with $M_1(g_1)=2^3 \left(1-8/(3 \pi)\right)$  and $M_1'(g_1)=2^7 \left(10/(3 \pi)-1\right)$.

\begin{rem}
\label{rem:remark5}
At this stage, we wish to mention that the number of arch configurations with $n$ arches is exactly
the same as that of tree-rooted planar maps with $n$ edges, as considered in \cite{AFZ24} so that
the two sets of configurations are in bijection. A precise connection between the two problems can be found
 in \cite{B07} but here, we shall make no use of any explicit bijection. Still, 
our results match precisely those already found in \cite{AFZ24}. Indeed, the block decomposition
of arch configurations that we will perform below leads to substitution relations 
\emph{which are exactly the same as those found in \cite{AFZ24}} for the decomposition of tree-rooted planar maps into 2-connected blocks. The two problems are therefore fully isomorphic and many of our results
can be found in \cite{AFZ24} in the tree-rooted map context.
\end{rem}

\bigskip
Let us now introduce the notion of blocks for arch configurations: we number by
$1,\ldots,2n$ the successive vertices along the straight segment from left to right. An arch configuration 
is said \emph{irreducible} if there is no proper subsegment $[i,i+1,...,i+2j]\subsetneq [1,2 ,..., 2n]$ 
such that the connections between the vertices labelled by this subsegment form a proper 
arch configuration with $j$ arches. 
Call $B(t)$ the generating function for irreducible arch configurations with a weight $t$ per arch (with a conventional
initial term $b_0=B(0)=1$). 
It is easily seen that the generating function $M_1(g)$ is related to $B(t)$ via \eqref{eq:M1}.
Indeed, let us start from a general arch configuration, as counted by $M_1(g)$, consider all the subsegments of the form 
$[i,i+1,...,i+2j]$ \emph{with $i>1$} which form a proper arch configuration and pick
the maximal ones (for the inclusion). Cutting out these subsegments and their connected arches, we end up with an irreducible arch configuration enumerated by $B(t)$. Note that, by imposing $i>1$, we always keep the component containing the first arch from the left.  Conversely, we may recover the initial configuration by inserting 
the removed maximal subsegments and their associated arch configuration. Given an irreducible arch configuration, one can perform one insertion for each segment between consecutive points or after the last point of the irreducible arch configuration: this gives $2p$ possible places for insertion if the irreducible arch configuration has $p$ arches. We thus obtain the substitution relation \eqref{eq:M1} between $M_1(g)$ and $B(t)$.
In particular, from our general analysis on Section~\ref{sec:gfduality}-Remark~\ref{rem:remark3}, even though we have no 
explicit expression for $B(t)$, we deduce the singular behavior
\begin{equation*}
B(t)=B(t_{\crit})-(t_{\crit}-t)B'(t_{\crit})-K_B(t_{\crit}-t)^2 \log(t_{\crit}-t)+ o\left((t_{\crit}-t)^2 \log(t_{\crit}-t)\right)
\end{equation*}
with $t_{\crit}= g_1\, M_1(g_1)^2= 4 \left(1-\frac{8}{3 \pi }\right)^2.$

If we iterate the cutting procedure within each of the removed maximal subsegments, we eventually obtain a decomposition of the original arch configuration into a number of blocks which are all irreducible arch configurations.
This defines the notion of blocks for an arch configuration and we may consider the generating fonction
$M_u(g)$ for arch configurations with a weight $g$ per arch and $u$ per block, which is precisely related to $B(t)$
by our canonical substitution relation \eqref{eq:mapsubst}. Figure ~\ref{fig:archescubic} displays an example of 
such decomposition into $5$ blocks. 

From \eqref{eq:ucritval} and \eqref{eq:gucr1}, we now get the explicit values (as already found in \cite{AFZ24}
in the context of tree-rooted maps)
\begin{equation}
u_{\crit}=\frac{9 \pi\, (4-\pi ) }{420\, \pi -81\, \pi ^2-512} =3.02217\cdots
\label{eq:ucritcubic}
\end{equation}
and
\begin{equation*}
g_c(u_{\crit})=\frac{\left(420\, \pi -81\, \pi ^2 - 512\right)^2}{576\, \pi^2(10-3 \pi )^2}=0.034288\cdots \ . 
\end{equation*}
From Remark~\ref{rem:remark3}, we deduce
\begin{equation*}
m_n^{(u_{\crit})}\propto \frac{g_c(u_{\crit})^{-n}}{n^{2-\gsp} (\log n)^{1/2}}\quad \hbox{with \ } \gsp=\frac{1}{2}\ ,
\end{equation*}
hence again $(1-\gs)(1-\gsp)=1$ as predicted by Liouville quantum duality.
\begin{figure}
  \centering
  \fig{.6}{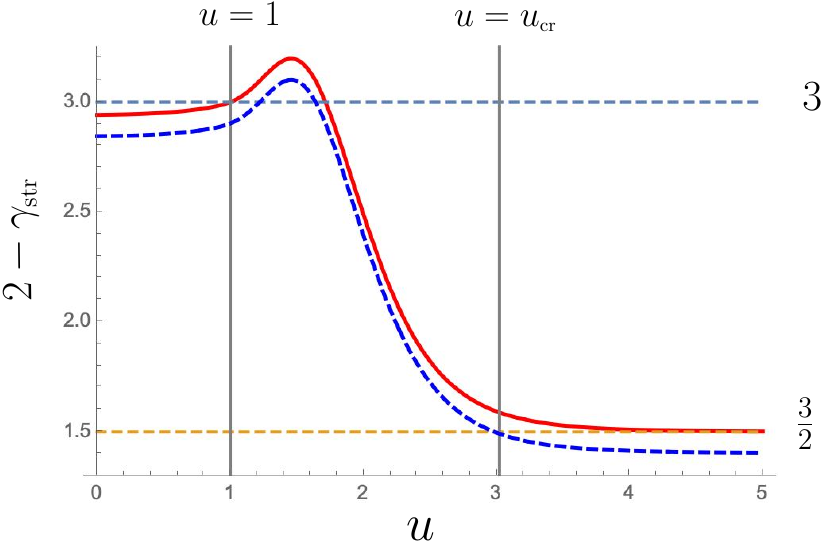}
   \caption{In red, the $(34,5)$-estimate of the exponent $2-\gst$ for the asymptotics \eqref{eq:genasymp1} of $m_n^{(u)}$ as a function of $u$. We indicated by vertical lines the values $u=1$ and $u=u_{\crit}$, as given by 
   \eqref{eq:ucritcubic}, and by horizontal dashed lines the
   expected values $3$ ($u<u_{\crit}$), $3/2$ ($u=u_{\crit}$ and $u>u_{\crit}$). The dashed blue line corresponds to using 
   the sequence $(t_n)_{n\geq 0}$ as input, with $t_n=(\log n)^{1/2} m^{(u)}_n$ for a better match in the case $u=u_{\crit}$.}
  \label{fig:cubicgamma}
\end{figure}

\medskip
For arbitrary $u$, we have the large $n$ asymptotics 
\begin{equation}
 m^{(u)}_n\propto \frac{g_*^{-n}}{n^{2-\gst}(\log n)^\eta}
  \label{eq:genasymp1}
 \end{equation}
with $g_*$ equal to $g_{\crit}(u)$ for $u< u_{\crit}$ and to $g_c(u)$ for $u\ge u_{\crit}$ and
$2-\gst=2-\gs=3$ for $u< u_{\crit}$, $2-\gst=3/2$ for $u\geq u_{\crit}$, while $\eta=1/2$ for $u= u_{\crit}$
and $\eta=0$ otherwise. Here again we may extract from the
first values of $m_n^{(u)}$ a numerical estimate of $2-\gst$.
Figure~\ref{fig:cubicgamma} displays the $(34,5)$-estimate of $2-\gst$. 
We note that our numerical estimate reproduces the
expected value $2-\gs=3$ for $u$ at and around $1$ and reaches the expected value $3/2$ for large $u$. The crossover between these two values is  
numerically smooth but, remarkably, the estimated value found for $u=u_{\crit}$ is fully consistent
with the expected value $2-\gsp=3/2$ after factoring out the $1/(\log n)^{1/2}$ correction (dashed blue line),
\emph{i.e.}, upon applying the procedure described in Appendix~\ref{app:numerics} the sequence $(t_n)_{n\geq 0}$ with
$t_n=(\log n)^{1/2}m_n^{(u)}$.
The values $\gs=-1$ and $\gsp=1/2$ obtained in this section correspond  to
the expected values \eqref{string} and \eqref{dualgammastring} for $\gamma$-LQG and ${\gamma'}-$LQG with $\gamma=\sqrt{2}=4/\gamma'$, as expected
for a model characterized by a central charge $c=-2$ \cite{DDGG23,DGG23}.
\subsection{Rooted Hamiltonian cycles on cubic maps with a second marked visited edge}
\label{sec:HP2points}
As a first example of $2$-point function, we may now consider rooted Hamiltonian cycles on cubic maps \emph{where we mark an additional visited edge}.
Again we adopt the convenient convention that a configuration where the marked visited edge is the root edge itself is counted twice.
In the arch configuration language, marking an additional edge corresponds to adding a bivalent vertex in the middle of a segment $[i,i+1]$ for $i\in \{1,\ldots, 2n-1\}$ or the segment before point $1$ or that after point $2n$ (these last two 
possibilities correspond to the marking of the root edge, counted twice as we just mentioned). Since, for a configuration with $n$ arches, there are $2n+1$ possible segments, the associated generating function $\widetilde{S}_1(g)$  is
given by 
\begin{equation}
\widetilde{S}_1(g)=\left(2g\frac{\partial\, }{\partial g}+1\right)M_1(g)\ .
\label{eq:tildeSdg}
\end{equation}
Note that we have set by convention $\widetilde{S}_1(0)=1$.

\begin{figure}
  \centering
  \fig{.8}{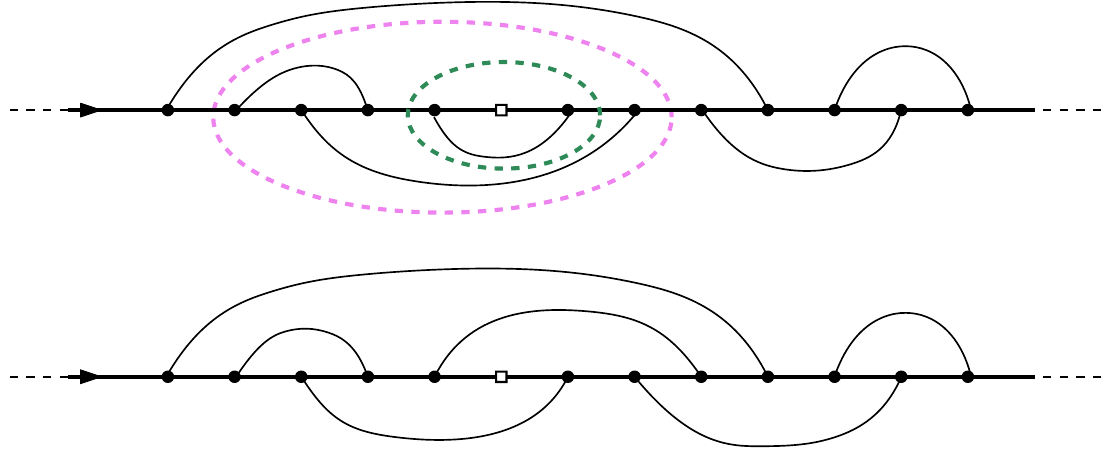}
   \caption{Top: an arch configuration with an additional bivalent vertex which is irreducible for the extended block prescription. The configuration contains two nested blocks (encircled by dashed colored lines) which separate the root edge (segment with the arrow) from the additional bivalent vertex. Such configurations are enumerated by $\widetilde{C}$. Bottom: an arch configuration with an additional bivalent vertex which is irreducible for the strict block prescription. Such configurations are enumerated by $C$.}
  \label{fig:irredcubic}
\end{figure}

Again, applied to the relation $M_1(g)=B(g\, M_1(g)^2)$, the above expression \eqref{eq:tildeSdg} immediately implies
the relation
\begin{equation}
\widetilde{S}_1(g)=M_1(g)^2\,\widetilde{C}\left(g\, M_1(g)^2\right)\quad \hbox{with \ }
\widetilde{C}(t)=\frac{1}{B(t)-2t\, B'(t)}\ .
\label{eq:tildeC}
\end{equation}
This latter substitution relation allows us to identify the quantity $\widetilde{C}(t)$ as the generating
function, with a weight $t$ per arch, for \emph{weakly} irreducible arch configurations with an additional bivalent vertex using the extended block prescription:
by this, we mean that a configuration can still contain sequences of irreducible blocks if these
blocks are nested and separate the bivalent vertex from the root edge (\emph{i.e.}, the bivalent vertex lies in
the deepest block of the sequences; see Figure~\ref{fig:irredcubic}-top). Indeed, starting from such an irreducible configuration with
$p$ arches, we can recover an arbitrary arch configuration by inserting a (possibly empty) arch configuration at 
each of the $2p+2$ segments separated by the $2p+1$ points on the straight segment (the $2p$ 
trivalent vertices and the additional bivalent one).

We may consider instead the generating
function  $C(t)$  for \emph{strongly} irreducible arch configurations with an additional bivalent vertex, defined by using now the strict block prescription: a configuration is strongly irreducible if, even after removing the 
additional bivalent vertex, it does not contain any irreducible proper block; see Figure~\ref{fig:irredcubic}-bottom. As opposed to $\widetilde{C}$ for which $\widetilde{C}(0)=1$, we choose the convention that $C(0)=0$.
We then have as before the relation
\begin{equation}
\widetilde{C}(t)=1+ C(t)\, \widetilde{C}(t)\quad \hbox{\emph{i.e.}, } C(t)=1-\frac{1}{\widetilde{C}(t)}=1-B(t)+2t\, B'(t)\ ,
\label{eq:relCCtilde}
\end{equation}
which can be understood combinatorially as follows. When going from the extended block to the strict block prescription, 
we see that a weakly irreducible arch configuration is nothing but a (possibly empty) sequence of (non empty) strongly irreducible ones, hence $\widetilde{C}(t)=1/(1-C(t))$.
To conclude, $C(t)$ in \eqref{eq:relCCtilde} has the same singular behavior 
as $B'(t)$, so that the generating function $\psi(\tau)=\tau^2\, C(\tau^2)$ of Section \ref{sec:2point} has a singularity at $\tau_\phi^2=t_{\crit}= 4 \left(1-\frac{8}{3 \pi }\right)^2$ of the form
\eqref{eq:betaonelog} corresponding to the case $\beta=1$ with a logarithmic singularity.
Again, we have $\beta=2\Delta-\gs$ with $\Delta=0$, so that the marking of an additional edge corresponds to a 2-point correlator  with quantum scaling exponent $\Delta=0$. 

If we now introduce a weight $u$ per irreducible block in strongly irreducible arch configuration and
call $S_u(g)=M_u(g)^2 C(g M_u(g)^2)$ the generating function for block-weighted arch configurations
with an additional bivalent vertex \emph{in the same block as the root edge}, we deduce from 
\eqref{eq:critlog2} that the large $n$ behavior of the $n$-th coefficient  $s^{(u_{\crit})}_n$ of $S_u(g)$ 
reads
\begin{equation*}
s^{(u_{\crit})}_n\propto \frac{g_c(u_{\crit})^{-n} (\log n)^{1/2} }{n^{1+2\Delta'-\gsp}} \quad \hbox{with}\quad
2\Delta'-\gsp=\frac{3}{2}\ .
\end{equation*} 
This corresponds to
\begin{equation*}
\Delta'=\gsp=\frac{1}{2}\ ,
\end{equation*}
so that, using $\gs=-1$ and $\Delta=0$, we have the duality relation $\Delta'=(\Delta-\gs)/(1-\gs)$ as expected.

\bigskip
\subsection{Open Hamiltonian paths on cubic maps} 
\label{sec:open}
\begin{figure}
  \centering
  \fig{.8}{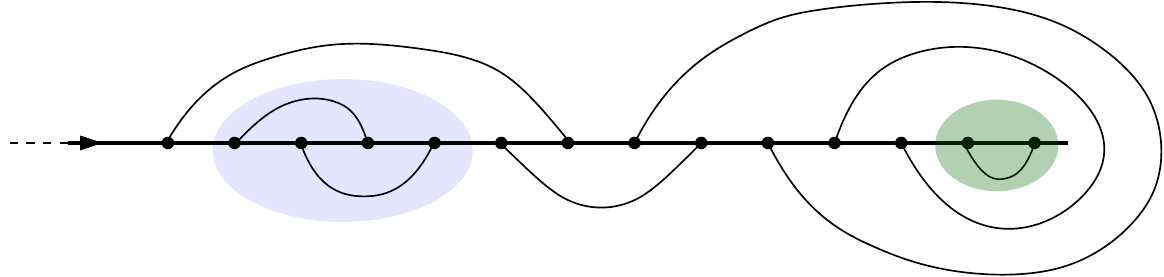}
   \caption{Representation of a Hamiltonian path on a cubic planar map as an open configuration of arches. 
   We indicated by colors its block decomposition, here with $2$ blocks which are irreducible regular arch configurations. The uncolored
   part corresponds to an irreducible open arch configuration, as enumerated by $\widetilde{\CM}$. The displayed configuration receives a weight $u^2 g^7$ in $\widetilde{\SM}_u$}
  \label{fig:openarchescubic}
\end{figure}

Another 2-point correlator is obtained by considering cubic planar maps equipped with an
oriented \emph{Hamiltonian path}, \emph{i.e.}, a path that visits all the vertices of the map 
exactly once but is no longer required to end where it started (its origin and endpoint need not being neighbors 
on the map). 
In the arch representation, this corresponds to having arches which may possibly wind around one (say the right) extremity of the straight segment, as displayed in Figure~\ref{fig:openarchescubic}. We will call \emph{open} such arch configurations 
and let $\widetilde{\SM}_1(g)$ be the associated generating function, with a weight $g$ per arch. We now have
\begin{equation*}
\widetilde{\SM}_1(g)=\sum_{n\geq 0}\widetilde{\sm}^{(1)}_n\ g^n \quad \hbox{with}\quad 
\widetilde{\sm}^{(1)}_n=\sum_{k=0}^{2n} {2n \choose k} \cat(n)=4^n\cat(n)\ .
\end{equation*}
We therefore obtain from \eqref{eq:asympcat} the large $n$ asymptotics
\begin{equation*}
\widetilde{\sm}_n^{(1)}\propto \frac{(g_1)^{-n}}{n^{1+2\Delta-\gs}}\quad \hbox{with}
\quad g_1=\frac{1}{16}\quad \hbox{and} \quad
\beta=2\Delta-\gs=\frac{1}{2}\ ,
\end{equation*} 
corresponding to
\begin{equation}
\gs=-1\ , \quad \Delta=-\frac{1}{4}\ .
\label{eq:gamdel}
\end{equation}

We may define the function $\widetilde{\CM}(t)$ through the substitution relation
\begin{equation*}
\widetilde{\SM}_1(g)=M_1(g)\, \widetilde{\CM}(g\, M_1(g)^2)\ .
\end{equation*}
The quantity $\widetilde{\CM}(t)$ is then the generating function of \emph{irreducible} open configurations,
now defined as configurations which do not contain a subsegment $[i,i+1,...,i+2j]\subset [1,2 ,..., 2n]$ 
such that the connections between the vertices labelled by this subsegment form a proper 
\emph{regular} (\emph{i.e.}, without winding arches) arch configuration with $j$ arches. Note that
an irreducible open configuration is either empty (corresponding to $\widetilde{\CM}(0)=1$) 
or necessarily contains winding arches. 

A general open arch configuration
is canonically decomposed into irreducible blocks, which are irreducible regular arch configurations,
and a single core made of a (possibly empty) irreducible open arch configuration. 
Associating a weight $u$ per irreducible block leads to a generating function
\begin{equation*}
{\SM}_u(g)=\sum_{n\geq 0}{\sm}^{(u)}_n\ g^n =M_u(g)\, \widetilde{\CM}(g\, M_u(g)^2)\ .
\end{equation*}
We have in particular ${\SM}_1(g)=\widetilde{\SM}_1(g)$, \emph{i.e.}, ${\sm}^{(1)}_n=\widetilde{\sm}^{(1)}_n$
for all $n$. 
For $u=u_{\crit}$, we then get by duality (see \eqref{eq:critlog} in the case $\beta=1/2$)
\begin{equation*}
{\sm}^{(u_{\crit})}_n\propto \frac{g_c(u_{\crit})^{-n}}{n^{1+2\Delta'-\gsp}(\log n)^{1/4}} \quad \hbox{with}\quad
2\Delta'-\gsp=\frac{\beta}{2}=\frac{1}{4}\ ,
\end{equation*}
corresponding to 
\begin{equation*}
\gsp=\frac{1}{2}\ ,\quad \Delta'=\frac{3}{8}=\frac{\Delta-\gs}{1-\gs}\ , 
\end{equation*}
in agreement with \eqref{eq:gamdel}.

\begin{figure}
  \centering
  \fig{.6}{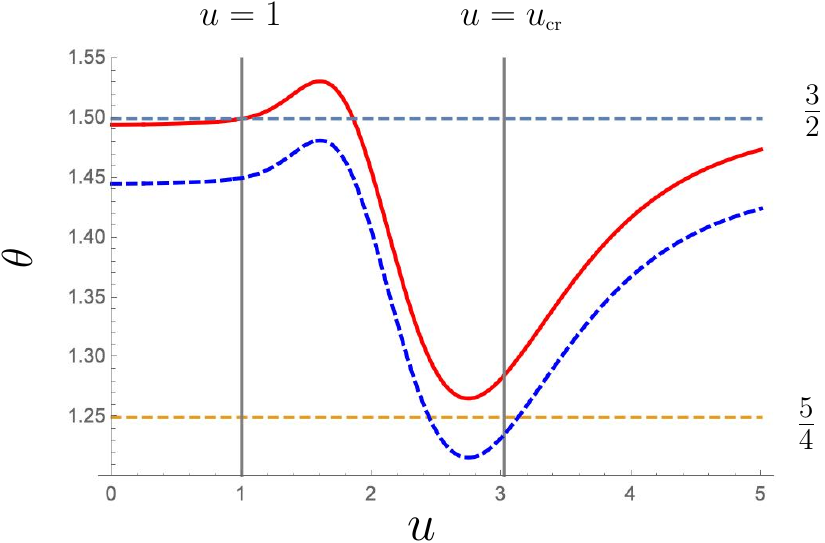}
   \caption{In red, the $(30,5)$-estimate of the exponent $\theta$ for the asymptotics \eqref{eq:genasymp2} of 
   ${\sm}_n^{(u)}$ as a function of $u$. We indicated by vertical lines the values $u=1$ and $u=u_{\crit}$, as given by 
   \eqref{eq:ucritcubic}, and by horizontal dashed lines the
   expected values $3/2$ ($u<u_{\crit}$ and $u>u_{\crit}$) and $5/4$ ($u=u_{\crit}$). The dashed blue line corresponds to using 
  the sequence $(t_n)_{n\geq 0}$ as input, with $t_n=(\log n)^{1/4} {\sm}^{(u)}_n$ for a better match in the case $u=u_{\crit}$.}
  \label{fig:cubicDelta}
\end{figure}

\medskip
For arbitrary $u$, we therefore have the large $n$ asymptotics 
\begin{equation}
{\sm}^{(u)}_n\propto \frac{g_*^{-n}}{n^{\theta}(\log n)^\eta}\ ,
\label{eq:genasymp2}
 \end{equation}
with $g_*$ equal to $g_{\crit}(u)$ for $u< u_{\crit}$ and to $g_c(u)$ for $u\ge u_{\crit}$, and
$\theta=3/2$ for $u< u_{\crit}$ and $u>u_{\crit}$, $\theta=5/4$ for $u=u_{\crit}$, while $\eta=1/4$ for $u= u_{\crit}$
and $\eta=0$ otherwise. Figure~\ref{fig:cubicDelta} displays the $(30,5)$-estimate of $\theta$. 

\bigskip\noindent
\subsection{Hamiltonian cycles on bicubic maps}
\label{sec:bicubic}
\begin{figure}
  \centering
  \fig{.8}{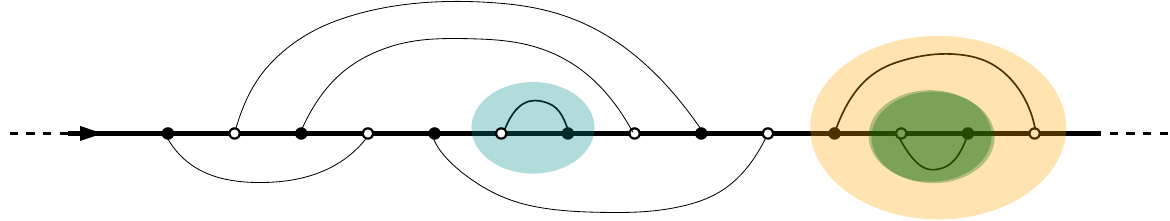}
   \caption{Representation of a (rooted) Hamiltonian cycle on a bicubic planar map as a system of non-crossing arches 
   linking black and white vertices whose colors alternate along a straight segment.}
  \label{fig:archesbicubic}
\end{figure}

We can repeat all the above discussion to describe \emph{bicubic} planar maps equipped with
rooted Hamiltonian cycle. Recall that a bicubic map is a cubic map whose vertices are colored
in black and white in such a way that any two neighboring vertices have different colors. In the 
equivalent formulation using arch configurations, we now have to color the vertices alternatively in black and white
along the straight segment and require that \emph{each arch connects vertices of different colors}; see 
Figure~\ref{fig:archesbicubic} for an illustration. We will call ``bicolored'' such arch configurations.

It has been realized in \cite{GKN99} that the problem of Hamiltonian cycles on bicubic maps lies in a different
universality class from that of Hamiltonian cycles on cubic maps; see also \cite{DDGG23,DGG23} for
a recent discussion on this subject.
We may now introduce the generating function $M_1(g)$ of Hamiltonian cycles on bicubic maps, 
which is also that of bicolored arch configurations, then $B(t)$ via \eqref{eq:M1} and 
$M_u(g)$ via \eqref{eq:mapsubst}, as well as $\widetilde{S}_1(g)$ via \eqref{eq:tildeSdg}, then $\widetilde{C}(t)$ via \eqref{eq:tildeC} and finally $C(t)$ via \eqref{eq:relCCtilde}. All these generating functions have exactly
the same interpretation as before in terms of (general, irreducible, block-weighted, $\ldots$) arch configurations 
except that we now impose that all the configurations that we encounter are bicolored.

\begin{figure}
  \centering
  \fig{.7}{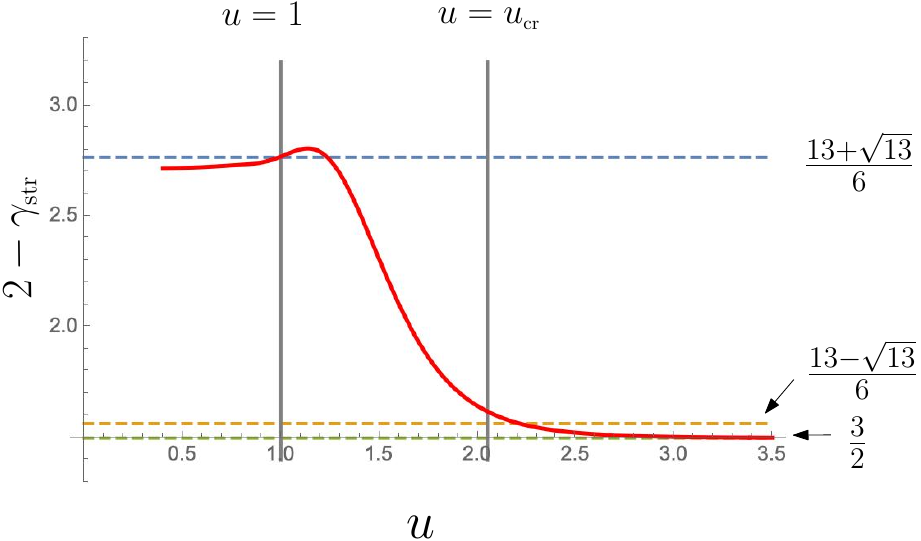}
   \caption{The $(34,5)$-estimate of the exponent $2-\gst$ for the asymptotics \eqref{eq:bicubicgamma} of $m_n^{(u)}$ as a function of $u$. The convergence algorithm becomes unstable for $u\lesssim 0.4$.
   The value of $u_{\crit}$ is estimated via \eqref{eq:ucritestim}.}
  \label{fig:bicubicgamma}
\end{figure}

With this latter constraint, the generating function $M_1(g)$ is not known, but we know its expansion in 
$g$ up to order $g^{34}$; see \cite[Table 3]{DGG23} for a complete list\footnote{We also set 
$m^{(1)}_0=1$.} of the values of $m^{(1)}_n$ for $n=1, 2,\ldots, 34$.
As explained in \cite{GKN99} (see also \cite{DDGG23,DGG23,DFG05}), it is now expected that
\begin{equation*}
m_n^{(1)}\propto \frac{(g_1)^{-n}}{n^{2-\gs}}\quad \hbox{with}\ \quad \gs=-\frac{\sqrt{13}+1}{6}\ ,
\end{equation*}
and with some unknown $g_1=g_{\crit}(1)$. 
The above value of $\gs=$ is that obtained via \eqref{string} for $\gamma$-LQG with $\gamma=(\sqrt{13}-1)/\sqrt{3}$, as expected for a model characterized by a central charge $c=-1$ \cite{GKN99,DDGG23,DGG23}.
The value of $g_1$ can be obtained numerically by considering
the sequence $(m_{n+1}^{(1)}/m_n^{(1)})$ which tends to $(g_1)^{-1}$ at large $n$. From the first values of 
$m_n^{(1)}$ for $n$ up to $34$, we estimate via some convergence algorithm
\begin{equation*}
g_1=0.098878\cdots\ .
\end{equation*}
As for the value of $u_{\crit}$, in view of the exact expression \eqref{eq:ucritval}, we can 
estimate it by considering the sequence
\begin{equation}
\frac{M_1^{[n]}(g_1)+2 g_1M_1^{[n]}{}'(g_1)}{M_1^{[n]}(g_1)(1-M_1^{[n]}(g_1))+2 g_1M_1^{[n]}{}'(g_1)}
\quad\hbox{with}\quad M_1^{[n]}(g)=\sum_{j=0}^n m_j^{(1)}g^j
\label{eq:M1n}
\end{equation}
for $n$ up to $34$ and evaluating its large $n$ limit. We get
\begin{equation}
u_{\crit}=2.053\cdots\ .
\label{eq:ucritestim}
\end{equation}
We predict by duality that 
\begin{equation*}
m_n^{(u_{\crit})}\propto \frac{g_c(u_{\crit})^{-n}}{n^{2-\gsp}} \quad \hbox{with}\ \quad \gsp=\frac{\sqrt{13}-1}{6}\ .
\end{equation*}
More generally, we predict
\begin{equation}
 m^{(u)}_n\propto \frac{g_*^{-n}}{n^{2-\gst}}
 \label{eq:bicubicgamma}
 \end{equation}
 with some $u$-dependent $g_*$ and with $\gst= \gs$ for $u<u_{\crit}$, $\gst= \gsp$ for $u=u_{\crit}$
 and $\gst=1/2$ for $u>u_{\crit}$. 
This prediction is numerically checked by computing the
 $(34,7)$-estimate of $\gst$: the results are presented in Figure~\ref{fig:bicubicgamma}.
 
\section{Meandric systems} 
\label{sec:meandres}
\subsection{Irreducible blocks}
\label{sec:irredmeandres}
\begin{figure}
  \centering
  \fig{.9}{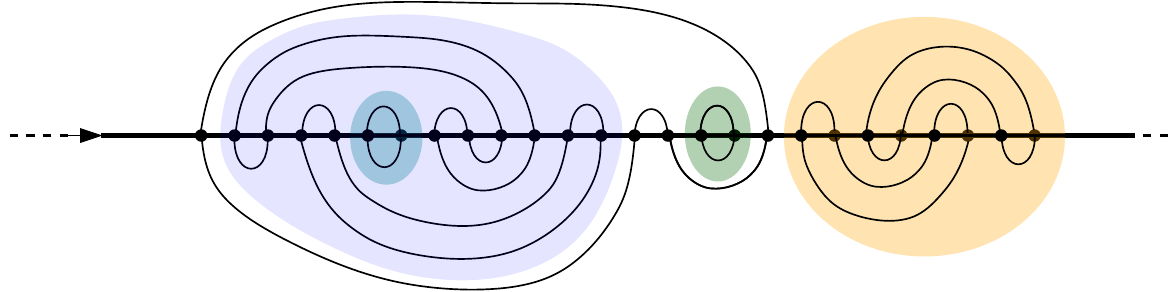}
   \caption{Representation of a (rooted) rigid Hamiltonian cycle on a quartic planar map as a meandric system, namely a pair of non-crossing arch
   configurations above and below a straight line, here in the $n=13$ case. We indicated by colors its block decomposition into $5$ irreducible meandric systems (including the uncolored one containing the root). This configuration has $7$ connected components (\emph{i.e.}, loops)
   and receives a weight $u^5 g^{13}$ in $M_u(g)$ and  $u^5 q^7 g^{13}$ in $M_u(g;q)$.}
  \label{fig:meandric}
\end{figure}

Another problem of block-weighted decorated maps corresponds to so-called meandric systems, which
are nothing but quartic  planar maps, \emph{i.e.}, maps on the Riemann sphere
whose all vertices have degree $4$, equipped with a rooted \emph{rigid} Hamiltonian cycle \cite{DGG23}. 
By rigid, we mean that, at each vertex, the Hamiltonian cycle goes straight so that the four edges around each
vertex alternate between visited and unvisited edges.  Assume that the map has $2n$ vertices, 
hence $4n$ edges, $2n$ of which are visited, and $2n+2$ faces. If we open the cycle at its root edge and stretch it into a straight line, we now obtain configurations made of two sets of $n$ non crossing arches each, connecting
$2n$ points along the segment both from above and from below; see Figure~\ref{fig:meandric}. This coding is clearly 
bijective and we will use the denomination \emph{meandric systems} both for the quartic maps with their
rigid Hamiltonian cycle and for the associated pairs of arch configurations \cite{BGP23,DGG23}. 
The number $m_n^{(1)}$ of meandric systems with a total of $2n$ arches ($n$ on each side) is given by
\begin{equation*}
m_n^{(1)}=\cat^2(n)
\end{equation*}
with $\cat(n)$ as in \eqref{eq:catalan}-\eqref{eq:asympcat}.
We deduce in particular, at large $n$,
\begin{equation*}
m_n^{(1)}\propto \frac{(g_1)^{-n}}{n^{2-\gs}}\quad \hbox{with}\quad g_1=\frac{1}{16}\quad \hbox{and} \quad
 \gs=-1\ ,
\end{equation*}
which again corresponds precisely to the case discussed in Remark~\ref{rem:remark3}.

The generating function $M_1(g)$ (with now a weight $g^{1/2}$ per arch) now reads
\begin{equation*}
M_1(g)=\sum_{n\geq 0}m_n^{(1)}g^n=\frac{\, _2F_1\left(-\frac{1}{2},-\frac{1}{2};1;16 g\right)-1}{4 g}\ .
\end{equation*}
It is not algebraic and has a singularity at $g=g_1=1/16$ of the form
\begin{equation*}
M_1(g)=M_1(g_1)-(g_1-g)M_1'(g_1)-\frac{128}{\pi}(g_1-g)^2\log(g_1-g)+ 
o\left((g_1-g)^2\log(g_1-g)\right)
\end{equation*}
with $M_1(g_1)=2^2 \left(4/\pi-1\right)$  and $M_1'(g_1)=2^6 \left(1-3/\pi \right)$.

Again, we have a natural notion of irreducibility: a meandric system is said irreducible if there is no proper 
subsegment $[i,i+1,...,i+2j]\subsetneq [1,2 ,..., 2n]$ 
such that the connections between the vertices labelled by this subsegment form a proper 
meandric system with $2j$ arches. We can then perform as before a decomposition of a meandric system
into irreducible blocks and give a weight $u$ to each block; see Figure~\ref{fig:meandric}.

\begin{figure}
  \centering
  \fig{.6}{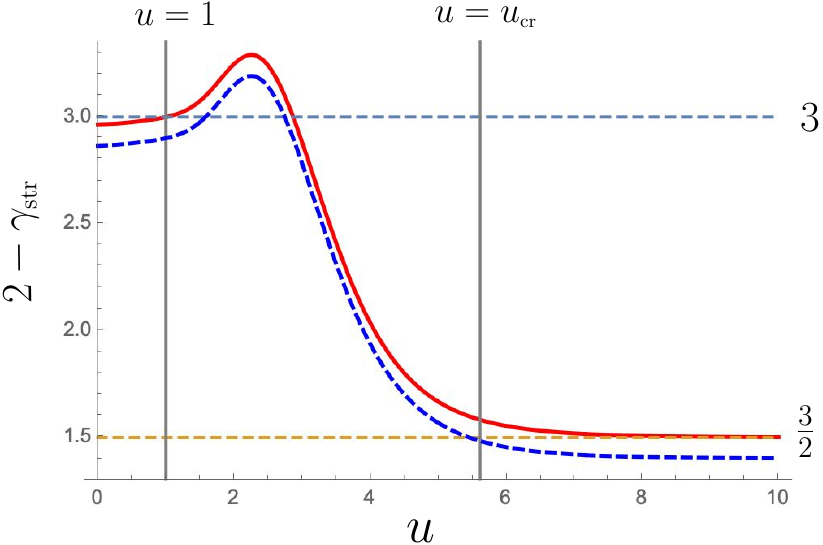}
   \caption{The $(30,5)$-estimate of the exponent $2-\gst$ for the asymptotics \eqref{eq:genasymp1} of $m_n^{(u)}$ as a function of $u$, here in the case of meandric systems. See the caption of Figure~\ref{fig:cubicgamma}.}
  \label{fig:quarticgamma}
\end{figure}

All the results obtained in Section~\ref{sec:Hamcubic} can be transposed \emph{mutatis mutandis}
to the present problem, so that
\begin{equation*}
m_n^{(u_{\crit})}\propto \frac{g_c(u_{\crit})^{-n}}{n^{2-\gsp} (\log n)^{1/2}}\quad \hbox{with \ } \gsp=\frac{1}{2}\ ,
\end{equation*}
with the only difference that $t_{\crit}$, $u_{\crit}$ and $g_c(u_{\crit})$ are now given by
\begin{equation}
t_{\crit}=\left(\frac{4}{\pi}-1\right)^2\ , \quad u_{\crit}=\frac{\pi\, (\pi-2)}{30 \pi  -3 \pi ^2-64}
\ , \quad g_c(u_{\crit})=\frac{\left(30 \pi  -3 \pi ^2-64\right)^2}{64 \pi ^2(\pi -3)^2}\ .
\label{eq:valquartic}
\end{equation}
Figure~\ref{fig:quarticgamma} is the analog for meandric systems of Figure~\ref{fig:cubicgamma}
for cubic maps carrying a Hamiltonian cycle.

\medskip
\subsection{Meandric systems weighted by both connected components and irreducible blocks} 
\label{sec:weightq}
The meandric systems can be classified according to their \emph{number of connected components}, defined as the number of connected components of the (in general disconnected) graph obtained by removing the straight segment; see Figure~\ref{fig:meandric} for an example.
Let us denote by $m_{n,k}^{(1)}$ the number of meandric systems with a total of $2n$ arches and $k$ connected components, with necessarily $1\leq k\leq n$.
We wish to assign a weight $q$ per connected components: this means that we consider the quantity
\begin{equation*}
m_n^{(1)}(q)= \sum_{k=1}^n m_{n,k}^{(1)}\, q^k
\end{equation*}
and the associated generating function
\begin{equation*}
M_1(g;q)= \sum_{n\geq 0} m_n^{(1)}(q)\, g^n\ .
\end{equation*}
We define similarly the quantities $m_n^{(u)}(q)$ and $M_u(g;q)$ where the meandric systems receive
in addition a weight $u$ per irreducible block.
\begin{figure}[h!]
  \centering
  \fig{.6}{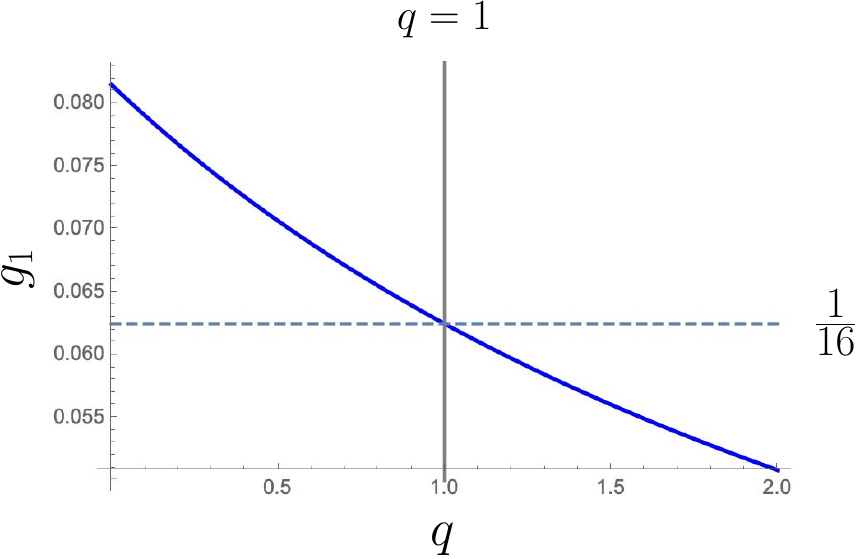}
   \caption{Numerical estimate of $g_1$ as a function of $q$. At $q=1$, the estimate matches the exact value $1/16$.}
  \label{fig:gcritmeanders}
\end{figure}

The previous analysis corresponds to the case $q=1$. For $0\leq q\leq 2$, 
it has been conjectured and verified numerically \cite{DGG00,DGJ00} (see also \cite{DFG05}) that\footnote{Note that,
for $q=0$, $m_n^{(1)}(0)$ should be understood as the 
$q\to 0$ limit of $m_n^{(1)}(q)/q$.}
\begin{equation}
\begin{split}
&m_n^{(1)}(q)\propto \frac{(g_1(q))^{-n}}{n^{2-\gs(q)}}\quad \hbox{with}\ \quad \gs(q)=\frac{c-1
-\sqrt{(25-c)(1-c)}}{12}\\
& \hbox{where}\quad c=-1-6 \frac{e^2}{1-e}\ , \quad e=\frac{1}{\pi}\arccos \frac{q}{2}\ , \\
\end{split}
\label{eq:gammaq}
\end{equation}
for some unknown $g_1(q)$. In particular, for
$q > 1$, we are in the desired regime \eqref{eq:alpharange}, $1<\alpha <2$, with
$\alpha=1-\gs$. More precisely, $\gs(q)$ increases from $-1$ to $-(1+\sqrt{13})/6$ when $q$ increases from $1$ to $2$.
\begin{figure}[h!]
  \centering
  \fig{.7}{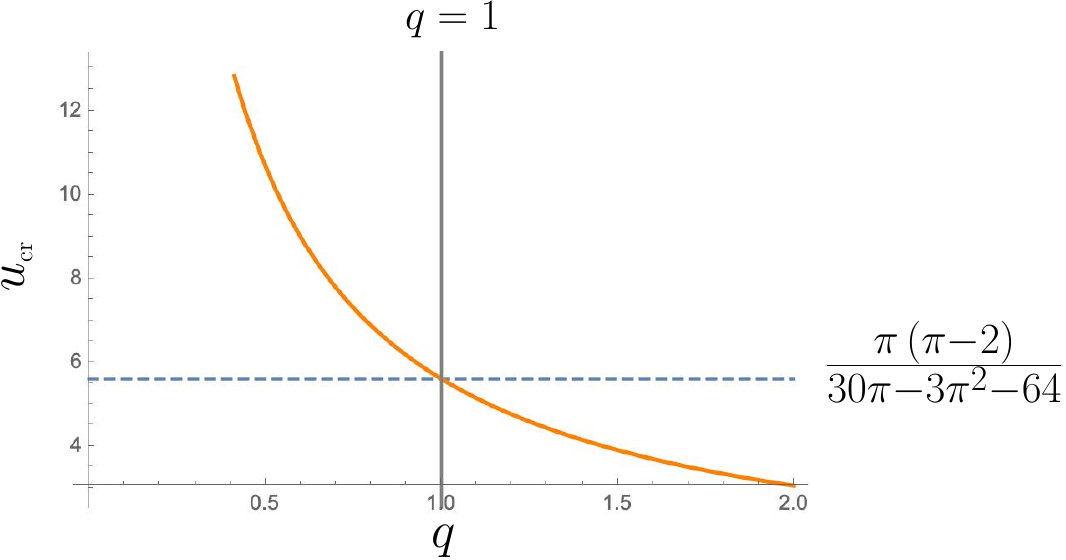}
   \caption{Numerical estimate of $u_{\crit}$ as a function of $q$. At $q=1$, the estimate matches the exact value
    given in \eqref{eq:valquartic}. The value of $u_{\crit}$ increases for decreasing $q$ and tends to $+\infty$ when $q\to 0$.}
  \label{fig:ucritmeanders}
  \end{figure}
  
For $1<q<2$, we predict from the duality relation $(1-\gs)(1-\gsp)=1$ that, at the critical point $u_{\crit}$ (which depends on $q$), we have 
\begin{equation}
m_n^{(u_{\crit})}(q)\propto \frac{(g_c(u_{\crit}))^{-n}}{n^{2-\gsp(q)}}\quad \hbox{with}\ \quad \gsp(q)=\frac{c-1
+\sqrt{(25-c)(1-c)}}{12}
\label{eq:gammaqprime}
\end{equation}
with $c$ as in \eqref{eq:gammaq}.

\bigskip 
The values of $m_{n,k}$ for $n=1,\ldots,20$ and $1\leq k\leq n$ are stored in the sequence A008828
of The On-Line Encyclopedia of Integer Sequences \cite{OEISA008828}. From these data, we can extract a 
numerical estimate of $g_1=g_1(q)$ as the large $n$ limit of $m_n^{(1)}(q)/m_{n+1}^{(1)}(q)$. This estimate is 
represented in Figure~\ref{fig:gcritmeanders} for $0\leq q \leq 2$. We can then estimate the value 
$u_{\crit}=u_{\crit}(q)$ as we did for bicubic maps, \emph{i.e.}, upon using the sequence defined in \eqref{eq:M1n}
(with $m_j^{(1)}\to m_j^{(1)}(q)$) and evaluating its large $n$ limit. The estimated value is 
displayed in Figure~\ref{fig:ucritmeanders}. Note that $u_{\crit}\to \infty$ for $q\to 0$. Indeed, the limit $q\to 0$
selects meandric systems with a single connected component. These configurations are automatically irreducible,
which implies that the system is subcritical for any finite value of $u$.

\begin{figure}[h!]
  \centering
  \fig{.7}{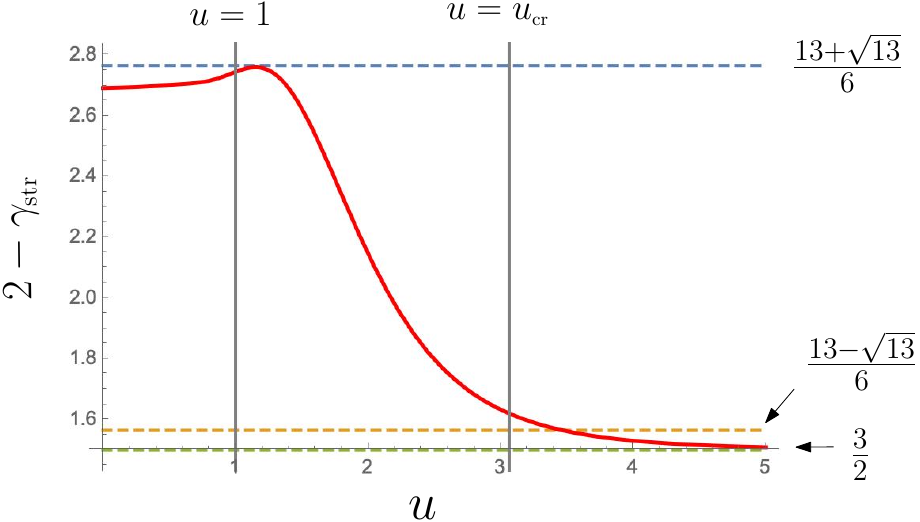}
   \caption{The $(20,5)$-estimate of the exponent $2-\gst$ for the asymptotics \eqref{eq:bicubicgamma} of $m_n^{(u)}=m_n^{(u)}(q)$ for $q=2$ as a function of $u$. We represented the expected values at $u=1$ and 
   $u=u_{\crit}$ as given by \eqref{eq:gammaq} and \eqref{eq:gammaqprime}, namely
   $2-\gs(2)=(13+\sqrt{13})/6$ and $2-\gsp(2)=(13-\sqrt{13})/6$ respectively.}
  \label{fig:q2meandersgamma}
\end{figure}

Finally, we displayed in Figure~\ref{fig:q2meandersgamma} the $(20,5)$-estimate of the exponent $2-\gst$ for the asymptotics \eqref{eq:bicubicgamma} of $m_n^{(u)}=m_n^{(u)}(q)$ for $q=2$ as a function of $u$. 
From \eqref{eq:gammaq} and \eqref{eq:gammaqprime}  with $q=2$ (\emph{i.e.}, $e=0$, $c=-1$), the expected value of 
$2-\gst$ is $2-\gs(2)=(13+\sqrt{13})/6$  for $u<u_{\crit}$, $2-\gsp(2)=(13-\sqrt{13})/6$ for 
 $u=u_{\crit}$ and $3/2$ for  $u>u_{\crit}$. Note that these values are the same 
 as those for bicubic maps. Our numerical estimate is fully compatible with the expected values, 
 even though the agreement is not as good as for bicubic maps; see Figure~\ref{fig:bicubicgamma}. This is due to 
 the fact that our $(N,5)$-estimate is limited to $N=20$ here as opposed to $N=34$ for bicubic maps.

\section{Duality and Hausdorff dimensions}
\label{sec:conclu}
In this paper, we used analytic combinatorics to derive a number of universal duality relations for the exponents characterizing the asymptotics of block-weighted decorated planar maps in their subcritical and critical phases.
We identified these relations as examples of Liouville quantum duality properties as predicted by LQG. 
In particular, we found that block-weighted decorated planar maps at their critical point $u=u_{\crit}$ 
provide a realization of $\gamma'$-LQG with $\gamma'>2$, the so-called ``dual branch of gravity''. 

Beyond critical exponents, it would be interesting to further explore the mass distribution in the various blocks
at $u=u_{\crit}$, with the aim of reproducing at the map level the construction of the measure in $\gamma'$-LQG
for $\gamma'>2$ presented in Section~\ref{sec:dualqmeasure}.
Here again, the analytic combinatorial approach seems to be very promising. 

Finally, we gave an explicit expression for the universal distance profile between points \emph{in
the same block} at $u=u_{\crit}$ in the particular case of pure quantun gravity (\emph{i.e.}, $c=0$, $\gamma^2=8/3$,
${\gamma'}^{\, 2}=6$),
for which we found a Hausdorff dimension $1/\nu=6$. 
The scaling argument presented in Section~\ref{sec:profile}, to heuristically explain the latter result \eqref{eq:scalingell},
can be generalized to other values of $\gamma'$. The Hausdorff dimension of \emph{a single (large) bubble} in the critical regime is given by
\begin{equation}
\widetilde{d}(\gamma'):= \alpha\, d(\gamma)=\frac{4}{\gamma^2}d(\gamma)\ ,
\label{eq:dtildeprime}
\end{equation}
where $d(\gamma)$ is the Hausdorff dimension of $\gamma<2$-LQG.
Here we used the relations established in this paper, namely $\alpha=1-\gs=4/\gamma^2=\gamma'/\gamma$.
The above relation can thus be recast in the dual form
\begin{equation}
\frac{\widetilde{d}(\gamma')}{\gamma'}= \frac{d(\gamma)}{\gamma}\ .
\label{eq:dtildevsd}
\end{equation}
The question of the value of Hausdorff dimension $d(\gamma)$ for
surfaces described by $\gamma$-LQG is still open. Notice that the formula conjectured by Ding and Gwynne \cite[(1.16)]{MR4076090} for $\gamma<2$ yields the form
$$\frac{d^{\textrm{Quad}}(\gamma)}{\gamma}=\frac{2}{\gamma}+\frac{\gamma}{2}+\frac{1}{\sqrt{6}}\ ,$$ 
which is invariant under duality. Because of \eqref{eq:dtildevsd}, the validity of $d(\gamma)=d^{\textrm{Quad}}(\gamma)$ would imply the identity $\widetilde{d}(\gamma')=d^{\textrm{Quad}}(\gamma')$, \emph{i.e.}, the extension to $\gamma'>2$ of the
conjectured formula.

The Hausdorff dimension $\widetilde{d}(\gamma')$ \eqref{eq:dtildeprime} of a single bubble should not
be confused with the Hausdorff dimension $D(\gamma')$ of the \emph{whole map} in the critical regime.
For instance, for maps without decoration (pure gravity), we have seen that $\widetilde{d}(\sqrt{6})=6$
whereas $D(\sqrt{6})=3$ \cite{FS24}.
We state that the latter dimension is given by the general formula
\begin{equation}
\label{eq:deprime}
D(\gamma')=\frac{\alpha}{\alpha-1}=\frac{1}{1-4/{\gamma'}^{\, 2}}=\frac{1}{1-\gamma^2/4}=\frac{1}{\gsp}\ ,
\end{equation}
associated with the rescaling of the distance function and height process on an $\alpha$-stable tree
\cite[Proposition 5.12]{ZSPhD}, in relation to the construction of the dual measure \eqref{dualmeasure}.
In contradistinction to the case of standard $\gamma$-LQG for $\gamma<2$, where the Liouville Hausdorff dimension
$d(\gamma)$ is unknown,
the Hausdorff dimension $D(\gamma')$ \eqref{eq:deprime} of the dual Liouville quantum gravity for $\gamma'>2$ can thus be explicitly 
determined.

\section*{Acknowledgements} We would like to thank Z\'ephyr Salvy for the crucial discussions we had at the start of this work, as well as Hugo Manet for fruitful exchanges. EG is partially supported by the ANR  grant CartesEtPlus ANR-23-CE48-0018.
\appendix
\section{Numerical estimates of critical exponents}
\label{app:numerics}
All the combinatorial quantities $t_n$ (implicitly depending on $u$, e.g.\ $m_n^{(u)}$ or $s_n^{(u)}$) that  we consider in this paper have their large $n$ asymptotics of the
form\footnote{In some cases, we have to incorporate in $t_n$ a power of $\log n$ to match this form.}
 \begin{equation}
 t_n\propto \frac{g_*^{-n}}{n^{\delta}}
 \label{eq:asympto}
 \end{equation}
 for some critical weight $g_*$ (equal to $g_c(u)$ or $g_{\crit}(u)$ depending 
 on whether $u$ is larger or smaller than $u_{\crit}$) and with some critical exponent $\delta$ depending
on the quantity $t_n$ at hand. 
 
 Now, from the knowledge of \emph{the first values of $t_n$} for $n=0,1,\ldots N$, we may obtain a numerical 
 estimate of the exponent $\delta$ as follows: we first construct from the sequence $t_n$ the sequence
 \begin{equation}
 \delta_n:=n^2\, \left(\frac{t_{n+2}\, t_n}{t_{n+1}^2}-1\right)\ ,
 \end{equation}
which, from \eqref{eq:asympto}, is such that 
\begin{equation*}
\delta_n\underset{n\to\infty}{\rightarrow} \delta \ . 
\end{equation*}
The convergence of $\delta_n$ to $\delta$ is in general quite slow, with typically $\delta_n=\delta+\O(1/n)$ at large $n$.
To improve our estimate of $\delta$, we have recourse to some series acceleration methods, \emph{i.e.}, 
build from $\delta_n$ new sequences for which the speed of convergence is faster. More precisely,
let us introduce the iterated finite difference operators $\Delta^p$, $p\in \N^*$, defined by
\begin{equation*}
\begin{split}
&(\Delta f)_N:=f_{N+1}-f_N\ ,\\
&(\Delta^2 f)_N:=(\Delta(\Delta f))_N=f_{N+2}-2f_{N+1}+f_N\ , \ \ldots  \\ 
\end{split}
\end{equation*}
and consider the new sequences 
\begin{equation*}
\tilde{\delta}^{(p)}_n:=\frac {1}{p!}\left(\Delta^p \hat{\delta}^{(p)}\right)_n \quad \hbox{with} \quad \hat{\delta}^{(p)}_n:=n^p \delta_n\ .
\end{equation*}
It is easily seen that the sequence $(\tilde{\delta}^{(p)}_n)_{n\geq 0}$ converges to $\delta$ with a speed that increases for increasing $p$ (with 
typically $\tilde{\delta}^{(p)}_n=\delta+\O(1/n^{p+1})$) and we thus obtain a better estimate of 
$\delta$ by taking a larger value of $p$. In practice, if we are limited to some maximal value $n=N$ for $t_n$, the
sequence $\tilde{\delta}^{(p)}_n$ is limited to a maximal value of $n$ equal to $N-2-p$ and the value of
$\tilde{\delta}^{(p)}_{N-2-p}$ is built from the values $t_m$ for $m$ in the range $m=N,N-1,\ldots N-2-p$, \emph{i.e.}, using smaller values of $m$ for larger $p$. The good choice of $p$ is therefore a compromise between letting $p$ increase to get a better acceleration of the convergence and keeping $p$ small enough to avoid using $t_m$ with a too small index $m$. We use typically a value of $p$ in the range $p=5,6,\ldots, 10$ for $N$ of order $35$ or so. We will
refer to $\tilde{\delta}^{(p)}_{N-2-p}$ as \emph{the $(N,p)$-estimate of $\delta$}.

\section{Distance profile in a simple block at $\boldsymbol{u=u_{\crit}}$: detailed calculation}
\label{app:profile}
Recall that the generating function $C(t)$ for rooted quadrangulations with an additional marked edge different from the root edge, with a weight $t$ per face, and without multiple edges, is given explicitly by \eqref{eq:CbarC} and \eqref{eq:barCpure}.  In particular, when $t$ tends to $t_{\crit}=4/27$ as
\begin{equation}
t=\frac{4}{27}\left(1-\eta^2 +\O(\eta^3)\right)\ ,
\label{eq:texp}
\end{equation}
we have the expansion
\begin{equation}
C(t)=\frac{5}{9}\left(1-\frac{8}{5\sqrt{3}}\eta+\O(\eta^2)\right)\ .
\label{eq:Ccritexp}
\end{equation}

Let us now set $u=u_{\crit}=9/5$ and consider, as in Section~\ref{sec:pure2p}, the generating function $S_{u_{\crit}}(g)$ for rooted quadrangulations with an additional marked edge, with a weight $g$ per face and $u_{\crit}$ per simple block, and \emph{such that the additional marked edge is in the same block as the root edge}. As we have seen in Section~\ref{sec:pure2p}, It is related to $C$ via the substitution relation
\begin{equation}
S_{u_{\crit}}(g)=C(g\, M_{u_{\crit}}(g)^2)\ .
\label{eq:appsubst}
\end{equation}
Now, for $ g$ approaching $g_c(u_{\crit})=25/432$, namely
\begin{equation*}
g=\frac{25}{432}(1-\epsilon^3)
\end{equation*}
with $\epsilon$ small, we have the expansions \eqref{eq:Mucritexp} and \eqref{eq:Sucrexp}. More precisely,
we have from \eqref{eq:MuEq}
\begin{equation*}
M_{u_{\crit}}(g)=\frac{8}{5}\left(1-\left(\frac{27}{32}\right)^{1/3}\!\!\!\epsilon^2 +\O(\epsilon^3)\right) , \quad g\, M_{u_{\crit}}(g)^2=\frac{4}{27}\left(1-\left(\frac{27}{4}\right)^{1/3}\!\!\!\epsilon^2 +\O(\epsilon^3)\right)  ,
\end{equation*}
so that, using \eqref{eq:texp}-\eqref{eq:Ccritexp} with $\eta=(27/4)^{1/6}\, \epsilon$, we get \eqref{eq:Sucrexp}, namely
\begin{equation}
S_{u_{\crit}}(g)=\frac{5}{9}\left(1-\left(\frac{256}{125}\right)^{1/3}\epsilon+\O(\epsilon^2)\right)\ .
\label{eq:Sucrexp2}
\end{equation}
\begin{figure}[h!]
  \centering
  \fig{.9}{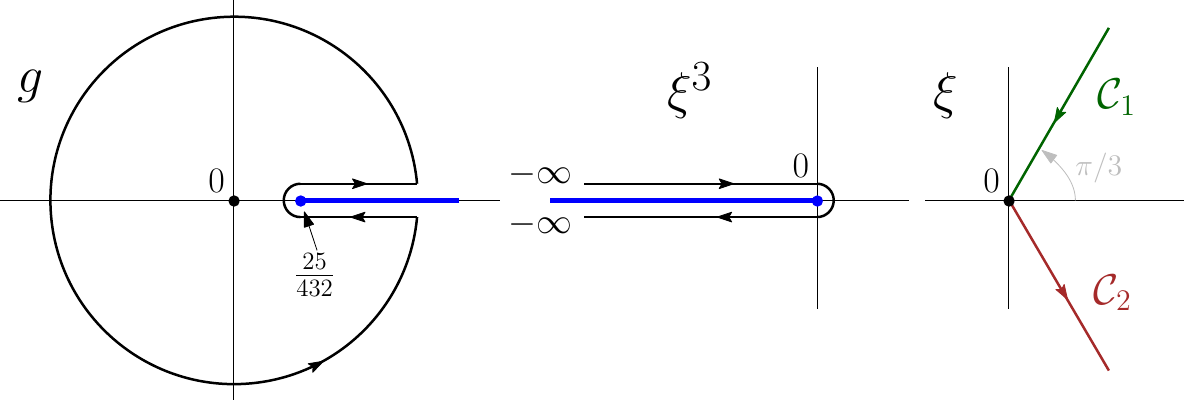}
   \caption{Left: in the integral \eqref{eq:contint}, we deform the contour into a circle of radius strictly larger than $g_c(u_{\crit})=25/432$ (which contributes $0$ at large $n$) with a notch that comes back near and to the left of 
 $g=25/432$. Middle: after the change of variable \eqref{eq:appgxi}, this contour in the $\xi^3$ complex plane
 encircles the real half line $\Re(\xi^3)\leq 0$. Right: in the $\xi$ complex plane, this corresponds 
 to travel as shown, with the two half-lines $\mathcal{C}_1$ and $\mathcal{C}_2$ at angle $\pm \pi/3$ from
 the real line.}
  \label{fig:contourxi}
\end{figure}
We can use this expansion to obtain the large $n$ asymptotics of the $n$-th coefficient of $S_{u_{\crit}}(g)$ using
the contour integral representation
\begin{equation}
[g^n] S_{u_{\crit}}(g)=\frac{1}{2\hbox{i}\pi}\oint \frac{dg}{g^{n+1}}S_{u_{\crit}}(g)\ .
\label{eq:contint}
\end{equation}
Setting
\begin{equation}
g=\frac{25}{432}\left(1-\frac{\xi^3}{n}\right)
\label{eq:appgxi}
\end{equation}
we obtain from \eqref{eq:Sucrexp2} with $\epsilon=\xi/n^{1/3}$ the large $n$ equivalent
\begin{equation*}
[g^n] S_{u_{\crit}}(g)\sim \left(\frac{432}{25}\right)^n \frac{1}{2\hbox{i}\pi} \int_{\mathcal{C}_1\cup\mathcal{C}_2} e^{\xi^3} d\xi\, \left(-\frac{3 \xi^2}{n}\right)
\frac{5}{9}\left\{1-\left(\frac{256}{125}\right)^{1/3}\frac{\xi}{n^{1/3}} \right\}
\end{equation*}
where $\mathcal{C}_1$ and $\mathcal{C}_2$ are the contours shown in Figure~\ref{fig:contourxi}-right. In particular, by symmetry, the term $1$ in the curly bracket
(which corresponds to a regular term in the expansion \eqref{eq:Sucrexp2} of $S_{u_{\crit}}(g)$ around $g_c(u_{\crit})$) 
does not contribute and, setting $\xi=\mu e^{\mathrm{i}\pi/3}$ along $\mathcal{C}_1$ and $\xi=\mu e^{-\mathrm{i}\pi/3}$
along $\mathcal{C}_2$, with $\mu$ real positive, we obtain finally
\begin{equation*}
\begin{split}
[g^n] S_{u_{\crit}}(g)&\sim \left(\frac{432}{25}\right)^n \frac{1}{2\hbox{i}\pi} \frac{1}{n^{4/3}} \int_0^\infty e^{-\mu^3} d\mu\, 3\mu^2 
\frac{5}{9}\left(\frac{256}{125}\right)^{1/3}\left(\mu e^{\mathrm{i}\pi/3}-\mu e^{-\mathrm{i}\pi/3}\right)\\
&\sim \left(\frac{432}{25}\right)^n\frac{1}{\pi\, n^{4/3}}\frac{5}{9}\left(\frac{256}{125}\right)^{1/3}\Gamma\left(\frac{4}{3}\right) 
\frac{\sqrt{3}}{2}\ .\\
\end{split}
\end{equation*}

In \cite{Minbus}, the authors introduced some refinement of $C(t)$, namely the generating function
$C_\ell(t)$ for rooted quadrangulations with an additional marked edge different from the root edge, with a weight $t$ per face, without multiple edges, and such that the \emph{distance between the additional marked edge and the root edge
is $\ell$}. By this, we mean more precisely that the graph distances
from the origin vertex of root edge to the extremities of the additional marked edge are $\ell-1$ and $\ell$ (the distances
of neighboring vertices on a quadrangulation necessarily differ by $1$).
We have of course
\begin{equation*}
C(t)=\sum_{\ell\geq 1} C_\ell(t)\ .
\end{equation*}
It was moreover shown \cite[Equation (3.33)]{Minbus} that for $t$ as in \eqref{eq:texp} and 
\begin{equation*}
\ell=\frac{L}{\eta^{1/2}}\ ,
\end{equation*}
we have\footnote{The correspondence with \cite{Minbus}
is via the identification of the variable $z$ in this reference with our variable $t$, and of
$g_\ell(z)$ with $C_\ell(t)$.} 
\begin{equation*}
C_\ell(t)\sim -\frac{8}{9} \eta^{3/2}\mathcal{F}'\left(L;\frac{3^{1/4}}{2^{1/2}}\right)
\quad \hbox{with}\quad  \mathcal{F}(L;\sigma)=\frac{2}{3}\sigma^2\left(1+\frac{3}{\sinh^2(\sigma\, L)}\right)\ ,
\end{equation*}
and $\mathcal{F}'(L;\sigma):=d\mathcal{F}(L;\sigma)/dL$.

Let us now introduce, for $\ell\geq 1$, the generating function $S^{(\ell)}_{u_{\crit}}(g)$ for quadrangulations with an additional marked edge, with a weight $g$ per face and $u_{\crit}$ per simple block, and \emph{such that the additional marked edge is in the same block as the root edge and at distance $\ell$ from it}.
We have of now the substitution relation
\begin{equation*}
S^{(\ell)}_{u_{\crit}}(g)= C_\ell(g\, M_{u_{\crit}}(g)^2)
\end{equation*}
since the substitution mechanism that leads to \eqref{eq:appsubst} does not modify the distances between the root
edge and any edge which lies in the same block as the root edge (clearly a path cannot ``lower its length'' by entering another block). Using
\begin{equation*}
[g^n] S^{(\ell)}_{u_{\crit}}(g)=\frac{1}{2\hbox{i}\pi}\oint \frac{dg}{g^{n+1}}S^{(\ell)}_{u_{\crit}}(g)\ ,
\end{equation*}
and taking $g$ as in \eqref{eq:appgxi}, which amounts to take $\eta=(27/4)^{1/6}\, \epsilon= (3^{1/2}/2^{1/3})\, \xi/n^{1/3}$
so that 
\begin{equation*}
\ell=n^{1/6}\, r \quad \hbox{with}\quad r=  \frac{2^{1/6}}{3^{1/4}}\frac{L}{\sqrt{\xi}}\ ,
\end{equation*}
we eventually obtain
\begin{equation*}
\begin{split}
[g^n] S^{(\ell)}_{u_{\crit}}(g)&\sim \left(\frac{432}{25}\right)^n \frac{1}{2\hbox{i}\pi} \int_{\mathcal{C}_1\cup\mathcal{C}_2}\!\!\!\!e^{\xi^3} d\xi\, \frac{3 \xi^2}{n}
\frac{8}{9}\left\{\left(\frac{3^{1/2}}{2^{1/3}} \right)^{3/2}\frac{\xi^{3/2}}{n^{1/2}}\mathcal{F}'\left(\frac{3^{1/4}}{2^{1/6}}\sqrt{\xi}\, r;\frac{3^{1/4}}{2^{1/2}}\right)
\right\}\\
&\sim \left(\frac{432}{25}\right)^n \frac{1}{2\hbox{i}\pi} \int_{\mathcal{C}_1\cup\mathcal{C}_2}\!\!\!\!e^{\xi^3} d\xi\, \frac{3 \xi^2}{n}
\frac{8}{9}\left\{\frac{3^{1/2}}{2^{1/3}} \frac{\xi}{n^{1/2}}\frac{d}{dr} \mathcal{F}\left(\frac{3^{1/4}}{2^{1/6}}\sqrt{\xi}\, r;\frac{3^{1/4}}{2^{1/2}}\right)
\right\}\\
&\sim \left(\frac{432}{25}\right)^n\frac{2^{5/3}}{3^{1/2}} \frac{1}{\pi\, n^{3/2}} \frac{d}{dr}
\Lambda(r)
\\
\end{split}
\end{equation*}
with
\begin{equation*}
\begin{split}
\Lambda(r)=\frac{1}{\hbox{i}}\int_0^\infty e^{-\mu^3} d\mu\, \mu^2 
&\left\{\mu e^{\mathrm{i}\pi/3} \mathcal{F}\left(\frac{3^{1/4}}{2^{1/6}}\sqrt{\mu}e^{\mathrm{i}\pi/6}\, r;\frac{3^{1/4}}{2^{1/2}}\right)\right.\\
&\quad\quad \left. -\mu e^{-\mathrm{i}\pi/3} \mathcal{F}\left(\frac{3^{1/4}}{2^{1/6}}\sqrt{\mu}e^{-\mathrm{i}\pi/6}\, r;\frac{3^{1/4}}{2^{1/2}}
\right)
\right\}\ .
\\
\end{split}
\end{equation*}
Introducing the distance profile
\begin{equation*}
\rho(r)=\lim_{n\to \infty} n^{1/6} \frac{[g^n] S^{(\lfloor n^{1/6}r\rfloor)}_{u_{\crit}}(g)}{[g^n] S_{u_{\crit}}(g)}\ , 
\end{equation*}
we obtain
\begin{equation*}
\rho(r)=\frac{d\, }{dr}\Phi(r)
\end{equation*}
with 
\begin{equation*}
\begin{split}
\Phi(r) =\frac{3}{\Gamma\left(\frac{4}{3}\right)}
\int_0^\infty e^{-\mu^3} d\mu\, \mu^3 
& \left\{e^{-\mathrm{i}\pi/6} \mathcal{F}\left(\frac{3^{1/4}}{2^{1/6}}\sqrt{\mu}e^{\mathrm{i}\pi/6}\, r;\frac{3^{1/4}}{2^{1/2}}\right)
\right.\\ &
\quad\quad\quad +\left. e^{\mathrm{i}\pi/6} \mathcal{F}\left(\frac{3^{1/4}}{2^{1/6}}\sqrt{\mu}e^{-\mathrm{i}\pi/6}\, r;\frac{3^{1/4}}{2^{1/2}}
\right)
\right\}\ .\\
\end{split}
\end{equation*}

After some simple manipulations, we end up with the expression \eqref{eq:secphiexp}:
\begin{equation*}
\begin{split}
\Phi(r) &=\frac{3}{\Gamma\left(\frac{4}{3}\right)}\int_0^\infty e^{-\mu^3} d\mu\, \mu^3 
\left\{1-6\frac{1-\co(\mu r^2)\, \coh(\mu r^2)+\frac{1}{\sqrt{3}}\si(\mu r^2)\, \sih(\mu r^2)}{\left(\co(\mu r^2)-\coh(\mu r^2)\right)^2}
\right\}
\\
& \hbox{with}\quad \co(x)=\cos\left(\frac{\sqrt{3x}}{2^{2/3}}\right) \ , \quad \coh(x)=\cosh\left(\frac{3\sqrt{x}}{2^{2/3}}\right) \ ,\\
& \phantom{with} \quad \si(x)=\sin\left(\frac{\sqrt{3x}}{2^{2/3}}\right) \ , \quad  \sih(x)=\sinh\left(\frac{3\sqrt{x}}{2^{2/3}}\right) \ .\\
\end{split}
\end{equation*}
As for $\rho(r)$, we obtain by differentiation its expression \eqref{eq:secrhoexp}:
\begin{equation*}
\rho(r) =\frac{3^2 2^{7/3}}{\Gamma\left(\frac{4}{3}\right)}\int_0^\infty e^{-\mu^3} d\mu\, \mu^{7/2} \sih(\mu r^2)
\frac{\co(\mu r^2)\left(\co(\mu r^2)+ \coh(\mu r^2)\right)-2}{\left(\co(\mu r^2)-\coh(\mu r^2)\right)^3}
\ .
\end{equation*}

%\begin{figure}
 % \centering
 % \fig{.65}{figure.pdf}
%  \caption{\small Caption}
%\end{figure}
%\pagebreak
\begin{table} %[htbp]
\end{table} %[htbp]
\newpage
\bibliographystyle{plain}%unsrt,alpha,plain}
\bibliography{duality}
\end{document}